\documentclass[11pt,a4paper]{article}
\pdfoutput=1
\usepackage{amssymb,amsmath,amsfonts, mathtools, mathrsfs}
\usepackage[utf8]{inputenc} 
\usepackage[dvipsnames]{xcolor}
\usepackage{braket}

\newif\ifnatbibsort\natbibsorttrue

\relax

\ifnatbibsort\RequirePackage[numbers,sort&compress]{natbib}\else\RequirePackage[numbers,compress]{natbib}\fi
\RequirePackage[colorlinks=true
  ,urlcolor=blue
  ,anchorcolor=blue
  ,citecolor=blue
  ,filecolor=blue
  ,linkcolor=blue
  ,menucolor=blue
  ,pagecolor=blue
  ,linktocpage=true
  ,pdfproducer=medialab
  ,pdfa=true
]{hyperref}

\usepackage{graphicx}
\usepackage{caption}
\usepackage{subfigure}
\usepackage{enumerate}
\usepackage{dsfont}
\usepackage{verbatim}
\usepackage{amsfonts,euscript,amssymb,braket}
\usepackage{tikz}
\usetikzlibrary{arrows,decorations.markings,patterns}
\usepackage{slashed}

\def\clock{{\count0=\time
    \divide\count0 60
    \ifnum\count0<10 0\fi\the\count0
    \multiply\count0 -60 \advance\count0 \time
    :\ifnum\count0<10 0\fi \the\count0
}}
\newcommand{\timestamp}{{\small\vbox{\hbox{\tt\jobname.tex}
      \hbox{\the\day/\the\month/\the\year, \clock}}}}

\newcommand{\bea}{\begin{eqnarray}}
\newcommand{\eea}{\end{eqnarray}}

\def\be{\begin{equation}}
\def\ee{\end{equation}}
\def\bea{\begin{eqnarray}}
\def\eea{\end{eqnarray}}
\def\Tr{{\rm Tr}}

\def\ket|#1>{| #1 \rangle}
\def\bra<#1|{\langle #1 |}
\def\<{\langle}
\def\>{\rangle}
\def\{{\lbrace}
\def\}{\rbrace}
\def\({\left(}
\def\){\right)}

\newcommand{\aver}[1]{\langle #1 \rangle}

\newcommand{\Zbb}{\mathbb{Z}}

\newcommand{\zb}{\bar{z}}

\def\barray{\begin{eqnarray}}
\def\earray{\end{eqnarray}}
\def\beq{\begin{equation}}
\def\eeq{\end{equation}}

\makeatletter
\let\old@startsection=\@startsection
\let\oldl@section=\l@section
\renewcommand{\@startsection}[6]{\old@startsection{#1}{#2}{#3}{#4}{#5}{#6\mathversion{bold}}}
\renewcommand{\l@section}[2]{\oldl@section{\mathversion{bold}#1}{#2}}
\makeatother

\numberwithin{equation}{section}

\newcommand{\mathcalJ}{\mathcal{J}}

\usepackage{color}

\def\ri {{\rm i}}

\def\e {{\rm e}}

\begin{document}
\renewcommand{\thefootnote}{\arabic{footnote}}

\overfullrule=0pt
\parskip=2pt
\parindent=12pt
\headheight=0in \headsep=0in \topmargin=0in \oddsidemargin=0in

\vspace{ -3cm} \thispagestyle{empty} \vspace{-1cm}
\begin{flushright} 
  \footnotesize
  \textcolor{red}{\phantom{print-report}}
\end{flushright}

\begin{center}
  \vspace{.0cm}

  {\Large\bf \mathversion{bold}
    Entanglement entropies of an interval
  }
  \\
  \vspace{.25cm}
  \noindent
      {\Large \bf \mathversion{bold}
	for the massless scalar field 
	in the presence of a boundary
      }

      \vspace{0.8cm} {
	Benoit Estienne$^{\,a}$,
	Yacine Ikhlef$^{\,a}$,
	Andrei Rotaru$^{\,a}$
	and
    	Erik Tonni$^{\,b}$
      }
      \vskip  0.7cm
      
      \small
	  {\em
	    $^{a}\,$Sorbonne Universit\'{e}, CNRS, Laboratoire de Physique Th\'{e}orique et Hautes \'{E}nergies, LPTHE, F-75005 Paris, France 
	    \\
	      \vspace{.25cm}
	    $^{b}\,$SISSA and INFN Sezione di Trieste, via Bonomea 265, 34136, Trieste, Italy 
	  }
	  \normalsize

\end{center}

\vspace{0.3cm}
\begin{abstract} 

  We study the entanglement entropies of an interval for the massless compact boson 
  either on the half line or on a finite segment, 
  when either Dirichlet or Neumann boundary conditions are imposed.
  In these boundary conformal field theory models,
 the method of the branch point twist fields is employed to obtain analytic expressions
 for the two-point functions of twist operators. In the decompactification regime, 
these analytic predictions in the continuum are compared with the lattice numerical results 
in massless harmonic chains for the corresponding entanglement entropies, finding good agreement.  
The application of these analytic results 
in the context of quantum quenches
is also discussed.

\end{abstract}

\newpage
\tableofcontents

\newpage
\section{Introduction}
\label{sec_intro}

Entanglement is a fascinating feature of quantum systems that is attracting the interest of a growing number of researchers from various fields of theoretical physics, 
including quantum gravity, condensed matter and quantum information 
(for further reading, we refer to the reviews \cite{Calabrese:2009qy, EislerPeschel:2009review, Casini:2009sr, Eisert:2008ur, Rangamani:2016dms}). 
The presence of  physical boundaries influences the features of entanglement in highly non-trivial ways 
that have been explored in various contexts gaining important insights
\cite{Calabrese:2004eu, Schollwok:2006aaa, Affleck:2009aa, Berthiere:2016ott,Taddia:2013txa,Affleck:1995ge,Laflorencie:2005duh}.
For instance, for the holographic entanglement entropy \cite{Ryu:2006bv, Ryu:2006ef}
the analysis of the boundary effects 
\cite{Karch:2000ct, Takayanagi:2011zk, Fujita:2011fp, Nozaki:2012qd} 
has recently provided a novel approach to the information paradox 
\cite{Penington:2019npb, Almheiri:2019psf, Almheiri:2019hni, Almheiri:2019qdq, Penington:2019kki}. In the context of topological phases of matter, the entanglement entropy can be used to detect and identify edge modes at a boundary \cite{Estienne:2019hmd,Estienne:2021qqe} or more generally at an interface \cite{Crepel:2018ycz,Crepel:2019gvb}.

The bipartite entanglement for a spatial bipartition can be studied by considering 
a quantum system whose space is bipartite into a region $A$ and its complement $B$ and 
whose Hilbert space $\mathcal{H}$ can be factorised accordingly as $\mathcal{H} = \mathcal{H}_A \otimes \mathcal{H}_B$.
Denoting by $\rho$ the density matrix of the system, the reduced density matrix $\rho_A\equiv \Tr_B \,\rho$ for the subsystem $A$ 
is introduced by taking its partial trace over $\mathcal{H}_B$.
When the system is in a pure state $|\Psi\rangle$, and therefore $\rho = | \Psi \rangle \langle \Psi |$, 
it is well known that  the entanglement entropy
\be
\label{ee-def-intro}
S_A \equiv  -\, \Tr \big(\rho_A  \log \rho_A\big)
\ee
is the unique measure of the bipartite entanglement.
It satisfies $S_A = S_B$ and all the other properties characterising an entanglement measure \cite{Plenio:2007zz, Hollands:2017dov}.
In order to evaluate (\ref{ee-def-intro}), it is often convenient to introduce the R\'enyi entropies 
\be
\label{renyi-ent-def-intro}
S_A^{(n)} \equiv  \frac{1}{1-n} \, \log \! \big( \textrm{Tr} \, \rho_A^n \big)
\ee
where the R\'enyi index $n$ takes integer values $n \geqslant 2$.
Since the normalization condition $\textrm{Tr} \, \rho_A = 1$ is assumed, 
by performing the analytic continuation $n \to 1$,
one finds $S_A^{(n)} \to S_A$,
and this is also known as the replica limit in the context of entanglement. 
The entanglement entropy (\ref{ee-def-intro}) and  the moments $\textrm{Tr} \, \rho_A^n$ of the reduced density matrix $\rho_A$ 
can be evaluated more directly also through the entanglement spectrum, i.e. the multiset made by the eigenvalues $ \lambda_j  \in (0,1)$ of 
the reduced density matrix. 
The largest eigenvalue $\lambda_{\textrm{\tiny max}}$ of the entanglement spectrum provides 
the single copy entanglement $S_A^{(\infty)}$ \cite{peschel-05,eisert-05,orus-06},
which can be obtained also as the limit $n \to \infty$ of the Rényi entropies
\be
\label{single copy-def-intro}
S_A^{(\infty)} \equiv  - \log(\lambda_{\textrm{\tiny max}}) = \lim_{n \to \infty} S_A^{(n)}\,.
\ee
The entanglement entropies correspond to the quantities (\ref{ee-def-intro}), (\ref{renyi-ent-def-intro}) and (\ref{single copy-def-intro}).

The entanglement entropies are usually evaluated 
through the replica method 
\cite{Callan:1994py, Holzhey:1994we, Calabrese:2004eu}, 
which naturally leads to write the moments of $\rho_A$ as follows
\be
\label{intro-moments-qft}
\textrm{Tr} \,\rho_A^n 
= 
\frac{\mathcal{Z}_n}{\mathcal{Z}^n}
\ee
where $\mathcal{Z}$ is the partition function of the model and $\mathcal{Z}_n$ is
the partition function of the same model defined on a  $n$-sheeted branched covering.
The determination of analytic expressions for (\ref{intro-moments-qft}) in quantum field theories
is typically a difficult task.

Focussing on critical systems in one spatial dimension, 
we can employ the powerful framework of Conformal Field Theory (CFT) \cite{Belavin:1984vu} 
and of Boundary Conformal Field Theory (BCFT) 
\cite{Cardy:1984bb, Cardy:1986gw, Cardy:1989ir, Cardy:1991tv, Cardy:2004hm}
whenever physical boundaries occur. 
In \cite{Calabrese:2004eu, Cardy:2007mb}, it has been shown that
the branch point twist fields \cite{Dixon:1986qv, Zamolodchikov:1987ae, Bershadsky:1986fv, Knizhnik:1987xp}
provide an insightful tool to investigate entanglement in quantum field theories. 
For instance, in one spatial dimension the R\'enyi entropies of an interval in the infinite line
is given by the two-point function of twist fields located at the endpoints of the boundary. 
For a CFT on the line and in its ground state, this two-point function depends only on the central charge of the model. 
For a BCFT on the half line and in its ground state,  
the R\'enyi entropies of an interval adjacent to the boundary
are given by the one-point function of a twist field \cite{Calabrese:2004eu}
and they encode also the boundary entropy introduced by Affleck and Ludwig \cite{Affleck:1991tk},
which characterises the boundary condition (b.c.) through the conformal boundary state.
This boundary entropy allows to explore the boundary renormalisation group flows
induced by the change of boundary conditions \cite{Friedan:2003yc, Casini:2016fgb}. 

The entanglement entropies of more complicated bipartitions encode 
more detailed information about the underlying quantum field theory. 
A paradigmatic example is given by the R\'enyi entropies of two disjoint intervals 
for a CFT on a line and in its ground state, 
which can be expressed as the four point function of twist fields, 
or equivalently as the partition function on a genus $n-1$ Riemann surface 
obtained as a  $n$-sheeted branched covering of the sphere through \eqref{intro-moments-qft}. 
The entanglement quantifiers encompass a significant amount of CFT data beyond the central charge 
\cite{Furukawa:2008uk, Caraglio:2008pk, Casini:2009vk, Calabrese:2009ez, Calabrese:2010he, 
Alba:2009ek, Fagotti:2010yr, Coser:2013qda, Hartman:2013mia, Coser:2015dvp, Grava:2021yjp}.
Some analytic results have been obtained
and they have been also checked through 
various numerical analysis in the proper lattice models.
The replica limit $n \to 1$ of these analytic expressions is typically very difficult;
hence the numerical extrapolation method discussed in \cite{Agon:2013iva, DeNobili:2015dla}
can be useful.

\begin{figure}[t!]
\vspace{-0cm}
\hspace{-.85cm}
  \includegraphics[width=1.1\textwidth]{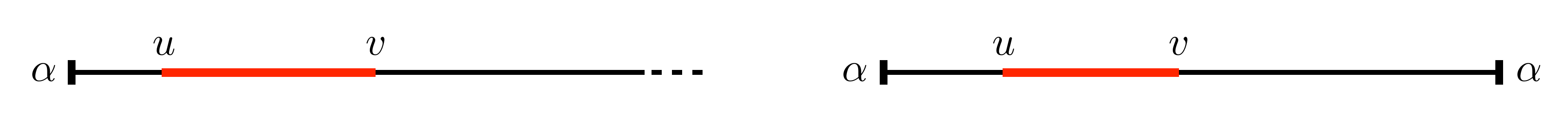}
\vspace{-.5cm}
\caption{Spatial bipartitions considered in this manuscript: the interval $A$ (red segment)
is either on the half line (left panel) or on the segment of finite length $L$ (right panel),
where the b.c. is labelled by $\alpha$. 
The same b.c. is imposed at both the endpoints of the segment. 
}
\label{figure-biparts-intro}
\end{figure}

All the challenges and characteristic features arising in a translation invariant space 
when the subsystem $A$ is made by the union of two disjoint regions
also occur when the model has a physical boundary and its
spatial bipartition is given by a region not adjacent to it. 
In the case of a BCFT on a segment where the same conformally invariant b.c. has been imposed 
at both its endpoints (or on the half line) and in its ground state, 
the R\'enyi entropies of an interval located at finite distance from the boundary (see Fig.\,\ref{figure-biparts-intro}) 
can be expressed as either the  partition function on a sphere with $n$ disks removed 
or as a two-point function of twist fields on the unit disk. 
These quantities probe a significant amount of BCFT data beyond the central charge and the boundary entropy.
For instance, the R\'enyi entropy  with $n=2$ encodes the entire annulus partition function \cite{Estienne:2021xzo}. 
However, for $n>2$ very few results are available in the literature. 
The case of an interval for the free massless Dirac field on the half line, which is the prototypical fermionic BCFT with $c=1$,
has been analyzed in \cite{Mintchev:2020uom} (see  \cite{Fagotti:2010cc} for a lattice computation) 
and later extended to an arbitrary number of intervals \cite{Rottoli:2022plr}. 
Another important BCFT with $c=1$ to explore is the massless compact boson either on the half line or on the segment. 
For this model, 
the R\'enyi entropies of an interval in the segment have been investigated in \cite{Bastianello:2019yyc},
where, in the case of Dirichlet b.c., 
an implicit expression has been found and evaluated numerically.
This result has been checked through lattice calculations in the XXZ chain \cite{Bastianello:2019ovv}.
In the same setup,  explicit analytic BCFT expressions for both Dirichlet b.c. and Neumann b.c. 
have been obtained for the special case of $n=2$ in \cite{Estienne:2021xzo}.

In this manuscript we consider the compact massless scalar field either on the half line or on the segment
where the same b.c. are imposed at both its endpoints.
Both Dirichlet b.c. and Neumann b.c. are investigated. 
By combining the twist field method \cite{Dixon:1986qv, Calabrese:2009ez}
with some results involving the branched covering of Riemann surfaces
\cite{Schnitzer:1986fi,Alvarez-Gaume:1986rcs,Alvarez-Gaume:1987wwg,Ginsparg:1986wr,Hamidi:1986vh,Dijkgraaf:1987vp,BERNARD1988251,ATICK19881},
we obtain analytic expressions for the R\'enyi entropies of an interval which is not adjacent to the boundary.
In the special case of $n=2$, the results of \cite{Estienne:2021xzo}  are recovered.
Furthermore, in the case of Dirichlet b.c., we have checked numerically that our expression agrees 
with the one obtained in \cite{Bastianello:2019yyc}.
In the decompactification limit, we have compared our analytic results with 
the corresponding numerical outcomes obtained for the harmonic chains in the massless regime,
finding a good agreement. 
Finally, the application of these analytic results
in the context of the BCFT approach to the quantum quenches developed in
\cite{Calabrese:2005in, Calabrese:2007mtj} has been also discussed.

The outline of this manuscript is as follows. 
Explicit formulas for the R\'enyi entropies are reported and discussed in Sec.\,\ref{sec_cft_results}.
In Sec.\,\ref{section_CFT_disk}  the BCFT computations underlying these results are described. 
 Sec.\,\ref{sec_HC_numerics} contains a comparison of our analytical results in the decompactification limit with the 
 lattice data obtained numerically for the corresponding semi-infinite and finite harmonic chains in the massless regime. 
In  Sec.\,\ref{sec-quenches} the application of our analytic expressions within the BCFT approach to the quantum quenches is discussed. 
In Sec.\,\ref{sec:conclusions} we summarise our results and mention some future directions. 
The Appendices\;\ref{subsec-XXZ-insight}, \ref{app_period_matrix}, \ref{app_quantum_part}, \ref{app-limits} and \ref{app-quenches}  
contain technical details and further discussions.

\section{BCFT expressions for the entanglement entropies}
\label{sec_cft_results}

In this section we discuss the procedure of our BCFT calculation and describe the main analytic results,
both at finite compactification radius $R$ and in the decompactification limit $R\rightarrow\infty$.

\subsection{Single interval either on the half line or on the segment}
\label{sec_main_BCFT_expression}

We are interested in calculating the entanglement entropy of an interval 
for a one dimensional quantum critical system in its ground state 
defined either on a segment of finite  length $L$ or on the half line.
In the latter case, we restrict to models where the same b.c. are  imposed at both ends of the system
and its limit $L \to \infty$ provides the former one. 
The critical point is assumed to be governed by a BCFT and  
we denote by $\alpha$ the label identifying the allowed conformally invariant boundary conditions.

We investigate the spatial bipartition of the system given by an interval $A=(u,v)$ and its complement 
when $A$ is not adjacent to the boundary,
namely when $0<u<v<L$, as shown in Fig.\,\ref{figure-biparts-intro}.  
The moments of the reduced density matrix $\rho_A$ read 
\be
\label{Tr-n-segment-00}
M_{A; \,\alpha}^{(n)} 
\equiv
\textrm{Tr} \, \rho_A^n 
=\,
  \langle  \tau_n(u)\,\tau^{\dag}_n(v) \rangle
  \ee
where $\tau_n$ and $\tau^{\dag}_n$ stand for the microscopic twist operators, which are defined e.g. on a lattice. 
In the scaling limit, the lattice operator $\tau_n$ can be expanded 
into a linear combination of scaling fields in the corresponding CFT model
and the most relevant among these primaries is the twist operator  $\mathcal{T}_n$,
a spinless primary field with scaling dimension \cite{Calabrese:2004eu}
\be
\label{Delta_n def}
\Delta_n = \frac{c}{12} \left(n- \frac{1}{n} \right) .
\ee
Hence \cite{Cardy:2007mb,Castro-Alvaredo:2009yqb,Estienne:2020txv}
 \be
 \label{eq_scaling_tau}
\tau_n  = C_n \, \epsilon^{\Delta_n}\, \mathcal{T}_n + \dots 
\ee
where the dots correspond to less relevant fields 
whose contribution to $\textrm{Tr} \, \rho_A^n$ matters when finite size corrections are taken into account. 
The prefactor $C_n$ is a non universal constant coming from the normalization of the microscopic operator $\tau_n$
and $\epsilon$ is a UV cutoff, like e.g. the lattice spacing.
Thus, the moments of the reduced density matrix $M_{A; \,\alpha}^{(n)} $ in (\ref{Tr-n-segment-00}) can be written as  
\be
\label{Tr-n-segment-0}
M_{A; \,\alpha}^{(n)} 
=\,
C^2_n \, \epsilon^{2\Delta_n}\,  \langle  \mathcal{T}_n(u)\,\mathcal{T}^{\dag}_n(v) \rangle_{_{\mathbb{S}}} 
\ee
where $\mathbb{S} = \{ w  \in \mathbb{C}\,; \, 0< \textrm{Re}(w) < L \}$ is the infinite strip (with imaginary time running along the imaginary axis) 
of width $L$ with the same boundary condition $\alpha$ imposed on both sides of the strip. 
Because of conformal invariance, the two-point function on the strip 
can be written in the following form
\be
\label{Tr-n-segment}
M_{A; \,\alpha}^{(n)} 
\equiv
\textrm{Tr} \, \rho_A^n 
=\,
C^2_n \; \frac{\mathcal{F}_n^{(\alpha)}( r ) }{\mathcal{P}(u,v)^{\Delta_n} }
\ee
where 
\be
\label{tilde-r-ratio-def}
\mathcal{P}(u,v)
\equiv
\frac{  
s(2u)\, s(2v)\, s(v-u)^2
}{
\epsilon^2 \, s(v+u)^2
} 
\;\;\qquad\;\;
r \equiv 
\left( \frac{s(v-u)}{s(v+u)} \right)^2
\qquad
s(w) \equiv \frac{2L}{\pi} \, \sin \! \left( \frac{\pi w }{2L} \right) 
\ee
and $\mathcal{F}_n^{(\alpha)}$ depends on the specific BCFT model, 
which is characterised also by the boundary conditions labelled by $\alpha$
and it is related to the partition functions on the Riemann sphere with $n$ boundary components. 
In particular, $\mathcal{F}_2^{(\alpha)}(r)$ is given by the partition function of the BCFT model on the annulus \cite{Estienne:2021xzo}.

From the moments (\ref{Tr-n-segment}), one obtains  the R\'enyi entropies (\ref{renyi-ent-def-intro})
of an interval $A$ for a BCFT on the strip $\mathbb{S} $
\be
\label{SA-n-BCFT}
S_{A;\alpha}^{(n)} 
= \,
\frac{\Delta_n}{n-1}\, \log \!\big[\mathcal{P}(u,v)\big]
+  
\frac{\log \! \big[ C^2_n\,\mathcal{F}_n^{(\alpha)}(r) \big]}{1-n}
\ee
whose analytic continuation $n\rightarrow 1$
provides the corresponding entanglement entropy (\ref{ee-def-intro}), which reads
\begin{equation}
\label{eq:SA-1-intro-definition}
    S_{A;\alpha}=\frac{c}{6}\log \!\big[\mathcal{P}(u,v)\big]+2 \,C'_1 + \mathcal{G}^{(\alpha)}_1(r)
\end{equation}
where we have defined $C'_1\equiv-\lim_{n\rightarrow 1} \frac{\log C_n}{n-1}$ and
\begin{equation}\label{eq:mathcalG-1-intro-definition}
\mathcal{G}^{(\alpha)}_1(r)
\equiv 
- \lim_{n\rightarrow 1} \partial_n \left(\log \mathcal{F}_n^{(\alpha)}(r) \right)
\end{equation}
while the corresponding single copy entanglement (\ref{single copy-def-intro}) is given by
\be
\label{SA-infty}
S_{A;\alpha}^{(\infty)} 
= \,
\frac{c}{12} \log \!\big[\mathcal{P}(u,v)\big]
+
\lim_{n \to \infty} \frac{\log \! \big[ C^2_n\,\mathcal{F}_n^{(\alpha)}(r) \big]}{1-n}\,.
\ee

Instead, when  $A $ is adjacent to the boundary, i.e. $u=0$, we have \cite{Calabrese:2004eu} 
\be
\label{SA-adjacent}
\textrm{Tr} \, \rho_A^n = C_n \, \epsilon^{\Delta_n}\, \frac{ g_{\alpha}^{1-n}  }{ s(2v) ^{\Delta_n}}
\ee
where $g_{\alpha}$ is related to the Affleck-Ludwig boundary entropy  \cite{Affleck:1991tk},
that plays a crucial role in the analysis of the boundary renormalisation group flows \cite{Friedan:2003yc, Casini:2016fgb}.

The expressions (\ref{Tr-n-segment-0}) and (\ref{Tr-n-segment}) for the moments of the reduced density matrix 
naturally lead to introduce two kinds of ratios that are UV finite.  
A first type of ratios can be defined for an assigned boundary condition $\alpha$ 
as follows
     \be
     \label{R-ratio-def}
     R_{A;\,\alpha}^{(n)}
     \equiv
     \frac{M_{A;\,\alpha} }{M_{A_u ; \,\alpha} \; M_{A_v ; \,\alpha}}
     \ee
     where $A_u \equiv [0,u]$ and $A_v \equiv [0,v]$
     (see Fig.\,\ref{figure-biparts-intro}).
    For the BCFT we are considering, 
     by using the expressions in (\ref{Tr-n-segment})-(\ref{tilde-r-ratio-def}), this ratio becomes
     \be
     \label{R-ratios-cft-def}
     R_{A;\,\alpha}^{(n)}
                    =
         g_{\alpha}^{2(n-1)} \, \frac{ \mathcal{F}_n^{(\alpha)}(r) }{ r^{\Delta_n} } \,.
     \ee
     From (\ref{R-ratio-def}) and (\ref{renyi-ent-def-intro}), the following UV finite combination of entanglement entropies
     can be introduced
     \be
     \label{MI-n-bdy-def}
     \mathcal{I}_{A;\alpha}^{(n)} \equiv S^{(n)}_{A_u;\alpha} + S^{(n)}_{A_v;\alpha}- S^{(n)}_{A;\alpha} \,.
     \ee
    For a BCFT on a segment, by using (\ref{SA-n-BCFT}), one finds that this UV finite combination becomes
          \be
          \label{MI}
     \mathcal{I}_{A;\alpha}^{(n)}
     =
      - \frac{c}{12} \left(1+\frac{1}{n} \right)\log (r)
     +
     \frac{\log \! \big[\mathcal{F}_n^{(\alpha)}(r) \big]}{n-1}
     +
   2 \log g_{\alpha}
     \ee
     which is a function of the harmonic ratio $r$ in \eqref{tilde-r-ratio-def}. 

     Another type of UV finite ratios can be defined only through the interval $A$, 
     without employing the entanglement entropies of intervals adjacent to the boundary;
     but it needs two different conformally invariant boundary conditions $\alpha_1$ and $\alpha_2$.
     From (\ref{Tr-n-segment-0}) and (\ref{Tr-n-segment}), 
     these ratios are 
     \be
     \label{M-over-M}
     \frac{M_{A; \,\alpha_1}^{(n)} }{ M_{A; \,\alpha_2}^{(n)} } 
     =
     \frac{\mathcal{F}_n^{(\alpha_1)}(r) }{ \mathcal{F}_n^{(\alpha_2)}(r) }
     \ee
     which naturally lead to the following difference of R\'enyi entropies
     \be
     \label{renyi-differences}
     S_{A; \alpha_1}^{(n)} - S_{A; \alpha_2}^{(n)} 
     =\,
    \frac{1}{1-n} \, \log\! \left(      \frac{\mathcal{F}_n^{(\alpha_1)}(r) }{ \mathcal{F}_n^{(\alpha_2)}(r) }\right)
     \ee
     where we have denoted by $S_{A; \alpha}^{(n)} $ the R\'enyi entropies (\ref{renyi-ent-def-intro}),
     in order to highlight its dependence on the boundary condition $\alpha$. 
     We remark that the above expressions hold for a BCFT on a segment and 
     that the corresponding ones for the BCFT on the half line 
     are obtained by taking $L \to \infty$.

The CFT quantities we are considering are related to the the two-point functions of twist fields
$\langle  \mathcal{T}_n(u)\, \mathcal{T}^{\dag}_n(v) \rangle_{_{\mathbb{S}}}$ 
on the strip $\mathbb{S} \equiv \{ w  \in \mathbb{C}\,; \, 0< \textrm{Re}(w) < L \}$.
Since the strip is conformally equivalent to the unit disk 
$\mathbb{D} \equiv \big\{ z \in \mathbb{C}\, ; |z|  \leqslant 1 \big\}$ via the map
$w  \mapsto z=\frac{s(w-u)}{s(w+u)}$, 
we have
\begin{equation} 
\label{eq:maptodisc}
    \langle  \mathcal{T}_n(u)\,\mathcal{T}^{\dag}_n(v) \rangle_{_{\mathbb{S}}}
  = 
  \,     \frac{ \langle  \mathcal{T}_n(0)\,\mathcal{T}^{\dag}_n(x) \rangle_{_\mathbb{D}} }{ s(u+v)^{2\Delta_n} }
  \;\; \qquad \;\;  
  x =  \frac{s(v-u)}{s(u + v )} 
\end{equation}
where the same boundary condition imposed on both the boundaries of $\mathbb{S}$
holds also on the boundary of $\mathbb{D}$. 
In the case of the BCFT on the  half line, 
we have to consider the right half plane $\mathbb{H} \equiv \big\{ z \in \mathbb{C}\, ; \textrm{Re}(z) \geqslant 0 \big\}$ instead of the strip.
The corresponding two-point functions of twist fields
can be studied by taking $L \to \infty$ in \eqref{eq:maptodisc}, finding
\begin{equation} 
\label{eq:rhpmaptodisk}
    \langle  \mathcal{T}_n(u)\,\mathcal{T}^{\dag}_n(v) \rangle_{_{\mathbb{H}}}
    =
  \, \frac{ \langle  \mathcal{T}_n(0)\,\mathcal{T}^{\dag}_n(x) \rangle_{_\mathbb{D}} }{ (u+v )^{2\Delta_n} }
  \;\;\; \qquad  \;\;\;  
  x =  \frac{v-u}{v+ u } \,. 
\end{equation}
Thus, the R\'enyi entropies we are interested in can be obtained from the two-point function of twist fields 
$\langle  \mathcal{T}_n(0)\,\mathcal{T}^{\dag}_n(x) \rangle_{_\mathbb{D}} $ on the unit disk, with $0<x<1$.
This two-point function is equivalent to the following ratio of partition functions \cite{Calabrese:2004eu}
\begin{equation}
 \langle  \mathcal{T}_n(0)\,\mathcal{T}^{\dag}_n(x) \rangle_{_\mathbb{D}} =   \mathcal{Z}_n(x)/\mathcal{Z}_1^n 
\end{equation}
where $\mathcal{Z}_n(x)$ is the BCFT partition function on the $n$-sheeted covering of the unit disk with branch points at $0$ and $x$, 
while $\mathcal{Z}_1$ is simply the BCFT partition function on the unit disk.

Few explicit analytic expressions for $\mathcal{F}_n^{(\alpha)}$ in (\ref{Tr-n-segment}) are available in the literature:
for the massless Dirac field, where $\mathcal{F}_n^{(\alpha)} = 1$ identically for any value of $n$ 
and independently of the (conformally invariant) boundary conditions \cite{Mintchev:2020uom}, 
and for the massless compactified scalar when $n=2$ and a generic conformally invariant boundary condition 
 \cite{Estienne:2021xzo}, 
in terms of the annulus partition function \cite{Estienne:2021xzo}. 
In this manuscript we extend the latter result for the massless compactified scalar to a generic value of the R\'enyi index $n$,
focussing on Dirichlet and Neumann boundary conditions.

\subsection{Entanglement entropies for the compact boson}
\label{sec_main_CFT_results_boson}

We consider a gapless one dimensional quantum system belonging to the Luttinger liquid universality class, 
whose low energy behaviour is captured by the massless compact real boson, which is a specific CFT with $c=1$.
A prototypical example is the spin-$\tfrac{1}{2}$ XXZ spin chain (see Appendix\,\ref{subsec-XXZ-insight}).
The Euclidean action of the massless compact real boson
on a generic Riemann surface $\mathcal{M}$ equipped with metric $g_{\mu\nu}$
whose target space is a circle of radius $R$ reads
\be
\label{action-boson}
  S[\phi] \,
  =  
  \frac{1}{8\pi}  \int_{\mathcal{M}}   g^{\mu \nu} \, \partial_{\mu} \phi \, \partial_{\nu} \phi\,  \sqrt{|g|} \,d^2x 
  \qquad 
  \phi \sim \phi + 2\pi R \,.
\ee
Considering the BCFT given by this model either on the half line or on the segment 
with the same b.c. at both its endpoints,
the main results of this manuscript are the analytic expressions for $\mathcal{F}_n^{(\alpha)}$ 
when $\alpha$ corresponds to either Dirichlet (D) or Neumann (N) b.c.,
which are respectively
\be
\label{Fn-main-res}
    \mathcal{F}_n^{\textrm{\tiny (D)}}(r)  = \frac{\Theta\big(\boldsymbol{\tau}(r)/R^2\big)}{\Theta\big(\boldsymbol{\tau}(r)\big)} 
    \;\;\;\; \qquad \;\;\;\;
    \mathcal{F}_n^{\textrm{\tiny (N)}}(r)  = \frac{\Theta\big(R^2\, \boldsymbol{\tau}(r)/4\big)}{\Theta\big(\boldsymbol{\tau}(r)\big)} 
\ee
where $ r \in (0,1)$, in terms of  the Siegel theta function defined as follows
\be
\label{siegel-def}
\Theta(\boldsymbol{h}) \,\equiv \! 
\sum_{\boldsymbol{m} \in \mathbb{Z}^{g}} \!
\e^{\ri \pi \, \boldsymbol{m}^{\textrm{\tiny t}} \cdot \boldsymbol{h} \cdot \boldsymbol{m} }
\ee
which is a function of a $g \times g $ symmetric matrix $\boldsymbol{h}$ having positive definite imaginary part. 
In our analysis, 
the matrix $\boldsymbol{\tau}(r)$ in (\ref{Fn-main-res})
is the $(n-1) \times (n-1)$ period matrix found in \cite{Calabrese:2009ez},
which occurs in the entanglement entropies of two disjoint intervals on the infinite line 
for the massless compact scalar in its ground state.
Its matrix elements are
\be
\label{tau-matrix-element}
\boldsymbol{\tau}_{i,j}(r)
\equiv\,
\ri \, \frac{2}{n} \sum_{k=1}^{n-1} \sin(\pi k/n)\, \frac{F_{k/n}(1-r)}{F_{k/n}(r)} \, \cos\!\big[2\pi k (i-j)/n\big] 
\ee
where $1 \leqslant i,j \leqslant n-1$
and $F_{k/n}(y) \equiv \, _2F_1(k/n, 1-k/n;1;y)$ is a particular hypergeometric function. 
The results of \cite{Estienne:2021xzo} are recovered by specialising 
the expressions in (\ref{Fn-main-res}) to $n=2$.

\begin{figure}[t!]
\vspace{-.6cm}
\hspace{-1cm}
  \includegraphics[width=1.15\textwidth]{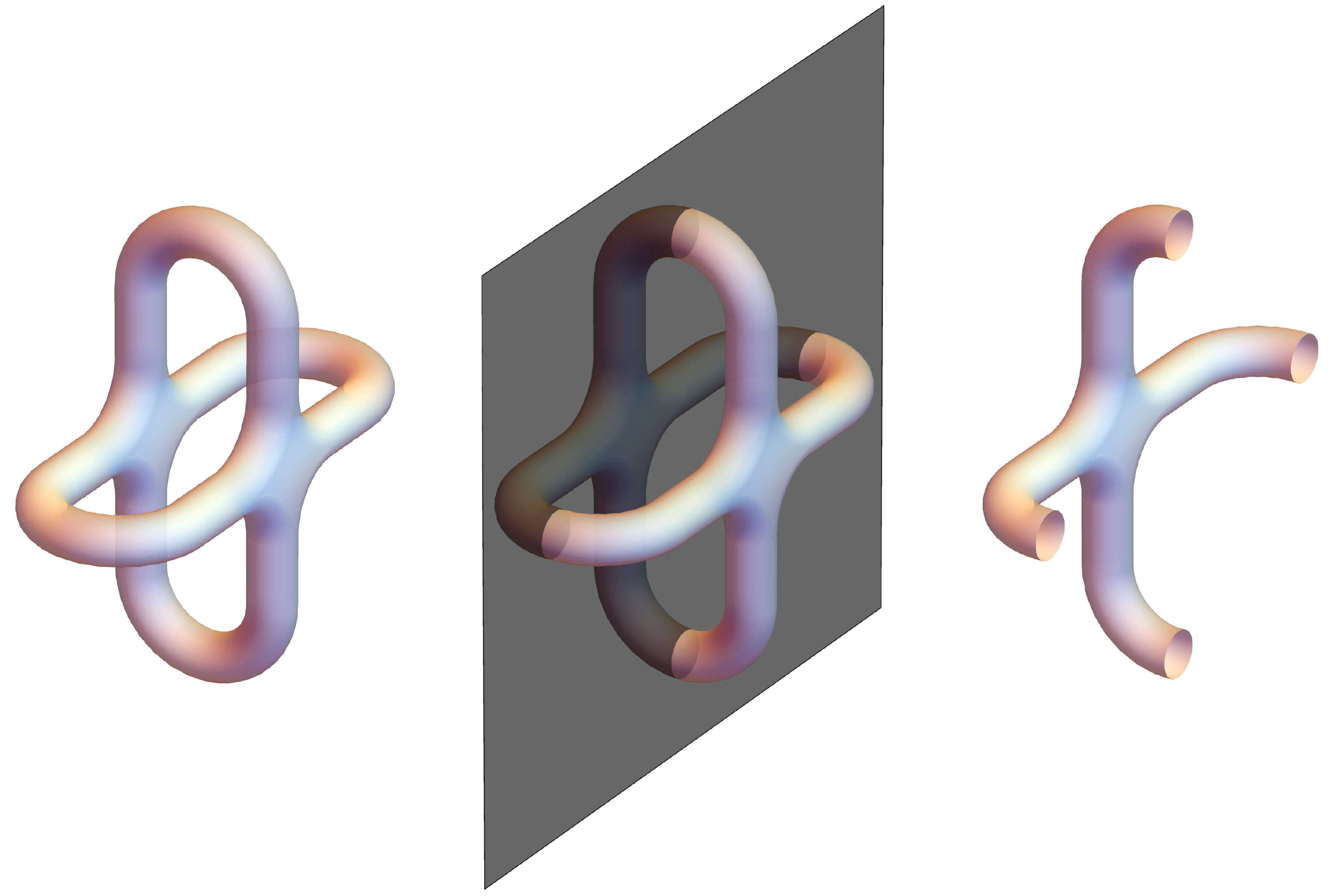}
\vspace{-.0cm}
\caption{
The Riemann surface $\mathscr{S}_4$ (right panel) is topologically a sphere with four boundaries.
It is obtained as one of the two halves of $\mathscr{M}_4$ for two equal intervals (left panel) 
determined by its reflection symmetry with respect to a plane (the black plane in the middle panel). 
}
\label{figure-riemann-surface-intro}
\end{figure}

When the bipartition on the infinite line is given by the union $A$ of two disjoint intervals \cite{Calabrese:2009ez}, 
the moments $\textrm{Tr} \, \rho_A^n$ are obtained as
the partition function of the model on  the $n$-sheeted Riemann surface $\mathcal{M} = \mathscr{M}_n$, 
which is a special Riemann surface with genus $g=n-1$ obtained through the replica construction
(see the Appendix\;A of \cite{Calabrese:2010he}).
For a CFT, this special Riemann surface 
is characterised by the harmonic ratio of the endpoints of the two intervals, 
which is a real parameter in $(0,1)$.
For instance, $\mathscr{M}_4$ has genus $g=3$ 
and it is shown in the left panel of Fig.\,\ref{figure-riemann-surface-intro} for the special case of two equal intervals.
Further analyses and generalisations of $\mathscr{M}_n$ have been discussed e.g. in 
\cite{Calabrese:2010he, Coser:2013qda, Grava:2021yjp}.

The method of the images allows to find the $n$-sheeted Riemann surface $\mathcal{M}=\mathscr{S}_n$ 
for a BCFT on the half line and in its ground state as follows. 
Consider $\mathscr{M}_n$ for two equal intervals 
(when $n=4$, see the left panel of Fig.\,\ref{figure-riemann-surface-intro}), 
which exhibits a $\mathbb{Z}_2$ symmetry (a reflection) with respect to a plane
(see the black plane in the middle panel of Fig.\,\ref{figure-riemann-surface-intro}).
The $n$-sheeted Riemann surface $\mathscr{S}_n$ 
corresponds to one of the two halves 
identified by this reflection plane, whose union gives $\mathscr{M}_n$.
Thus, $\mathscr{S}_n$ has the topology of a sphere with $n$ boundaries, 
which has genus $g=0$ and Euler characteristic $\chi = 2 -n$
(for $n=4$, see the right panel of Fig.\,\ref{figure-riemann-surface-intro}).

Upon examining (\ref{Tr-n-segment}), (\ref{tilde-r-ratio-def}), and (\ref{Fn-main-res}), 
one can readily observe that the BCFT expressions for the moments of the reduced density matrix 
remain unchanged when $u$ and $v$ are exchanged, which corresponds to the exchange of $\mathcal{T}_n$ with $\mathcal{T}_n^\dagger$. 
This symmetry exchange is a consistency check for our analytic results.

Furthermore, since the entire system is in a pure state, it is well know that $S_A^{(n)} = S_B^{(n)}$ for any value of $n$.
In our analysis, this is verified because the computation of the $S_A^{(n)}$ and of $S_B^{(n)}$
involve the same $n$-sheeted branched covering of the strip (or of the half plane).

By employing (\ref{intro-moments-qft}), (\ref{Tr-n-segment-0}), (\ref{Tr-n-segment}) and (\ref{Fn-main-res}), 
we observe that the following relation occurs between the partition functions corresponding to Dirichlet b.c. and Neumann b.c.
\be
\label{N-D-relation-Z}
\mathcal{Z}_n^{\textrm{\tiny (N)}}(R) = \mathcal{Z}_n^{\textrm{\tiny (D)}}(2/R) \,.
\ee
This provides a non trivial consistency check of the dependence on the compactification radius $R$ in (\ref{Fn-main-res});
indeed, for the compact massless boson this T-duality relation is expected to hold, 
as discussed in \cite{Blumenhagen:2009zz} for the annulus (i.e. for the $n=2$ case)
and in  \cite{Alvarez:1996up} for a generic Riemann surfaces with boundaries.

As a further consistency check, in the Appendix\;\ref{app-limits-compact} 
we recover the known result for $S_A^{(n)}$
when the interval $A$ is adjacent to the boundary \cite{Calabrese:2004eu} 
as a limiting case.

Summarising, for the massless compact scalar field on the segment with the same b.c. at its endpoints, which is a BCFT with $c=1$,
the moments of the reduced density matrix of an interval not adjacent to the boundary (see the right panel of Fig.\,\ref{figure-biparts-intro})
are given by (\ref{Tr-n-segment}) with $c=1$ and $\mathcal{F}_n^{(\alpha)}(r)$ replaced by 
the function in (\ref{Fn-main-res}) associated to the proper boundary condition.
The corresponding expressions for the interval in the half line (see the left panel of Fig.\,\ref{figure-biparts-intro})
can be obtained by taking the limit $L \to +\infty$.

The R\'enyi entropies for the massless scalar field in the case of Dirichlet b.c.
and the spatial bipartition in the right panel of Fig.\,\ref{figure-biparts-intro}
have been already studied in  \cite{Bastianello:2019yyc}, where implicit results have been found. 
This analysis has been developed further in \cite{Bastianello:2019ovv} for inhomogeneous systems. 
In order to apply the outcomes of these works, linear integral equations must be solved and,
since analytic solutions have not been found, approximate results can be obtained numerically
by discretising the interval $A$, as discussed in \cite{Bastianello:2019yyc}.
This provides an important benchmark for the analytic BCFT expression we obtain for Dirichlet b.c.:
we checked\footnote{We are grateful to Alvise Bastianello for having shared with us his Mathematica notebook 
for the numerical evaluation of the expression obtained in \cite{Bastianello:2019yyc}.} 
that it  is compatible with the one obtained numerically in \cite{Bastianello:2019yyc}.
In particular, for $n\leqslant 5$  and various sizes for the interval,
we found numerical agreement between 
the period matrix $\boldsymbol\tau$ in (\ref{tau-matrix-element}) and the matrix $\mathcal{M}$ of \cite{Bastianello:2019yyc},
once the difference in the notations has been taken into account.
Also the compatibility for the R\'enyi entropies 
when $n\leqslant 5$ and for various values of the compactification radius $R\in(0.5,3)$
has been checked. 
As $n$ increases, higher accuracy in the discretisation procedure is needed to find agreement with our BCFT result in the continuum.

\subsubsection{Decompactification regime}
\label{subsec-decom-regime}

The decompactification regime as defined as 
the limit of large compactification radius $R \to \infty$ and it corresponds to the case 
where the target space is the infinite line. Taking the limit $R \to \infty$ and rescaling by appropriate powers of $R$ to obtain a finite result, 
we find that the moments $\textrm{Tr} \rho_A^n$ of the interval $A=[u,v]$ on the strip of length $L$
for Dirichlet b.c. and Neumann b.c. become respectively 
\be
\label{Mn-D-N-dec-segment}
M_{A; \,\textrm{\tiny D} }^{(n)} 
  =
  \frac{ C_n^2  }{\mathcal{P}(u,v)^{\Delta_n}}\; \widetilde{\mathcal{F}}_n^{\textrm{\tiny (D)}} (r) 
  \;\;\qquad\;\;
  M_{A; \,\textrm{\tiny N} }^{(n)} 
  =
  \frac{ C_n^2 }{\mathcal{P}(u,v)^{\Delta_n}}\;  \widetilde{\mathcal{F}}_n^{\textrm{\tiny (N)}} (r) 
\ee
where we have introduced
\be
\label{F-DN-dec-def}
    \widetilde{\mathcal{F}}_n^{\textrm{\tiny (D)}} (r) \equiv \frac{1}{ \sqrt{ \prod_{k=1}^{n-1} F_{k/n}(1-r)}  } 
    \;\;\; \qquad  \;\;\;  
    \widetilde{\mathcal{F}}_n^{\textrm{\tiny (N)}} (r) \equiv  \frac{1}{ \sqrt{ \prod_{k=1}^{n-1} F_{k/n}(r)}  } 
\ee
and $C_n$ are non-universal constants. The moments $\textrm{Tr} \rho_A^n$ for the interval on the half line with either Dirichlet b.c. and Neumann b.c. 
are obtained by taking $L\rightarrow \infty$ in (\ref{Mn-D-N-dec-segment}). 
Notice that the constant $C_n$ is assumed to relate the twist fields in the $\mathbb{Z}_n$ orbifold of the non-compact boson BCFT 
and the lattice twist operators in a harmonic chain model, that is properly regularized  both in the UV and in the  IR.
 In the case of Dirichlet b.c., $C_n$ has the same UV origin as in \eqref{eq_scaling_tau},
while the case of Neumann boundary conditions is more subtle
because the lattice model requires an IR regularization due to the occurrence of the zero mode.

By using (\ref{Mn-D-N-dec-segment}), the UV finite combination (\ref{renyi-differences}) in this regime reads
\be
\label{renyi-diff-dec-2F1}
     S_{A; \textrm{\tiny D}}^{(n)} - S_{A; \textrm{\tiny N}}^{(n)}
     =\,
    \frac{1}{2(n-1)} \,\sum_{k=1}^{n-1} \log\! \left(  \frac{F_{k/n}(1-r) }{ F_{k/n}(r) } \right) .
\ee

In Sec.\,\ref{sec_HC_numerics} 
the analytic expressions in (\ref{Mn-D-N-dec-segment}) and (\ref{renyi-diff-dec-2F1}) are compared with
the corresponding results obtained numerically in the proper harmonic chains,
finding a good agreement.

\section{Entanglement entropies in the unit disk}
\label{section_CFT_disk}

In this section, we discuss the main BCFT calculation of this manuscript. 
As anticipated in Sec.\,\ref{sec_cft_results}, we employ the mirror trick to evaluate the partition function $\mathcal{Z}(\mathscr{S}_n)$ 
for the compact boson on the specific surface $\mathscr{S}_n$ occurring in our problem because of the replica construction. 
Since the action is quadratic, the partition function factorizes into a classical part and a quantum part, which are evaluated separately.

\subsection{Partition function}

\subsubsection{Mirror trick}
\label{sec_mirror_trick}

In Sec.\,\ref{sec_main_CFT_results_boson} we have qualitatively discussed that 
evaluating the R\'enyi entropies for the bipartitions in Fig.\,\ref{figure-biparts-intro}
corresponds to compute a partition function on a surface $\mathcal{M}=\mathscr{S}_n$ with boundary 
which is topologically equivalent to a sphere with $n$ equal disks removed
(for $n=4$, see the right panel of Fig.\,\ref{figure-riemann-surface-intro}). 
In a BCFT, this partition function can be obtained also as 
the two-point function $\langle  \mathcal{T}_n(0)\,\mathcal{T}^{\dag}_n(x) \rangle_{_\mathbb{D}}$ of twist fields on the unit disk $\mathbb{D}$,
placed at the origin and at $x$, with $0<x<1$. 

The replica construction introduces the branched covering $\mathcal{M} = \mathscr{D}_n$,
which is obtained by joining cyclically $n$ copies of $\mathbb{D}$ through the cut along the interval $(0,x)$.
A standard approach to investigate partition functions on a  Riemann surface $\mathcal{M}$ with boundaries 
is to introduce the so-called \emph{double} of $\mathcal{M}$,
that we denote by $D(\mathcal{M})$ \cite{Schweigert:2000ix}.
The double of $\mathcal{M}$ is a compact Riemann surface endowed with an antiholomorphic involutive map $\sigma$ 
(called a real structure) such that $\mathcal{M} = D(\mathcal{M})/\sigma $
and the boundary $\partial \mathcal{M}$ corresponds to the fixed points of $\sigma$.  
For instance, the double of the upper half plane is the whole plane with $\sigma(z) = \bar{z}$
and the double of the right half plane is the entire plane with real structure given by the reflection w.r.t. the vertical axis. 
The double of the unit disk $\mathbb{D}$ is the Riemann sphere $\mathbb{CP}^1$ 
with real structure $\sigma(z) = 1/\bar{z}$.
This construction is allowed when $\partial \mathcal{M}$ is analytic w.r.t. the complex structure induced by the metric
and this condition is verified for $\mathscr{D}_n$.
Thus, $D(\mathscr{D}_n)$ is simply the $n$-sheeted covering of the Riemann sphere $\mathbb{CP}^1$ 
with branch points at $0,x, 1/x$ and $\infty$ and the antiholomorphic involution is (a lift of) $\sigma(z) =  1/\bar{z}$.  
The case of the interval in the right half plane is considered in Fig.\,\ref{figure-riemann-surface-intro}.

\begin{figure}[t!]
\vspace{-.2cm}
\hspace{.6cm}
  \includegraphics[width=.9\textwidth]{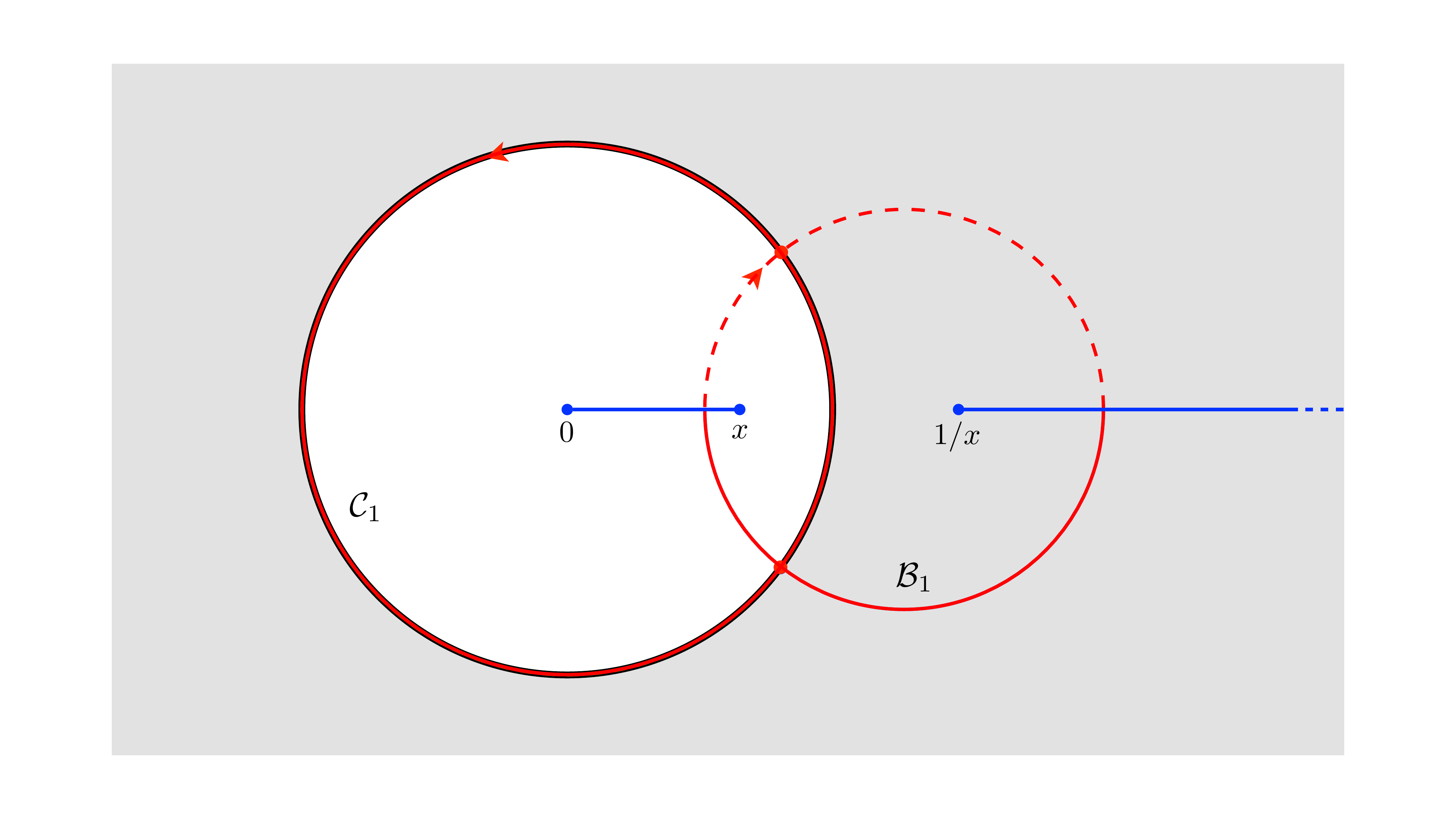}
\vspace{.2cm}
\caption{
The cycles (red curves) discussed in Sec.\,\ref{sec_mirror_trick}. 
Solid curves lie in the first sheet, while dashed line stand for curves on the second sheet. 
The unit disk $\mathbb{D}$ corresponds to the white region.
The twist fields are located in the origin and in $x$.
}
\label{figure-cycles-disk}
\end{figure}

The first step of our analysis consists in 
constructing a canonical homology basis for the genus $n-1$ surface $D(\mathscr{D}_n)$ that is compatible with the involution $\sigma$. 
Consider the cycles $\mathcal{C}_1$ and $\mathcal{B}_1$ shown in Fig.\,\ref{figure-cycles-disk}. 
The contour $\mathcal{C}_1$ is any cycle in the same homology class as the the unit circle. 
In particular, its homology class is invariant (even) under $\sigma$. 
On the other hand, the homology class of $\mathcal{B}_1$ is odd under $\sigma$;
indeed $\mathcal{B}_1$ can be chosen so that  $\sigma(\mathcal{B}_1)$ is the same loop as $\mathcal{B}_1$, but with the opposite orientation.
These contours can be replicated on each sheet through the deck transformation $f$ sending sheet $j$ to $j+1$,
i.e. $\mathcal{C}_{j+1} = f (\mathcal{C}_{j})$ and $\mathcal{B}_{j+1} = f (\mathcal{B}_{j})$, for $1\leqslant j \leqslant n-1$. 
Since the deck transformation commutes with the real structure $\sigma$, the (homology class of the) cycles $\mathcal{B}_{j}$ and $\mathcal{C}_{j}$ 
are respectively odd and even under $\sigma$. 
Although the contours  $\mathcal{C}_j$ and $\mathcal{B}_j$ for $1 \leqslant j \leqslant n-1$ generate the whole homology group,
they do not form a canonical basis because $\mathcal{B}_j$ intersects both $\mathcal{C}_j$ and $\mathcal{C}_{j+1}$. 
More precisely their intersection numbers are $\sharp(\mathcal{C}_i, \mathcal{B}_j) = \delta_{i,j} - \delta_{i,j+1}$.  
However, the cycles $\mathcal{A}_j = \mathcal{C}_1 + \cdots + \mathcal{C}_j$ are such that $\sharp(\mathcal{A}_i, \mathcal{B}_j) = \delta_{i,j}$ (and $\sharp(\mathcal{A}_i, \mathcal{A}_j) = 0$) and therefore  $(\mathcal{A}_j, \mathcal{B}_j)$ is a canonical homology basis. 
Furthermore, this basis satisfies $ \sigma(\mathcal{A}_j) =  \mathcal{A}_j$ and $\sigma(\mathcal{B}_j) = -  \mathcal{B}_j$ (up to smooth deformations).
Thus, under $\sigma$, each $\mathcal{A}_j$ is invariant, while each $\mathcal{B}_j$ just changes its orientation. 
The period matrix corresponding to this canonical homology basis is (\ref{tau-matrix-element})
and its derivation is discussed in Appendix\;\ref{app_period_matrix}.

We are interested in the partition function of the real compact massless boson $\phi \sim \phi + 2\pi R$
on a compact Riemann surface $\mathcal{M}$ with boundaries.
The action (\ref{action-boson}) for this BCFT can be written also as follows
\be
\label{action-boson-section}
  S[\phi] 
 \, =\, 
  \frac{1}{8\pi}  \int_{\mathcal{M}}   g(d\phi, d\phi)  d\mu  
 \, =\,  
 \frac{1}{8\pi}  \int_{ \mathcal{M}}   d\phi \wedge \star \,d\phi
 \,=\,
 \frac{\ri}{4\pi}  \int   \partial \phi \wedge  \bar{\partial}\phi
\ee
where $g$ is the metric tensor on $\mathcal{M}$,
whose volume form is $d\mu = \sqrt{|g|} \,d^2x$,
and $\star$ is the Hodge star operator. In the last equality we introduced the Dolbeault operators $\partial = \frac{1}{2} \left( d + i \star d \right)$ and $\bar{\partial} =  \frac{1}{2} \left( d - i \star d \right)$. 
The same b.c.  are imposed on all the boundary components of  $\partial \mathcal{M}$
and in our case they are either of Dirichlet or Neumann type.
The action (\ref{action-boson-section}) is invariant under Weyl rescaling $g \to \e^{\varphi} g$.
Such invariance can be made more manifest by introducing the decomposition $d \phi = \partial \phi + \bar{\partial} \phi$
and using that $\star \,dz = -\ri dz$, $\star\, d\bar{z} = \ri d\bar{z}$,
which imply  $  \star\, d\phi =  -\ri \partial \phi + \ri \bar{\partial} \phi $.
This leads to the last expression in (\ref{action-boson-section}), that  is manifestly independent of the metric (in a given conformal class). 
The partition function $\mathcal{Z} = \int [D\phi] \,\e^{-S[\phi]}$
is obtained by performing the path integration over all field configurations satisfying the appropriate b.c. on $\partial \mathcal{M}$,
which are either of Neumann or Dirichlet type  in our analysis. 
A non compact real boson $\phi $ takes values in $\mathbb{R}$,
while in the compact case that we are considering  the real field $\phi$ takes values in the circle of radius $R$, 
i.e. in $ \mathbb{R}/ (2\pi R\,\mathbb{Z})$. 
In the latter case, the field  configurations
can be classified through their windings  (or instanton sectors),
meaning that
\begin{align}
  \label{eq_winding}
  \int_{\mathcal{P}_j}  d\phi \,=\, 2\pi n_j R
  \;\;\; \qquad \;\;\; 
  n_j \in \mathbb{Z}
\end{align}
where the integer $n_j$ determines the winding number corresponding to the path $\mathcal{P}_j$,
that can be either a non contractible cycle on the Riemann surface
or an open path connecting  two of its boundary components.

Following the standard procedure to deal with these windings discussed in \cite{Dixon:1986qv},
in the path integral for the partition function we decompose the field 
$\phi=\phi_{\textrm{\tiny cl}} + \phi_{\textrm{\tiny qu}}$
into the classical field $\phi_{\textrm{\tiny cl}}$ and a quantum part $\phi_{\textrm{\tiny qu}}$.
The classical solution $\phi_{\textrm{\tiny cl}}$ 
is a harmonic function $\phi_{\textrm{\tiny cl}} : \mathcal{M} \to \mathbb{R}/(2\pi R \,\mathbb{Z})$ satisfying \eqref{eq_winding}
(hence it depends on the winding vector $\boldsymbol{n}$, whose  $j$-th element is $n_j$),
while the quantum field $\phi_{\textrm{\tiny qu}}$ has no windings. 

The quadratic form of the action (\ref{action-boson-section}) combined with the fact that $\phi_{\textrm{\tiny cl}} $ is a solution of the equation of motion
lead to decomposition $S[\phi] = S[\phi_{\textrm{\tiny cl}}] + S[\phi_{\textrm{\tiny cl}}]$, 
where only the classical term $S[\phi_{\textrm{\tiny cl}}]$ depends on the winding vector. 
This implies that the partition function on the Riemann surface $\mathcal{M} $ factorises as follows
\be
\label{Z_factorization}
\mathcal{Z}(\mathcal{M} ) = \mathcal{Z}_{\textrm{\tiny cl}}(\mathcal{M};R) \,\mathcal{Z}_{\textrm{\tiny qu}}(\mathcal{M} ) 
\;\;\;\qquad\;\;\;
\ee
where the classical and the quantum terms are defined respectively as
\be
\label{Z-cl-Z-qu}
\mathcal{Z}_{\textrm{\tiny cl}}(\mathcal{M};R) \equiv \sum_{\boldsymbol{n}} \e^{-S[\phi_{\textrm{\tiny cl}}] } 
\;\;\;\qquad\;\;\;
\mathcal{Z}_{\textrm{\tiny qu}}(\mathcal{M}) \equiv  \int [D\phi_{\textrm{\tiny qu}}] \, \e^{-S[\phi_{\textrm{\tiny qu}}]} \,.
\ee
We remark that the quantum term $\mathcal{Z}_{\textrm{\tiny qu}}(\mathcal{M}) $ is independent of the compactification radius $R$.

\subsubsection{Classical term}

We consider first the classical part $\mathcal{Z}_{\textrm{\tiny cl}}(\mathcal{M};R) $ 
of the partition function in (\ref{Z_factorization}) when the Dirichlet b.c.  $\phi= \phi_0$ is imposed, 
where $ \phi_0$ is a constant in $ \mathbb{R}/ (2\pi R\,\mathbb{Z})$. 
Crucially, we impose the same constant  $\phi_0$ on all the components of $\partial \mathcal{M}$ 
(see Appendix \ref{subsec-XXZ-insight}). 
By exploiting the $U(1)$ invariance of the theory,  we  can assume that $\phi_0 =0$ without loss of generality. 
Thus, the classical solutions of this Dirichlet problem are harmonic functions $\phi_{\textrm{\tiny cl}}$ that vanish on the boundary. 
Combining the vanishing (mod $2\pi R$)  of $\phi_{\textrm{\tiny cl}}$ on $\partial \mathcal{M}$
with the fact that any harmonic function is (locally) the real part of an analytic function, 
$\phi_{\textrm{\tiny cl}}$  can be extended to the double $D(\mathcal{M})$ through the Schwarz reflection principle via the condition 
$\phi_{\textrm{\tiny cl}}(\sigma(p)) = - \phi_{\textrm{\tiny cl}}(p)$.
Hence the classical solutions $\phi_{\textrm{\tiny cl}}$ on $\mathcal{M}$ satisfying vanishing Dirichlet b.c. 
are in one-to-one correspondence with the classical solutions on the double $D(\mathcal{M})$ that are odd under $\sigma$. 
This implies that $d\phi_{\textrm{\tiny cl}}$ is a harmonic form on $D(\mathcal{M})$ 
satisfying $ \sigma^* d\phi_{\textrm{\tiny cl}} = - d\phi_{\textrm{\tiny cl}}$, where $\sigma^\ast$ denote the pullback by $\sigma$. \\
On the other hand, the even harmonic forms on $D(\mathcal{M})$ are such that $ \sigma^* d\phi_{\textrm{\tiny cl}}=  d\phi_{\textrm{\tiny cl}}$
and their restriction to $\mathcal{M}$ provide the classical solutions satisfying Neumann b.c. for all the components of  $\partial \mathcal{M}$.
In this case $\phi_{\textrm{\tiny cl}}$ can be extended to $D(\mathcal{M})$  via $\phi_{\textrm{\tiny cl}}(\sigma(p)) = \phi_{\textrm{\tiny cl}}(p)$.
Alternatively, we can consider the dual field $\theta_{\textrm{\tiny cl}}$, which is defined by $d\theta_{\textrm{\tiny cl}} = \star \,d\phi_{\textrm{\tiny cl}}$.
When $\phi_{\textrm{\tiny cl}}$ satisfies Neumann b.c.,  the dual field $\theta_{\textrm{\tiny cl}}$ obeys Dirichlet b.c.;  hence $\sigma^* d\theta_{\textrm{\tiny cl}} = - d\theta_{\textrm{\tiny cl}}$. 
Since  $\sigma_{\textrm{\tiny cl}}$ is an orientation reversing isometry, it anticommutes with the Hodge star operator and therefore $\sigma^* d\phi_{\textrm{\tiny cl}} = d\phi_{\textrm{\tiny cl}}$.

In the computation of the moments $\textrm{Tr}\, \rho_A^n$ for the bipartitions shown in Fig.\,\ref{figure-biparts-intro},
the Riemann surface $\mathcal{M} = \mathscr{D}_n$ is topologically equivalent to a sphere with $n$ disks removed
(for $n=4$, see the right panel of Fig.\,\ref{figure-riemann-surface-intro}).
Its double $D( \mathscr{D}_n)$ is the compact Riemann sphere of genus $g = n-1$ described in Sec.\,\ref{sec_mirror_trick}.
When $n=4$, the double $D( \mathscr{D}_n)$ is topologically equivalent to the Riemann surface 
shown in the left panel of Fig.\,\ref{figure-riemann-surface-intro}.
Consider the canonical homology basis  $(\mathcal{A}_j, \mathcal{B}_j)$ discussed in Sec.\,\ref{sec_mirror_trick}.
It is a standard result of Hodge theory \cite{farkas1992riemann} that there exists a unique dual basis $(\alpha_j, \beta_j)$ of real harmonic one-forms
such that 
\be
\label{can-homo-basis}
  \oint_{\mathcal{A}_i} \alpha_j =  \delta_{i,j} \qquad \oint_{\mathcal{B}_i} \alpha_j = 0 
  \qquad \textrm{and} \qquad
    \oint_{\mathcal{A}_i} \beta_j = 0 \qquad \oint_{\mathcal{B}_i} \beta_j = \delta_{i,j} \,.
\ee
Since the homology class of $\mathcal{A}_i$ and $\mathcal{B}_i$ are respectively even and odd under sigma, the relations \eqref{can-homo-basis} are also satisfied by $\sigma^* \alpha_j$ and $-\sigma^* \beta_j$.
Hence, uniqueness implies that $\sigma^* \alpha_j = \alpha_j$ and $\sigma^* \beta_j = - \beta_j$,
meaning that, under $\sigma$, the one-forms $\alpha_j$ are even, while $\beta_j$ are odd. 
For Dirichlet b.c. such that $\phi$ takes the same value (modulo $2\pi R$) on each boundary component of $\partial \mathscr{D}_n$,
we have that
\be
\label{Diriclet-B-plus-cycle}
  \int_{\mathcal{B}_j^+} d\phi_{\textrm{\tiny cl}} = 2\pi R \, m_j  
  \;\;\qquad \;\;
  m_j \in \mathbb{Z} 
\ee
where $  j \in \big\{1, \dots, n -1\big\}$ and
 $\mathcal{B}_j^+$ is defined as the part of  $\mathcal{B}_j$ that lies inside $\mathscr{D}_n$. 
Hence, $\mathcal{B}_j^+$ is a path connecting the $j$-th component  to the $(j+1)$-th component of  $\partial \mathscr{D}_n$ (see Fig.\,\ref{figure-cycles-disk}).
Thus, on $D( \mathscr{D}_n)$ we have that
\begin{equation}\label{eq:winding_Bj}
    \oint_{\mathcal{B}_j} d\phi_{\textrm{\tiny cl}} = 4\pi R \,m_j \quad\quad m_j \in \mathbb{Z} \,.
\end{equation}
Moreover, since $\phi$ is constant on each boundary component, $ \oint_{\mathcal{A}_j} d\phi_{\textrm{\tiny cl}} = 0 $ for all the allowed values of $j$.
The analytic continuation of $d\phi_{\textrm{\tiny cl}}$ to $D( \mathscr{D}_n)$ involves only the odd harmonic forms $\beta_j$ as follows
\be
  d\phi_{\textrm{\tiny cl}} = 4\pi R \sum_j m_j \beta_j  \,.
\ee

The Riemann bilinear relation \cite{farkas1992riemann}
provides the value of the action (\ref{action-boson-section}) for this classical solution.
It reads
\be
\label{S-cl-value}
S[\phi_{\textrm{\tiny cl}}]
  =
  \frac{1}{8\pi}  \int_{\mathscr{D}_n}   d\phi_{\textrm{\tiny cl}} \wedge \star \,d\phi_{\textrm{\tiny cl}} 
\,=\,
 \frac{1}{16\pi}  \int_{D(\mathscr{D}_n)}   d\phi_{\textrm{\tiny cl}} \wedge \star\, d\phi_{\textrm{\tiny cl}} 
  \,=\,
     \pi R^2 \,  {\bf m}^{\textrm{\tiny t}}  \cdot    \boldsymbol{\tau}_2^{-1} \! \cdot \boldsymbol{m}
\ee
where $\boldsymbol{\tau}_2 \equiv \textrm{Im}(\boldsymbol{\tau})$ is the imaginary part of the 
$(n-1)\times (n-1)$ period matrix $\boldsymbol{\tau}$ 
of  $D(\mathscr{D}_n)$ in the canonical homology basis $(\mathcal{A}_j, \mathcal{B}_j)$. 
In our case the period matrix is purely imaginary.
Indeed, given a basis of the holomorphic one-forms $\omega_k$ such that $\oint_{\mathcal{A}_j}  \omega_k = \delta_{j,k}$, 
we have $\sigma^* \omega_k = \overline{\omega}_k$ and therefore 
\be
 \overline{ \boldsymbol{\tau}}_{i,j} = \int_{\mathcal{B}_i} \bar{\omega}_j 
   =  \int_{\sigma(\mathcal{B}_i)}\omega_j = - \,\boldsymbol{\tau}_{i,j} \,.
\ee
From (\ref{S-cl-value}) and the first expression in (\ref{Z-cl-Z-qu}), 
for vanishing Dirichlet b.c. one obtains
\be
\label{Zn-classical-D}
\mathcal{Z}^{\textrm{\tiny (D)}}_{n, \,\textrm{\tiny cl}}
    =  \,
    \Theta\big(\!- \!R^2\, \boldsymbol{\tau}^{-1}\big)
\ee
in terms of the Siegel theta function (\ref{siegel-def}) 
and of the period matrix defined by (\ref{tau-matrix-element}),
whose derivation is discussed in the Appendix\;\ref{app_period_matrix}.

The case where Neumann b.c. are imposed on all the $n$ boundary components of $\mathscr{D}_n$
can be addressed by adapting the steps for Dirichlet b.c. described above. 
For Neumann b.c., the classical solutions can be extended to  $D(\mathscr{D}_n)$ 
through the requirement $\phi_{\textrm{\tiny cl}}\circ \sigma = \phi_{\textrm{\tiny cl}}$,
as already mentioned. 
Given the canonical homology base introduced above (see Sec.\,\ref{sec_mirror_trick} and (\ref{can-homo-basis})),
these classical solutions correspond to harmonic forms $d\phi_{\textrm{\tiny cl}}$ satisfying
\be
\label{neumann-A-B-cycles}
   \int_{\mathcal{A}_j} d\phi_{\textrm{\tiny cl}} =2 \pi n_j R
   \;\qquad \;
   \int_{\mathcal{B}_j} d\phi_{\textrm{\tiny cl}} =0 
   \;\;\;\;\qquad\;\;\; \;
   n_j \in \mathbb{Z} 
\ee
where $j  \in \big\{ 1, \dots, n -1\big\}$ and the absence of winding over $\mathcal{B}_j$ follows from the fact that $ d\phi_{\textrm{\tiny cl}}$ is even under $\sigma$. Thus we have 
\be
  d\phi_{\textrm{\tiny cl}} = 2\pi R \sum_j n_j \alpha_j  \,.
\ee
Again, the Riemann bilinear relation allows to compute the action (\ref{action-boson-section}) for these classical solutions
and the result is
\be
S[\phi_{\textrm{\tiny cl}}] 
= 
 \frac{\pi R^2}{4} \; 
{\bf n}^{\textrm{\tiny t}}  \cdot \boldsymbol{\tau} \cdot  \boldsymbol{\tau} _2^{-1}  \cdot \overline{\boldsymbol{\tau} } \cdot  {\bf n}  
= -  \,
\ri \, \frac{\pi R^2}{4} \; {\bf n}^{\textrm{\tiny t}}  \! \cdot \boldsymbol{\tau} \cdot  {\bf n} 
\ee
where the last step has been obtained by using that $\boldsymbol{\tau}$ is pure imaginary, i.e. $\boldsymbol{\tau} = \ri \, \boldsymbol{\tau}_2$. 
Finally, we find
\be
\label{Zn-classical-N}
    \mathcal{Z}^{\textrm{\tiny (N)}}_{n, \,\textrm{\tiny cl}}
    =  \,
    \Theta\big(R^2\, \boldsymbol{\tau}/4\big)
\ee
in terms of the Siegel theta function (\ref{siegel-def}).

\subsubsection{Quantum term}

The quantum part of the partition function in (\ref{Z_factorization}) and (\ref{Z-cl-Z-qu}) for $\mathcal{M} = \mathscr{D}_n$
is independent of the compactification radius $R$.
Rather than determining the quantum determinant of the Green function of the Laplacian on $\mathscr{D}_n$,
we find it more convenient to adapt the analysis discussed in  \cite{Calabrese:2009ez}, 
which is based on the method introduced in \cite{Dixon:1986qv}.
In Sec.\,\ref{sec_mirror_trick} the field $\phi$ has been decomposed into the sum $\phi=\phi_{\textrm{\tiny cl}} + \phi_{\textrm{\tiny qu}}$
and in the following analysis of the quantum term $\phi_{\textrm{\tiny qu}}$ is denoted just by $\phi$ to enlighten the expressions.
Since $\mathcal{M} = \mathscr{D}_n$ is made by $n$ copies of the unit disk joined cyclically along the cut $(0,x)$,
the path integral in (\ref{Z-cl-Z-qu}) can be rewritten by introducing a field $\phi_j$ on the $j$-th copy, with $1\leqslant j \leqslant n$.
The total action reads $S[\phi] = \sum_{j} S[\phi_j]$ and the fields on the consecutive copies are coupled through their boundary condition along the cut.

Following \cite{Casini:2005rm},
it is useful to perform a discrete Fourier transform for the $n$ bosonic fields in the 
different replicas and introduce
\be
\label{phi-tilde-k-def}
  \tilde{\phi}_k = \frac{1}{\sqrt{n}} \sum_{j=1}^n \e^{- 2\pi \ri  \,k \,j/n} \,\phi_j
  \;\;\;\;\qquad\;\;\;
  1 \leqslant k \leqslant n
\ee
which is a complex combination of fields; hence it is more convenient to replace the real bosons $\phi_j$ with the complex bosons $\Phi_j$
throughout the computation. 
The result for the real field is obtained by taking the square root of the final expression. 
The fields introduced through the transformation (\ref{phi-tilde-k-def}) are decoupled.
However, the coupling of  the original fields $\phi_j$ through the cut 
imposes the following twist condition around the origin
\be
\label{twist-field-condition}
  \tilde{\phi}_k(\e^{2i\pi} z, \e^{-2i \pi} \bar{z}) 
  = \e^{2\pi \ri\, k/n} \, \tilde{\phi}_k(z, \bar{z}) 
\ee
and a similar one around the branch point at $x$, with the phase factor $\e^{2\pi \ri\, k/n}$ in the r.h.s. replaced by its complex conjugate.

The partition function of a complex scalar on the Riemann sphere satisfying the above twisted boundary conditions around four branch points 
for an assigned value of $k/n$ has been studied in \cite{Dixon:1986qv}.
In the case of the unit disk $\mathbb{D}$ and of two branch points we are dealing with,
this analysis tells us that the corresponding partition function can be written as 
the two-point function of particular twist fields $\mathcal{T}_{k/n}$ and $\mathcal{T}^{\dag}_{k/n}$
placed at the endpoints of the branch cut. 
This leads us to write the quantum part of the partition function as the following product  (up to normalization)
\be
  \label{eq_factorization_quantum_part}
  \mathcal{Z}_{n, \,\textrm{\tiny qu}}^2
  = 
  \prod_{k=1}^{n-1}  \langle \mathcal{T}_{k/n}(0)\, \mathcal{T}^{\dag}_{k/n}(x) \rangle_{_\mathbb{D}}
\ee
where the mode corresponding to $k=n$ does not contribute because the corresponding twist field is the identity operator.

A method to determine $\langle \mathcal{T}_{k/n}(0)\, \mathcal{T}^{\dag}_{k/n}(x) \rangle_{_\mathbb{D}}$ was developed in \cite{Dixon:1986qv}
and it is based on the the expectation value of the stress-energy tensor $T(z)$ in the presence of the twist fields. 
In Appendix \ref{app_quantum_part} we have adapted the analysis of \cite{Dixon:1986qv}  to the specific cases under investigation
and the main results are presented below. 
We find that
\bea
  \partial_x \log 
  \langle \mathcal{T}_{k/n}(0)\, \mathcal{T}^{\dag}_{k/n}(x) \rangle_{_\mathbb{D}}
   &=&
   \textrm{Res}_{z \to x}  
   \frac{ 
   \langle T(z)\,\mathcal{T}_{k/n}(0)\, \mathcal{T}^{\dag}_{k/n}(x) \rangle_{_\mathbb{D}}
   }{  
   \langle \mathcal{T}_{k/n}(0)\, \mathcal{T}^{\dag}_{k/n}(x) \rangle_{_\mathbb{D}}
   }
   \\
   \label{ode-2point-twist}
   \rule{0pt}{.7cm}
   &=&
      - \, 2 h_{k/n} \left( \frac{1}{x} + \frac{1}{x-1/\bar{x}} \right)  - \partial_x  \log\! \big[ E^{(\alpha)}_{k/n}(x) \big]
\eea
where
\be
\label{h-E-k-defs}
h_{k/n} \equiv\frac{1}{2} \;\frac{k}{n} \left(1 - \frac{k}{n} \right)
\;\;\;\qquad\;\;\;\;\;
  E^{(\alpha)} _{k/n}(x)  \equiv 
  \left\{\begin{array}{ll}
  F_{k/n}\big(1-|x|^2\big) \hspace{.6cm} & \textrm{Dirichlet b.c.}
  \\
  \rule{0pt}{.5cm}
    F_{k/n}\big(|x|^2\big) & \textrm{Neumann b.c.}
  \end{array}
  \right.
\ee
 (we remind that $F_{k/n}(y) \equiv \, _2F_1(k/n, 1-k/n;1;y)$). 
 As for the quantum part of the partition function \eqref{eq_factorization_quantum_part},
 this leads to 
  \be
  \label{eq_derivative_quantum_part}
 \partial_x \log \mathcal{Z}_{n, \,\textrm{\tiny qu}} =  - \, \Delta_n \left( \frac{1}{x} + \frac{1}{x-1/\bar{x}} \right)  -  \partial_x  \log   E^{(\alpha)}_{n}(x) 
\ee
 where we used that  $\sum_{k=1}^{n-1} h_{k/n} = \Delta_n= \tfrac{1}{12} (n-1/n)$ (see \eqref{Delta_n def} with $c=1$)
 and we introduced 
 \begin{align}
   E_n^{(\alpha)}(x) \equiv  \sqrt{\prod_{k=1}^{n-1} E^{(\alpha)}_{k/n}(x)} \;.
\end{align}
From the relations reported in Appendix\;C of \cite{Calabrese:2009ez}, the function $ E_n^{(\alpha)}(x)$ 
can be expressed as a Siegel theta function as follows
\be
\label{theta-prod-identities}
  \sqrt{ \prod_{k=1}^{n-1} F_{k/n}(|x|^2)} 
  \,=\,\Theta\big(\boldsymbol{\tau}(|x|)\big) 
  \;\;\qquad\;\;
    \sqrt{ \prod_{k=1}^{n-1} F_{k/n}(1-|x|^2)} 
  \,=\,\Theta\big(\!-\!\boldsymbol{\tau}(|x|)^{-1}\big) 
\ee
in terms of the Siegel theta (\ref{siegel-def}) and of the period matrix $\boldsymbol{\tau}(x)$ defined in (\ref{tau-matrix-element}). 
Then, integrating \eqref{eq_derivative_quantum_part},
 for the quantum part of the partition function we get
\be
\label{Zn-quantum-final}
    \mathcal{Z}_{n, \,\textrm{\tiny qu}}(x)
  \propto
    \frac{1}{\mathcal{P}(x)^{ \Delta_n } \, E_n^{(\alpha)}(x)}
    \;\;\;\qquad\;\;\;
      \mathcal{P}(x) \equiv  |x|^2 \big(1 - |x|^2\big)
\ee
where
\be
\label{E-alpha-final}
  E_n^{(\alpha)}(x)  \equiv 
  \left\{\begin{array}{ll}
  \Theta\big(\boldsymbol{\tau}(|x|)\big) \hspace{1.5cm} &  \textrm{ Dirichlet b.c.}
  \\
  \rule{0pt}{.5cm}
    \Theta\big(\!-\!\boldsymbol{\tau}(|x|)^{-1}\big) &  \textrm{ Neumann b.c.}
  \end{array}
  \right.
\ee
and the overall constant, which can depend both on $n$ and on the b.c., will be fixed later.


\subsubsection{Two-point functions of twist fields} 
\label{subsec-finiteR-two-point-normalization}


Combining the classical part and the quantum part of the partition function, 
given by (\ref{Zn-classical-D})-(\ref{Zn-classical-N}) and (\ref{Zn-quantum-final})-(\ref{E-alpha-final}) respectively, 
we find that the two-point functions of the twist fields on the unit disk $\mathbb{D}$ 
for the compactified massless scalar field with either Dirichlet b.c. or Neumann b.c. read respectively
\bea
\label{two-point-twist-disk-D}
  \langle \mathcal{T}_{n}(0)\, \mathcal{T}^{\dag}_{n}(x) \rangle_{_\mathbb{D}}^{\textrm{\tiny (D)}}
  &  \propto&
  \frac{ 1 }{\mathcal{P}(x)^{\Delta_n}}\;
  \frac{\Theta\big(\!- \!R^2\, \boldsymbol{\tau}(x)^{-1}\big)}{ \Theta\big(\!-\!\boldsymbol{\tau}(x)^{-1}\big) }
  \,=\,
  \frac{ 1 }{R^{ n-1} \, \mathcal{P}(x)^{\Delta_n}}\;
  \frac{\Theta\big( \boldsymbol{\tau}(x)/R^2\big)}{ \Theta\big(\boldsymbol{\tau}(x)\big) }
  \\
  \rule{0pt}{.8cm}
  \label{two-point-twist-disk-N}
    \langle \mathcal{T}_{n}(0)\, \mathcal{T}^{\dag}_{n}(x) \rangle_{_\mathbb{D}}^{\textrm{\tiny (N)}}
  &  \propto&
  \frac{ 1  }{\mathcal{P}(x)^{\Delta_n}}\;
  \frac{\Theta\big(R^2\, \boldsymbol{\tau}(x)/4\big)}{ \Theta\big(\boldsymbol{\tau}(x)\big) }
\eea
where $\mathcal{P}(x)$ has been defined in \eqref{Zn-quantum-final} and the last expression of (\ref{two-point-twist-disk-D}) 
has been obtained by employing the following identity 
\be
\label{theta-prod-id}
 \Theta\big(\!-\!\boldsymbol{\tau}(x)^{-1}\big) 
 = 
 \sqrt{ \textrm{det} \big(\! -\! \ri \boldsymbol{\tau}(x) \big) }\,  \Theta\big( \boldsymbol{\tau}(x) \big) 
\ee
which involves the Siegel theta function (\ref{siegel-def}) and the period matrix (\ref{tau-matrix-element}).

In the limit $x\rightarrow 0$, for (\ref{tau-matrix-element}) we have
$\boldsymbol\tau(x)_{i,j} \rightarrow + \ri \infty$;
hence $\Theta(\eta \boldsymbol{\tau}(x))\rightarrow 1$ for any constant $\eta >0$,
which implies that  $\mathcal{F}_n^{(\alpha)}(x) \to 1$ as $x \to 0$ for any finite value of $R$.
By applying this observation to (\ref{two-point-twist-disk-D}) and (\ref{two-point-twist-disk-N}), at the leading order we have that
\be
\label{2pt-sphere-norm-cond}
   \langle \mathcal{T}_{n}(0)\, \mathcal{T}^{\dag}_{n}(x) \rangle_{_\mathbb{D}}^{(\alpha)}
  \sim \frac{1}{|x|^{2\Delta_n}} 
  \;\;\;\qquad\;\;\;
  x \to 0
  \;\;\;\qquad\;\;\;
  \alpha \in \big\{ \textrm{D}, \textrm{N} \big\}
\ee
which fixes the overall normalizations in (\ref{two-point-twist-disk-D}) and (\ref{two-point-twist-disk-N}). 
Thus, the final result for the two-point functions of twist fields on the unit disk 
with either Dirichlet b.c. or Neumann b.c. read respectively
\be
\label{two-point-twist-disk-final-finiteR}
  \langle \mathcal{T}_{n}(0)\, \mathcal{T}^{\dag}_{n}(x) \rangle_{_\mathbb{D}}^{\textrm{\tiny (D)}}
  = \, \frac{ 1  }{\mathcal{P}(x)^{\Delta_n}}\;
  \frac{\Theta\big( \boldsymbol{\tau}(x)/R^2\big)}{ \Theta\big(\boldsymbol{\tau}(x)\big) }
\;\qquad\;
 \langle \mathcal{T}_{n}(0)\, \mathcal{T}^{\dag}_{n}(x) \rangle_{_\mathbb{D}}^{\textrm{\tiny (N)}}
    =
    \frac{ 1  }{\mathcal{P}(x)^{\Delta_n}}\;
  \frac{\Theta\big(R^2\, \boldsymbol{\tau}(x)/4\big)}{ \Theta\big(\boldsymbol{\tau}(x)\big) }\,.
\ee
Finally, from (\ref{two-point-twist-disk-final-finiteR}) one obtains also
the two-point functions of twist fields in the infinite strip  and in the right half plane,
which are given by (\ref{eq:maptodisc}) and (\ref{eq:rhpmaptodisk}) respectively,
as discussed in the final part of Sec.\,\ref{sec_main_BCFT_expression}.

\subsection{Decompactification regime}
\label{subsec:decompactification_derivations}

An important regime to explore is given by the decompactification limit $R\to \infty$.

Taking this limit in (\ref{Fn-main-res}) does not provide well defined finite expressions. 
A similar problem  already arises for the conformal boundary states of the massless compact boson.
Indeed, the boundary states corresponding to Dirichlet b.c. and Neumann b.c. 
can be constructed through the $|(m,n)\rangle\rangle$ Ishibashi states as follows \cite{Oshikawa:1996dj}
\be
| \textrm{N} \rangle_R =\sqrt{\frac{R}{2}} \, \sum_{n \in \mathbb{Z}}|(0, n)\rangle \! \rangle
\;\;\;\qquad\;\;\;
\left| \textrm{D}\right\rangle_R =\sqrt{\frac{1}{R}} \, \sum_{m \in \mathbb{Z}}|(m, 0)\rangle \! \rangle
\ee
which do not have a well defined behaviour as $R\rightarrow \infty$.
For these boundary states, a formal regularization scheme for the BCFT data of the massless compact boson 
(spectrum of primary fields, boundary states and structure constants) 
has been implemented \cite{Restuccia:2013tba,Runkel:2001ng}
to construct a well defined decompactification limit.
However, extending this procedure to the $\mathbb{Z}_n$ orbifold of the massless compact boson is beyond the scopes of our work.

Well defined expressions in the decompactification limit can be obtained as follows.
By using that  $\Theta\big(R^2\, \boldsymbol{\tau}(x)/4\big) \to 1$ 
and $\Theta\big(\!- \!R^2\, \boldsymbol{\tau}(x)^{-1}\big) \to 1$ as $R \to \infty$ 
and disregarding proportionality constants for the moment, 
it is straightforward to find that the two-point functions of twist fields in 
(\ref{two-point-twist-disk-D}) and (\ref{two-point-twist-disk-N}) become respectively
\be
\label{two-point-twist-disk-ND-dec}
  \langle \mathcal{T}_{n}(0)\, \mathcal{T}^{\dag}_{n}(x) \rangle_{_\mathbb{D}}^{\textrm{\tiny (D)}}
  \propto
  \frac{ 1}{\mathcal{P}(x)^{\Delta_n}}\; \widetilde{\mathcal{F}}_n^{\textrm{\tiny (D)}}  \big(x^2\big)
\;\;\qquad\;\;
    \langle \mathcal{T}_{n}(0)\, \mathcal{T}^{\dag}_{n}(x) \rangle_{_\mathbb{D}}^{\textrm{\tiny (N)}}
  \propto
  \frac{1 }{\mathcal{P}(x)^{\Delta_n}}\; \widetilde{\mathcal{F}}_n^{\textrm{\tiny (N)}} \big(x^2\big)
\ee
where the identities (\ref{theta-prod-identities}) and the functions (\ref{F-DN-dec-def}) have been employed. 
The BCFT normalization in (\ref{two-point-twist-disk-ND-dec}) needs a careful and slightly technical discussion of the $x\rightarrow 0$ limit 
that we report in the Appendix \ref{app:normalization-decomp}.
The final results for the two-point functions on the unit disk 
with either Dirichlet b.c. or Neumann b.c. in the decompactification regime are respectively
\be
\label{two-point-twist-disk-ND-dec-normalized}
  \langle \mathcal{T}_{n}(0)\, \mathcal{T}^{\dag}_{n}(x) \rangle_{_\mathbb{D}}^{\textrm{\tiny (D)}}
  =
  \frac{ 1}{\mathcal{P}(x)^{\Delta_n}}\; \widetilde{\mathcal{F}}_n^{\textrm{\tiny (D)}}  \big(x^2\big)
\;\;\qquad\;\;
    \langle \mathcal{T}_{n}(0)\, \mathcal{T}^{\dag}_{n}(x) \rangle_{_\mathbb{D}}^{\textrm{\tiny (N)}}
  =
  \frac{1 }{\mathcal{P}(x)^{\Delta_n}}\; \widetilde{\mathcal{F}}_n^{\textrm{\tiny (N)}} \big(x^2\big) \,.
\ee
which provide the results reported in Sec.\;\ref{subsec-decom-regime}.

As for the R\'enyi entropies of an interval in the segment, from (\ref{SA-n-BCFT}) we have 
 \be
 \label{ren-dec-segment-rep}
S_{A; \alpha}^{(n)} 
= \,
\frac{\Delta_n}{n-1}\, \log \!\big[ \mathcal{P}(u,v)\big]
-
\frac{\log\!\big(C_n\big)}{n-1}
+
\frac{1}{2(n-1)} \sum_{k=1}^{n-1} \log \!\big[ F_{k/n}(y_\alpha) \big]
\ee   
where we have introduced $y_{\textrm{\tiny D}} \equiv 1-r$ for Dirichlet b.c. 
and  $y_{\textrm{\tiny N}} \equiv r$ for Neumann b.c.,
and we remind that $F_{k/n}(y)$ has been defined in the text below (\ref{tau-matrix-element}).
The corresponding result for the interval on the half line is obtained by taking the limit $L\rightarrow \infty$ in (\ref{ren-dec-segment-rep}).

Finally, by using (\ref{MI}) and  (\ref{ren-dec-segment-rep}), we find  the following UV finite quantity
\be
          \label{MI-half line-dec}
     \mathcal{I}_{A; \alpha}^{(n)}
     =
     \frac{\Delta_n}{1-n} \,\log (r)
     +
\frac{1}{2(1-n)} \sum_{k=1}^{n-1} \log \!\big[ F_{k/n}(y_\alpha) \big]
\ee
where we have also employed that $g_{\textrm{\tiny D}} = g_{\textrm{\tiny N}}=1$ in our conventions, 
as shown in Sec.\,\ref{app:normalization-decomp} (see (\ref{eq:ludwig_affleck_noncompact}))
and in agreement with the existing results  \cite{Recknagel:2013uja}.

The analytic continuation $n \to 1$ of (\ref{ren-dec-segment-rep})
can be studied by employing the following result obtained in \cite{Calabrese:2009ez}
\be
\label{D1-def}
\lim_{n \to 1} \partial_n \Bigg( \sum_{k=1}^{n-1} \log \!\big[ F_{k/n}(y) \big] \Bigg)
=
\int_{-\ri \infty}^{\ri \infty} \frac{\pi z \, \log [ F_z(y)]}{[\sin(\pi z )]^2} \; \frac{dz}{ \ri }
\equiv
-\,\mathcal{D}_1'(y)
\ee
where the integral along the imaginary axis defining $\mathcal{D}_1'(y)$ is evaluated numerically.
This leads to the following result for the entanglement entropy of the interval in the segment 
\be
\label{eq:EE-allBC-allgeometries}
S_{A; \alpha}
= \,
\frac{1}{6}\, \log \!\big[ \mathcal{P}(u,v)\big]
-
\frac{\mathcal{D}_1'(y_\alpha)}{2}
+
2C'_1
\ee
where the non-universal constant shift is given by $C'_1\equiv -\lim_{n\rightarrow 1} \frac{\log{C_n}}{n-1}$.
The  entanglement entropy of an interval on the half line is obtained by taking the $L\rightarrow \infty$ 
of (\ref{eq:EE-allBC-allgeometries}).

From the UV finite quantity (\ref{MI-half line-dec}), we find it worth introducing
\be
\label{MI-hat-def}
     \widehat{\mathcal{I}}_{A;\alpha}^{(n)}
     \equiv
     \mathcal{I}_{A; \alpha}^{(n)}
     -
     \frac{\Delta_n}{1-n} \log (r) 
\ee
whose analytic continuation $n \to 1$ can be written in terms of (\ref{D1-def}) as follows
\be
\label{MI-hat-def-n=1}
     \widehat{\mathcal{I}}_{A;\alpha}^{(1)}
     =
\frac{\mathcal{D}_1'(y_\alpha)}{2} \,.
\ee

The limit $n \to \infty$ of (\ref{ren-dec-segment-rep}) 
provides the single copy entanglement entropy in the decompactification regime. 
By introducing 
\be
\label{sc-cft-explicit-0}
\lim_{n \to \infty} \,\frac{1}{n-1} \sum_{k=1}^{n-1} \log \!\big[ F_{k/n}(y) \big]
=
\int_0^1 \log \!\big[ F_{\kappa}(y) \big] d\kappa
\equiv 
\mathcal{S}(y)
\ee
where the integral defining $\mathcal{S}(y)$ can be evaluated numerically,
for the single copy entanglement entropy of the interval in the segment
one finds 
\be
\label{sc-cft-explicit-1}
S_{A; \alpha}^{(\infty)}
= \,
\frac{1}{12}\, \log \!\big[ \mathcal{P}(u,v)\big]
+
\frac{\mathcal{S}(y_\alpha)}{2}
+ 2
C'_{\infty}
\ee
where $C'_{\infty}\equiv -\lim_{n\rightarrow \infty} \frac{\log C_{n}}{n-1}$. 
The limit $L\rightarrow\infty$ of (\ref{sc-cft-explicit-1}) gives
the single copy entanglement entropy of the interval in the half line 
in the decompactification regime.

\section{Numerical results from harmonic chains} 
\label{sec_HC_numerics}

In this section we compare the BCFT results reported
in Sec.\,\ref{subsec-decom-regime} and Sec.\,\ref{subsec:decompactification_derivations} for the decompactification regime
with the entanglement entropies of a block of consecutive sites
in the spatial bipartitions shown in Fig.\,\ref{figure-biparts-intro}
for harmonic chains defined either on the semi-infinite line or on the segment,
when either Dirichlet b.c. or Neumann b.c. are imposed.

The Hamiltonian of a finite harmonic chain with nearest neighbour spring-like interactions 
made by $N-1$ sites in the interior and two sites at its endpoints reads
\be
\label{HC ham}
\widehat{H} = \sum_{i=0}^{N} \left(
\frac{1}{2m}\,\hat{p}_i^2+\frac{m\omega^2}{2}\,\hat{q}_i^2 \right)+\sum_{i=0}^{N-1}\frac{\kappa}{2}(\hat{q}_{i+1} -\hat{q}_i)^2
\ee
in terms of the position and the momentum operators  $\hat{q}_i$ and $\hat{p}_i$,
that are Hermitian operators satisfying the canonical commutation relations
$[\hat{q}_i , \hat{q}_j]=[\hat{p}_i , \hat{p}_j] = 0$ 
and $[\hat{q}_i , \hat{p}_j]= \textrm{i} \delta_{i,j}$
(we set $\hbar =1$).
At the endpoints of the harmonic chain we impose the same boundary condition, 
which is either Dirichlet b.c.
\be
\label{dirichlet-bc-hc}
\hat{q}_0 = \hat{q}_{N}  = 0
\ee
or Neumann b.c.
\be
\label{neumann-bc-hc}
\hat{q}_1 - \hat{q}_0 = 0 
\;\;\;\qquad\;\;\;
\hat{q}_N - \hat{q}_{N-1} = 0  \,.
\ee

We consider these quadratic systems in their ground state, that is a Gaussian state. 
Since these are free systems, the crucial objects needed to perform our numerical analysis 
are the correlation matrices $\mathsf{Q}$ and $\mathsf{P}$,
whose generic elements are the two-point correlators in the ground state,
which are given by $ \langle \hat{q}_i \hat{q}_j  \rangle $ and $\langle \hat{p}_i \hat{p}_j  \rangle$ respectively
\cite{Peschel:2003rdm, Audenaert:2002xfl, Botero04, Plenio:2004he, Cramer:2005mx, Schuch_2006, 
EislerPeschel:2009review, Casini:2009sr, Eisert:2008ur}.

For the Dirichlet b.c. (\ref{dirichlet-bc-hc}), the generic elements of the correlation matrices $\mathsf{Q}$ and $\mathsf{P}$
are respectively \cite{Lievens:2007nt}
\bea
\label{qq-dirichlet}
& &
\langle \hat{q}_i \hat{q}_j  \rangle 
=
\frac{1}{N} \sum_{k=1}^{N-1} \frac{1}{m \omega_k} \, 
\sin(\pi k\, i/N)  \, \sin(\pi k\, j/N) 
\\
\rule{0pt}{.8cm}
\label{pp-dirichlet}
& &
\langle \hat{p}_i \hat{p}_j  \rangle 
=
\frac{1}{N} \sum_{k=1}^{N-1} m \omega_k \, 
\sin(\pi k \, i/N)  \, \sin(\pi k \, j/N) 
\eea
where the dispersion relation reads
\be
\omega_k \equiv 
\sqrt{\omega^2 +\frac{4\kappa}{m}\, \big[ \sin(\pi k/(2N)) \big]^2} \,>\,\omega
\;\;\;\qquad\;\;\;
1 \leqslant k \leqslant N -1  \,.
\ee
In the massless regime (i.e. when $\omega = 0$) and in the thermodynamic limit $N \to \infty$,
these correlators simplify respectively to \cite{Calabrese:2012nk}
\bea
\label{corr qq dirichlet thermo}
& &
\langle \hat{q}_i \hat{q}_j  \rangle =
\frac{1}{2\pi \, \sqrt{\kappa m}} 
\Big(  \psi(1/2+i+j) -  \psi(1/2+i-j)  \Big)
\\
\label{corr pp dirichlet thermo}
\rule{0pt}{.9cm}
& &
\langle \hat{p}_i \hat{p}_j  \rangle =
\frac{2\sqrt{\kappa m}}{\pi}  \left( \frac{1}{ 4 (i+j)^2-1} - \frac{1}{4 (i-j)^2-1} \right) 
\eea
where $\psi(z)$ is the digamma function.
These correlators can be employed to investigate the semi-infinite massless harmonic chain with Dirichlet b.c. imposed at its origin.

In the case of Neumann b.c. (\ref{neumann-bc-hc}),
the generic elements of the correlation matrices 
$\mathsf{Q}$ and $\mathsf{P}$ read respectively \cite{Berthiere:2019lks, Jain:2021ppx}
\be
\label{eq:qq-pp-neumann}
\langle \hat{q}_i \hat{q}_j  \rangle 
=
\frac{1}{2} \sum_{k=1}^{N-1} \frac{1}{m  \omega_k} \,  V_{i,k} \, V_{j,k} 
\;\;\;\qquad\;\;\;
\langle \hat{p}_i \hat{p}_j  \rangle 
=
\frac{1}{2} \sum_{k=1}^{N-1} m  \omega_k \,  V_{i,k} \, V_{j,k} 
\ee
where
\be
\label{omega-def-neumann}
\omega_k \equiv \sqrt{\omega^2+ \frac{4\kappa}{m} \, \big[ \sin(\theta_k/2)\big]^2  }
\;\;\;\qquad\;\;\;
\theta_k \equiv \frac{\pi ( k-1)}{N-1}
\ee
and
\be
\label{V-matrices-def-neumann}
V_{i,k}= \sqrt{\frac{2-\delta_{k,1}}{N-1}} \, \cos\!\big(\theta_k (i-1/2)\big) \,.
\ee
The correlators in (\ref{eq:qq-pp-neumann}) can be written as
$\langle \hat{q}_i \hat{q}_j  \rangle = \tfrac{1}{2} \,M^{(1)}_{i,j}$ 
and $\langle \hat{p}_i \hat{p}_j  \rangle = \tfrac{1}{2}\, M^{(-1)}_{i,j}$,
where $  M^{(\eta)}_{i,j}$ is defined as follows
\begin{equation}
  M^{(\eta)}_{i,j}
  \equiv
  \frac{1}{(N-1)\,(m\omega)^{\eta}}
  +
  \frac{2}{N-1}  \sum_{k=2}^{N-1}\frac{ \cos[\theta_k(i-1/2)] \,\cos[\theta_k(j-1/2)] }{ \big(2\sqrt{\kappa m}\,\big)^\eta \, \big\{  m\omega^2/(4\kappa)+ [\sin(\theta_k/2)]^2 \big\}^{\eta/2} } \;.
\end{equation}
In the thermodynamic limit $N \to \infty$, this expression becomes
\bea
\label{pp4}
& & \hspace{-2cm}
\mathcal{I}^{(\eta)}_{i,j}
\,\equiv\,
\frac{2}{\pi \, \big(2\sqrt{\kappa m}\,\big)^\eta }  
\int_0^\pi 
\frac{ \cos[\theta(i-1/2)] \,\cos[\theta(j-1/2)]}{ \big\{  m\omega^2/(4\kappa)+ [\sin(\theta/2)]^2 \big\}^{\eta/2} }
\; d\theta
\\
\rule{0pt}{.9cm}
& & \hspace{-1.3cm}
=\,
\frac{1}{\pi  \big(\sqrt{2\kappa m}\,\big)^\eta \, \big[ m\omega^2/(2\kappa) + 1 \big]^{\eta/2}} \, 
\int_0^\pi \!
\frac{ \cos[\theta(i+j-1)] + \cos[\theta(i-j)]}{\big\{ 1- \big[ m\omega^2 /(2\kappa) + 1\big]^{-1} \cos(\theta) \big\}^{\eta/2}}
\; d\theta \,.
\nonumber
\eea
By employing the following formula \cite{Gradshteyn:2007:Table}
\be
\label{F-func-int-def}
F(a,b ; n,\tilde{a}^2) 
\,\equiv
\frac{1}{\pi}
\int_{0}^{\pi} \frac{\cos (n \theta)}{(1-a \cos \theta)^{b}} \, d \theta
\,=\,
\frac{2^b\,\Gamma(n+b)}{a^b\,\Gamma(n+1)  \Gamma(b)}\; \tilde{a}^{n+b}\;
     {}_{2} F_{1} \big(b, n+b \,; n+1\, ; \tilde{a}^{2}\big)
     \ee
     with  $\tilde{a} \equiv \big(1 -\sqrt{1-a^2} \,\big)/a$,
     the integral in the last step of (\ref{pp4}) can be performed, finding 
     \be
     \label{I-cal-def}
     \mathcal{I}^{(\eta)}_{i,j}
     \,=\,
     \frac{1}{\big(\sqrt{2\kappa m}\, \tilde{\omega} \big)^\eta } \, 
     \Big[\,
       F\big(  1/\tilde{\omega}^2 ,  \eta/2\, ; i + j -1,\tilde{a}^2\big)
       +
       F\big(  1/\tilde{\omega}^2 ,  \eta/2\, ; i - j,\tilde{a}^2  \big)
       \,\Big]
     \ee
     where $ \tilde{\omega}^2 \equiv m\omega^2/(2\kappa) + 1$.
     This observation allows us to write the analytic expressions of the correlators in (\ref{eq:qq-pp-neumann})
     in the thermodynamic limit in terms of (\ref{I-cal-def}) 
     as follows
     \be
     \label{qq-pp-neumann}
     \langle \hat{q}_i \hat{q}_j  \rangle  = \frac{1}{2} \, \mathcal{I}^{(1)}_{i,j}
     \;\;\;\qquad\;\;\;
     \langle \hat{p}_i \hat{p}_j  \rangle  = \frac{1}{2} \, \mathcal{I}^{(-1)}_{i,j} \,.
     \ee

     In the massless limit $\omega \to 0$, these expressions become respectively
     \bea
     \label{corr qq neumann thermo}
     & &
     \hspace{-.2cm}
     \langle \hat{q}_i \hat{q}_j  \rangle 
     =
     -\frac{1}{2\pi \sqrt{\kappa m}} 
     \Big(  
     \psi \big( |i-j| +\tfrac{1}{2} \big) +  \psi \big(i+j + \tfrac{1}{2}\big) 
     + \log(m\omega^2/\kappa) +4\gamma + \log(4) + 2\, \psi\big(\tfrac{1}{2}\big) 
     \Big)
     \nonumber
     \\
     & &
     \\
     \label{corr pp neumann thermo}
     & & \hspace{-.2cm}
     \langle \hat{p}_i \hat{p}_j  \rangle =
     - \frac{\sqrt{\kappa m}}{2\pi}  \left( \frac{1}{(i-j)^2-1/4} + \frac{1}{(i+j-1)^2-1/4} \right) 
     \eea
     where  $\gamma=-\psi(1)\simeq 0.5772$  is the Euler-Mascheroni constant. 
     In all our numerical analyses we have set $\kappa=m=1$.

     We remark that all 
     the finite correlation matrices $\mathsf{Q}$ and $\mathsf{P}$ 
     introduced above are symmetric matrices satisfying 
       $\mathsf{Q} \,\mathsf{P} = \tfrac{1}{4}\,\boldsymbol{1}$, 
       where $\boldsymbol{1}$ is the identity matrix, as expected for the ground state.

     An important feature to highlight is the occurrence of the zero mode,
     which is forbidden by the Dirichlet b.c. (\ref{dirichlet-bc-hc}),
     while it is  instead allowed by the Neumann b.c. (\ref{neumann-bc-hc}).
     The zero mode has a significant impact on the behaviour of the correlators in the massless limit $\omega \to 0$.
     Indeed, while the correlators (\ref{qq-dirichlet}) and (\ref{pp-dirichlet}) satisfying Dirichlet b.c. yield finite results in this limit, 
     the ones (\ref{eq:qq-pp-neumann}) characterised by Neumann b.c. are divergent because of the term 
     corresponding to $k=1$ in $\langle \hat{q}_i \hat{q}_j \rangle$.
     This crucial feature makes the analysis of the massless regime more complicated when Neumann b.c. hold.
     The occurrence of the zero mode induces to introduce a small but non-vanishing mass as regulator,
     but this introduces effects in the entanglement entropies that are challenging 
     to quantify through analytic methods \cite{Yazdi:2016cxn}.

     The entanglement entropies $S_A^{(n)}$ of a block $A$ made by $N_A$ consecutive sites can be computed 
     through a well established method 
     \cite{Peschel:2003rdm, Audenaert:2002xfl, Botero04, Plenio:2004he, Cramer:2005mx, Schuch_2006, 
EislerPeschel:2009review, Casini:2009sr, Eisert:2008ur}.
     The first step consists in introducing 
     the reduced correlation matrices $\mathsf{Q}_A$ and $\mathsf{P}_A$, 
     whose generic elements are respectively $\langle \hat{q}_i \hat{q}_j  \rangle $ and $\langle \hat{p}_i \hat{p}_j  \rangle $, 
     with  $i, j \in A$.
     Then, the R\'enyi entropies $S_A^{(n)}$ are obtained as follows
     \be
     \label{renyi-HC-mu}
     S_A^{(n)}
     =
     \frac{1}{n-1}\sum_{j=1}^{N_A}
     \log \left[
       \left(\mu_{j}+\frac{1}{2}\right)^{n} \! - \left(\mu_{j}-\frac{1}{2}\right)^{n}
       \,\right]
     \ee
     where $\{ \mu_1^2,\dots,\mu_{N_A}^2 \}$ is the spectrum of the $N_A \times N_A$ matrix $\mathsf{Q}_A\,\mathsf{P}_A$
     and provides the symplectic eigenvalues $\{ \mu_1,\dots,\mu_{N_A} \}$
     of the covariance matrix $\mathsf{Q}_A \oplus \mathsf{P}_A$.
     The limits $n \to 1$ and $n \to \infty$ of (\ref{renyi-HC-mu}) give respectively 
     the entanglement entropy
     \be
          \label{ee-HC-mu}
     S_A
     =
    \sum_{j=1}^{N_A}
     \log \left[
       \left(\mu_{j}+\frac{1}{2}\right)  \log\! \left(\mu_{j}+\frac{1}{2}\right)
       - 
      \left(\mu_{j}-\frac{1}{2}\right)  \log\! \left(\mu_{j}-\frac{1}{2}\right)
       \,\right]
     \ee
     and the single copy entanglement 
     \be
     \label{SCE-HC-mu}
     S_A^{(\infty)}
     =
    \sum_{j=1}^{N_A}
     \log \! \left(\mu_{j}+\frac{1}{2}\right) .
     \ee

     In the following we report some numerical results for the entanglement entropies of a block $A$ made by 
     $N_A$ consecutive sites providing the bipartitions shown in Fig.\,\ref{figure-biparts-intro},
     when either Dirichlet b.c. or Neumann b.c. are imposed and the whole harmonic chain is in its ground state. 
     This leads to four possible setups for the harmonic chain: 
     either an infinite chain on the semi-infinite line 
     or a finite chain made by $N$ consecutive sites on the segment,
     and either Dirichlet b.c. or Neumann b.c. (we stress that, in the case of the segment, 
     the same b.c. is chosen at both its boundaries).     
     The continuum limit of the lattice results for the semi-infinite chain and for the segment
     are compared against the corresponding BCFT expressions for the
     massless scalar field in the decompactification regime, 
     either on the right half plane $\mathbb{H}$ 
     or on the strip $\mathbb{S}$ respectively.

     \begin{figure}[t!]
\vspace{-.5cm}
\hspace{-1.6cm}
  \includegraphics[width=1.16\textwidth]{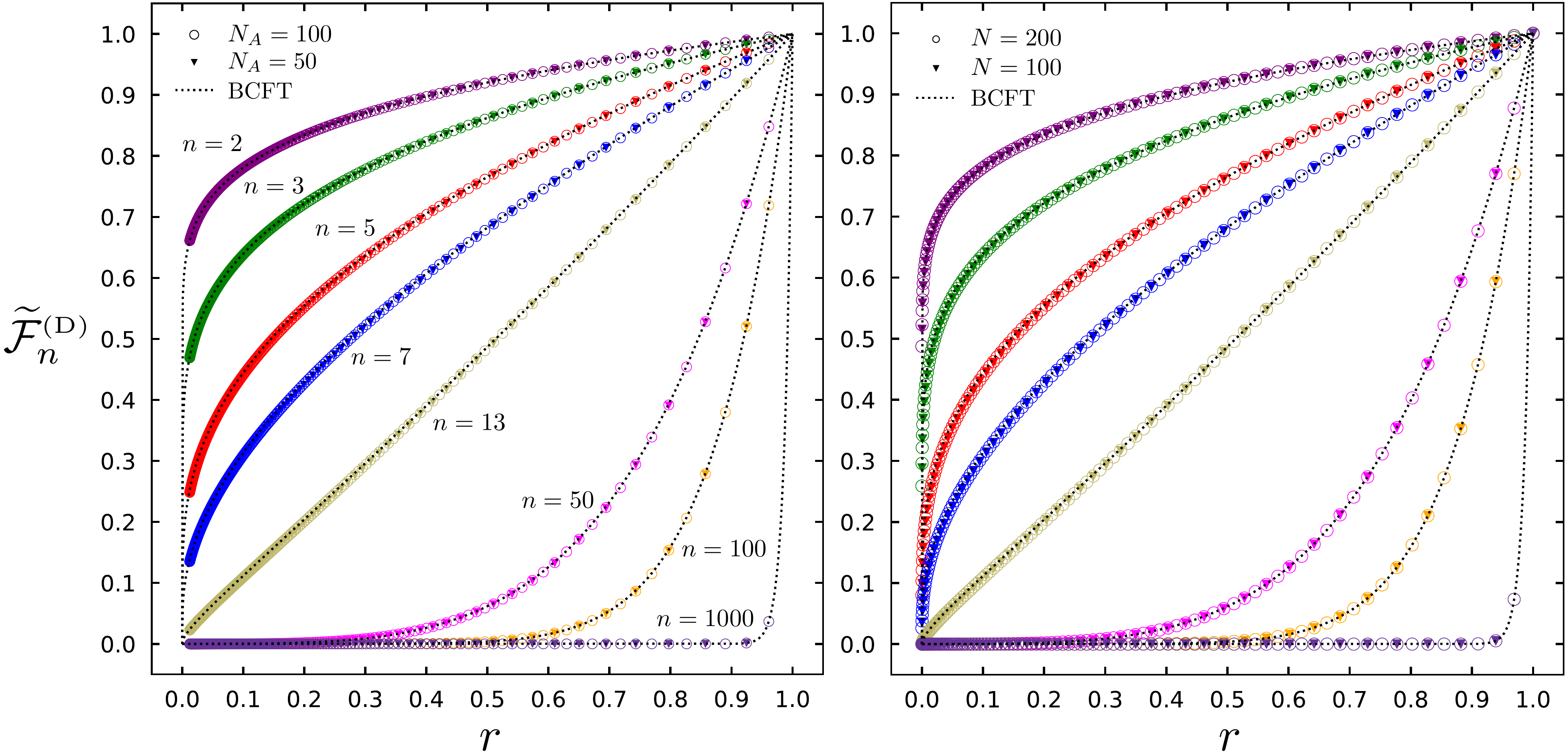}
  \caption{ 
  The function $\widetilde{\mathcal{F}}_n^{\textrm{\tiny (D)}}$ 
  for the bipartitions in Fig.\,\ref{figure-biparts-intro} with Dirichlet boundary conditions. 
  The black dotted lines correspond to the BCFT result given by the first expression in (\ref{F-DN-dec-def}).
  The data points have been obtained for the block of $N_A$ consecutive sites 
  either in the semi-infinite chain (left panel) or on the finite chain made by $N$ sites (right panel)
  with Dirichlet b.c. (see Sec.\,\ref{sec_HC_numerics}).
  }
  \label{fig:FnDirichlet}
\end{figure}

Unless stated otherwise, for the semi-infinite chains the interval size $N_A$ is kept fixed
  while its distance from the boundary is varied in such a way that $r$ covers the whole range $r\in(0,1)$. 
  Instead, in the finite chains of even size $N$, the whole range $r\in(0,1)$ is spanned
  by keeping one endpoint of $A$ fixed in the middle of the chain   
  (this is not ambiguous for even values of $N$)
  while the interval grows towards one of the boundaries of the finite chain.

  In the case of Dirichlet b.c., our lattice data are taken employing 
  the correlators (\ref{qq-dirichlet}) with $\omega = 0$  and (\ref{corr qq dirichlet thermo}),
  which are well defined expressions. 
    Instead, when Neumann b.c. are imposed,  
     the first correlator in (\ref{eq:qq-pp-neumann}) diverges in the massless limit $\omega\rightarrow 0$
     because of the occurrence of the zero mode corresponding to $k=1$.
     To avoid this issue, we set the mass parameter $\omega$ to a small non-vanishing value.
     In our analysis we have chosen $\omega N_A\sim10^{-10}$ for the semi-infinite chains
     and $\omega N \sim10^{-10}$ for the finite chains, 
     which are much smaller than the other scales. 
     This procedure is the standard one in the case of periodic b.c. (or infinite chain), where the zero mode occurs as well.

     For the harmonic chains on the semi-infinite line, 
     when Dirichlet b.c. are chosen we have observed that block sizes $N_A\in\{50,100\}$ are large enough to obtain
  a nice agreement with the BCFT predictions.
  Instead, for Neumann b.c. large sizes for the blocks are typically needed:
  we used $N_A\in\{200,400\}$ for the semi-infinite chains
  and $N\in\{100,200\}$ for the finite chains.
     
     We find it worth remarking that, in all the figures of this manuscript, 
the lattice data corresponding to Dirichlet b.c. have not been shifted to be compared with the BCFT curves for all the UV finite quantities considered. 
Instead,  for the ones corresponding to Neumann b.c., 
we have to introduce a constant shift that we are not able to determine analytically
which depend on the R\'enyi index and the lattice zero-mode regulator.
 It would be interesting to establish a quantitative relation between this shift and $\omega$, if it exists 
(see e.g. \cite{Yazdi:2016cxn,Jain:2021ppx} for some results in this direction).

\begin{figure}[t!]
\vspace{-.5cm}
\hspace{-1.6cm}
  \includegraphics[width=1.16\textwidth]{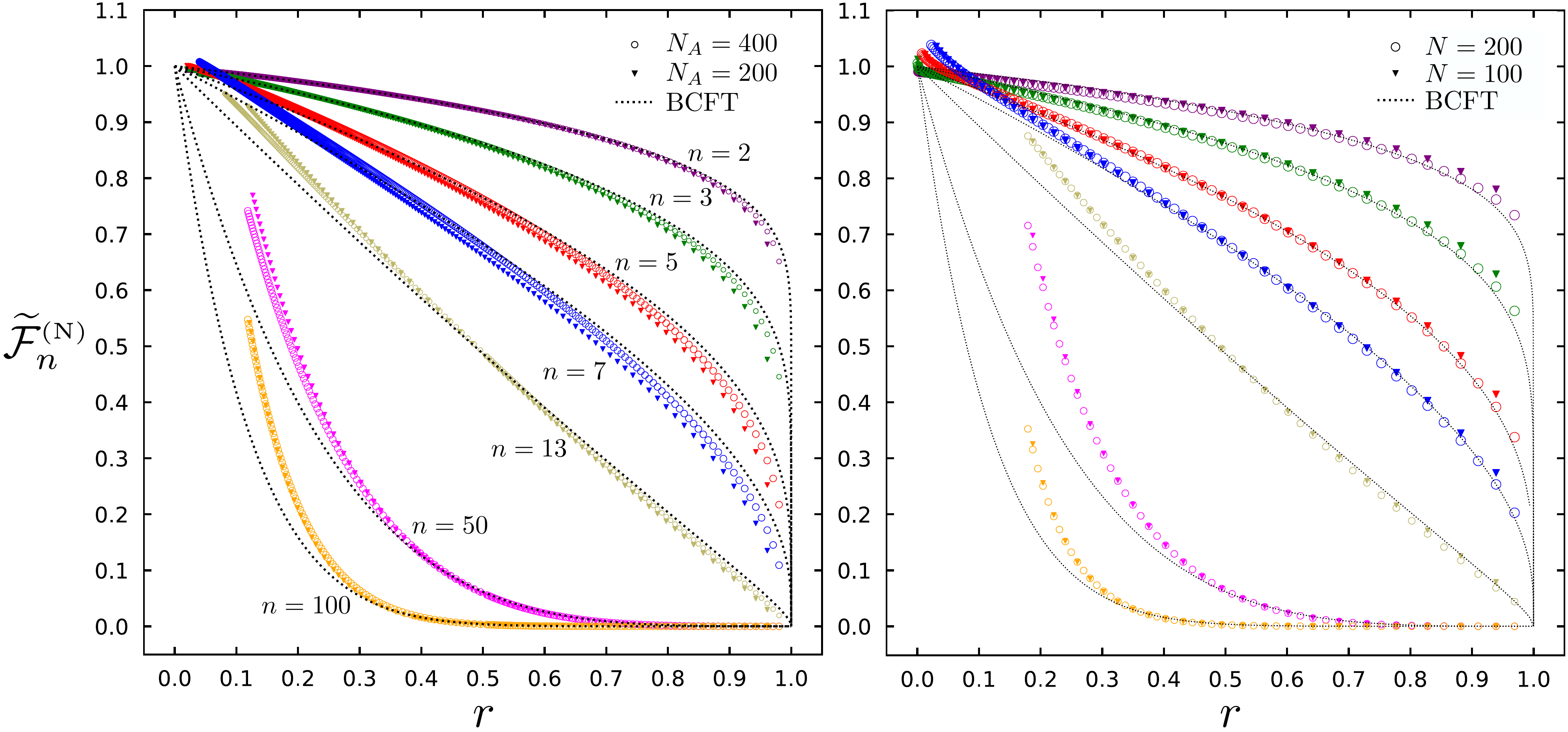}
 \caption{ 
   The function $\widetilde{\mathcal{F}}_n^{\textrm{\tiny (N)}}$ 
  for the bipartitions in Fig.\,\ref{figure-biparts-intro} with Neumann boundary conditions. 
  The black dotted lines correspond to the BCFT result given by the second expression in (\ref{F-DN-dec-def}).
  The data points have been obtained for the block of $N_A$ consecutive sites 
  either in the semi-infinite chain (left panel) or on the finite chain made by $N$ sites (right panel)
  with Neumann b.c., as discussed in Sec.\,\ref{sec_HC_numerics}.
  }
  \label{fig:FnNeumann}
\end{figure}

\begin{figure}[t!]
\vspace{-.5cm}
\hspace{-1.6cm}
  \includegraphics[width=1.16\textwidth]{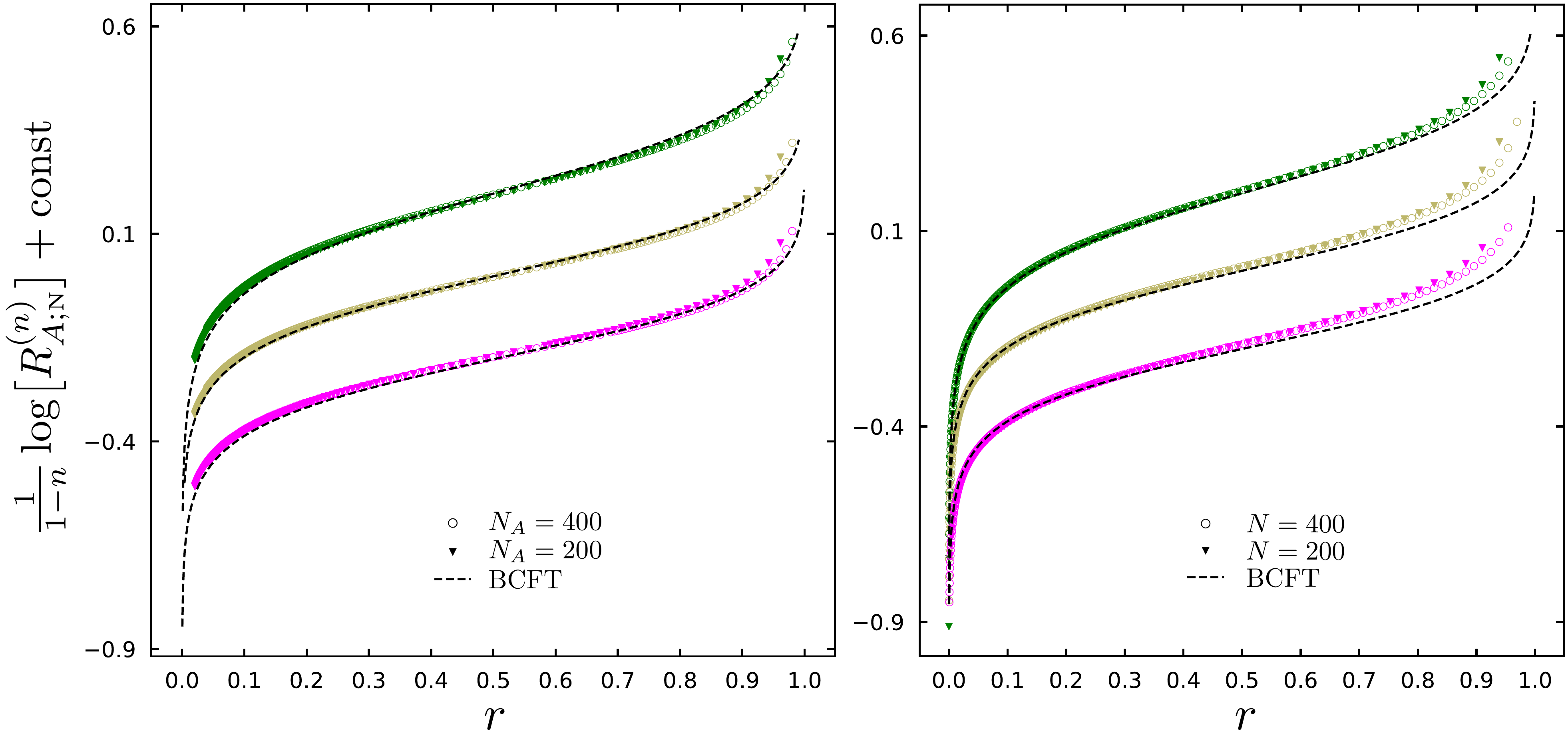}
 \caption{ 
   The UV finite ratio (\ref{R-ratio-def})
  for the bipartitions shown in Fig.\,\ref{figure-biparts-intro} with Neumann b.c.,
  either in the semi-infinite chains (left panel) or in the finite chains (right panel),
  and $n \in \{3, 13, 50\}$, with the same colour code of Fig.\,\ref{fig:FnNeumann}. 
  Artificial vertical shifts of the curves for different values of $n$ have been introduced 
  for better visibility. }
  \label{fig:logRnNeumann}
\end{figure}

    The subsystems $A = [u, v]$, $A_u = [0, u]$, and $A_v = [0, v]$ are used to construct the ratios (\ref{R-ratio-def}), 
    which represent combinations of entanglement entropies in (\ref{MI-n-bdy-def}).
    The expressions for the UV finite ratios (\ref{R-ratio-def}) in BCFT are obtained by combining (\ref{R-ratios-cft-def}) and (\ref{F-DN-dec-def}). 
    Additionally, they suggest to consider $r^{\Delta_n} R^{(n)}_A$ for both finite and semi-infinite chains.
     From (\ref{tilde-r-ratio-def}), we have that the ratio $r$ is 
     $r=[s(N_v-N_u)/s(N_v+N_u)]^2 $ for the finite chains
     and $r=[(N_v-N_u)/(N_v+N_u)]^2 $ for the semi-infinite chains.
The numerical data for these quantities 
and the corresponding BCFT expressions
are displayed in Fig.\,\ref{fig:FnDirichlet} for Dirichlet b.c.,
while the ones for Neumann b.c. are shown 
in Fig.\,\ref{fig:FnNeumann} and Fig.\,\ref{fig:logRnNeumann}.
In the case of Dirichlet b.c.,
a remarkable agreement between the lattice data points in the scaling limit
and the BCFT predictions is observed, 
for any value  of $r$ considered 
and even for very large values of the R\'enyi index $n$.
Instead, when Neumann b.c. are imposed,
we obtain a nice agreement for $n \leqslant 7$ and $r \in (0.5, 1)$.
In order to understand these discrepancies, 
for Neumann b.c. we have reported also the UV finite ratio (\ref{R-ratio-def}) 
in the cases of $n=3$, $n=13$ and $n=50$ (see Fig.\,\ref{fig:logRnNeumann}),
with the same colour code adopted in Fig.\,\ref{fig:FnNeumann}.
 To get a better visibility of the data, the curves for different values of $n$ have been vertically displaced.
The relation between the BCFT expressions of the quantities considered 
in Fig.\,\ref{fig:FnNeumann} and Fig.\,\ref{fig:logRnNeumann} is given in (\ref{R-ratios-cft-def}).
The agreement between the lattice data points and the BCFT curves in Fig.\,\ref{fig:logRnNeumann}
suggests that
larger blocks and system sizes are needed in Fig.\,\ref{fig:FnNeumann}
to obtain a better match with the BCFT predictions.

\begin{figure}[b!]
\vspace{-1.5cm}
\hspace{-1.6cm}
  \includegraphics[width=1.16\textwidth]{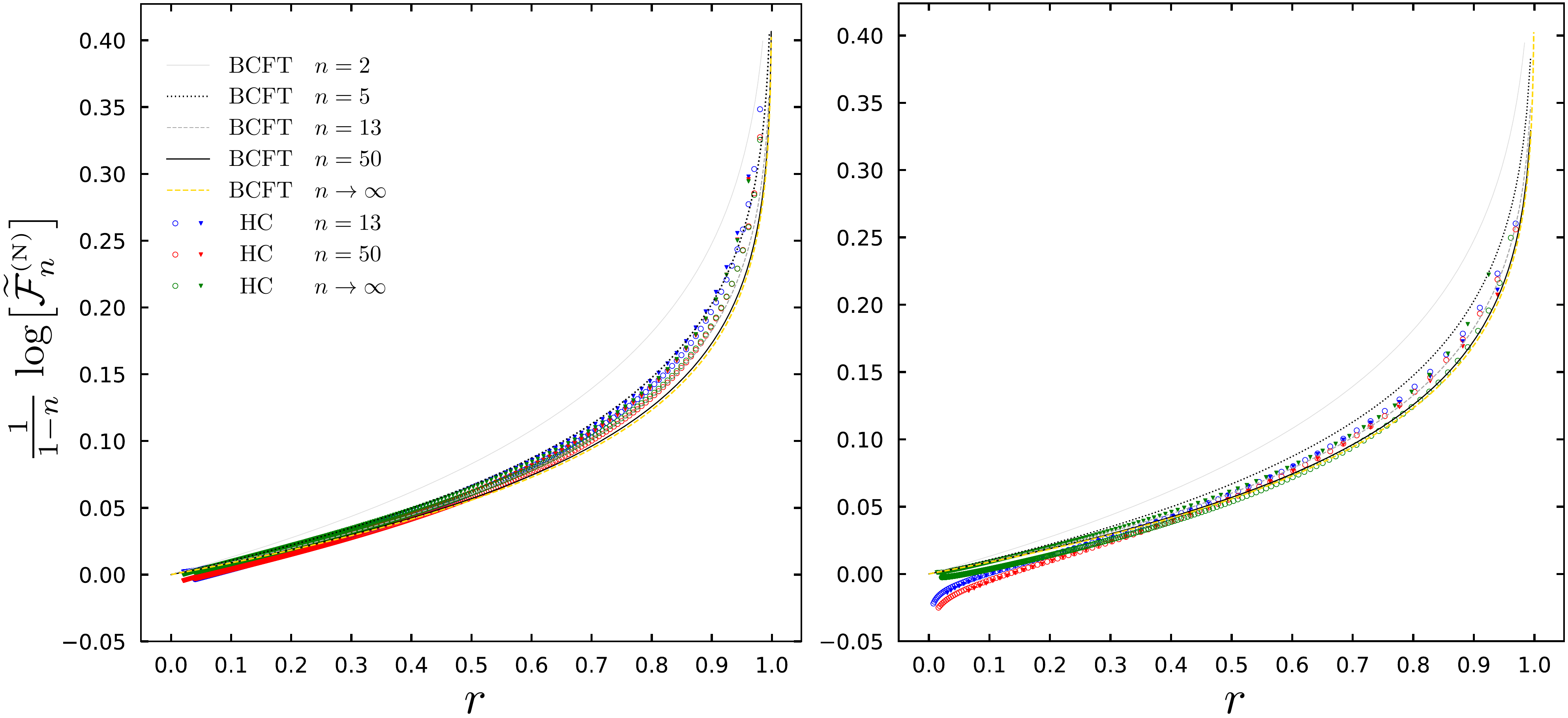}
 \caption{
  Single copy entanglement entropy for the bipartitions in Fig.\,\ref{figure-biparts-intro} (see (\ref{sc-hc-combination-figs}))
 when Neumann b.c. are imposed.
 The same harmonic chains of Fig.\,\ref{fig:FnNeumann} have been employed. 
  }
  \label{fig:FnNeumannLargeRenyi}
\end{figure}

The occurrence of the zero mode in the case of Neumann b.c. could be another possible source of the 
the discrepancy observed in Fig.\,\ref{fig:FnNeumann}.
As $n$ is increased to larger values at fixed $N$ or $N_A$, 
one needs smaller values of $\omega$ to get a convergence of the lattice data points (up to shift). 
Notice that the data for $\widetilde{\mathcal{F}}_n^{\textrm{\tiny (N)}}(r)$ is particularly sensitive to these effects, 
in contrast with the quantities shown in Fig.\,\ref{fig:logRnNeumann} and Fig.\,\ref{fig:FnNeumannLargeRenyi},
which contain the same essential information. 
We leave the quantitative analysis of such discrepancies for future work.

\begin{figure}[t!]
\vspace{-.5cm}
\hspace{-1.6cm}
  \includegraphics[width=1.16\textwidth]{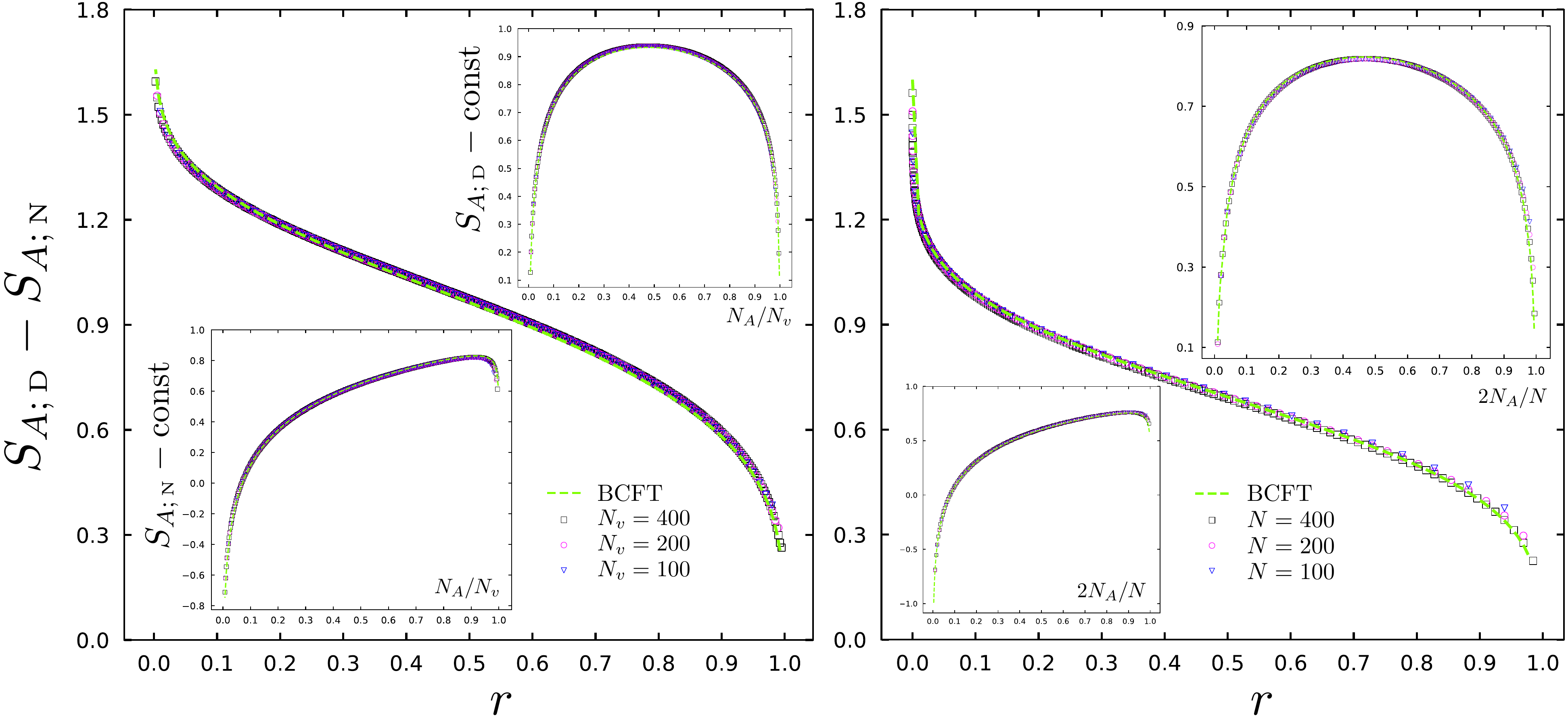}
  \caption{ 
     The difference between the entanglement entropy for Dirichlet b.c. and Neumann b.c. for the bipartitions of Fig.\,\ref{figure-biparts-intro},
     either in the semi-infinite chains (left panel) or in the finite chains (right panel).
     The insets show the entanglement entropy for the two different boundary conditions. 
     The BCFT curves have been obtained from (\ref{eq:EE-allBC-allgeometries}).
}
  \label{fig:FnEntropyDiff}
\end{figure}

In Sec.\,\ref{sec_main_BCFT_expression} we have observed that 
the difference (\ref{renyi-differences}) between the entanglement entropies corresponding to two different conformally invariant 
boundary conditions  is another interesting UV finite quantity to investigate
and it is a function of the harmonic ratio $r$.
For the massless scalar field in the decompactification regime that we are exploring, 
we consider (\ref{renyi-diff-dec-2F1}) and its analytic continuation $n \to 1$, 
which can be easily written by using (\ref{eq:EE-allBC-allgeometries}).
The results of our analyses for this UV finite quantity are shown in Fig.\,\ref{fig:FnEntropyDiff}.

The collection of the lattice data has been performed by setting $N_v$ to a constant value in both setups: 
we have chosen $N_v\in\{100,200,400\}$ for the semi-infinite chains,
while we took $N_v=N/2$ with $N\in\{100,200,400\}$  for the finite chains.
Then, we  varied $N_u\in\{1,\dots,N_v-1\}$ and plotted the entanglement entropy in terms of $N_A/N_v$ 
and the entropy difference in terms of the corresponding cross ratio $r$  in each case.
For Neumann b.c., we have set $\omega N_v \sim 10^{-10} $ in both the semi-infinite and finite chains.
In the case of Dirichlet b.c. we have considered $\omega=0$, but we checked that 
introducing a small non-vanishing $\omega$, such that $\omega N_v \sim 10^{-10}$,
does not lead to changes that can be observed. 
The agreement between the lattice data points and the corresponding BCFT predictions is excellent. 
In the insets of Fig.\,\ref{fig:FnEntropyDiff}
we have reported $S_{A; \alpha} - \textrm{const}$, where the constant value that has been subtracted
is $\tfrac{1}{6}\log (2N_v)$ for the semi-infinite chains (left panel)
and $\tfrac{1}{3}\log (2N/\pi)$ for finite chains (right panel).

\begin{figure}[t!]
\vspace{-.5cm}
\hspace{-1.6cm}
  \includegraphics[width=1.16\textwidth]{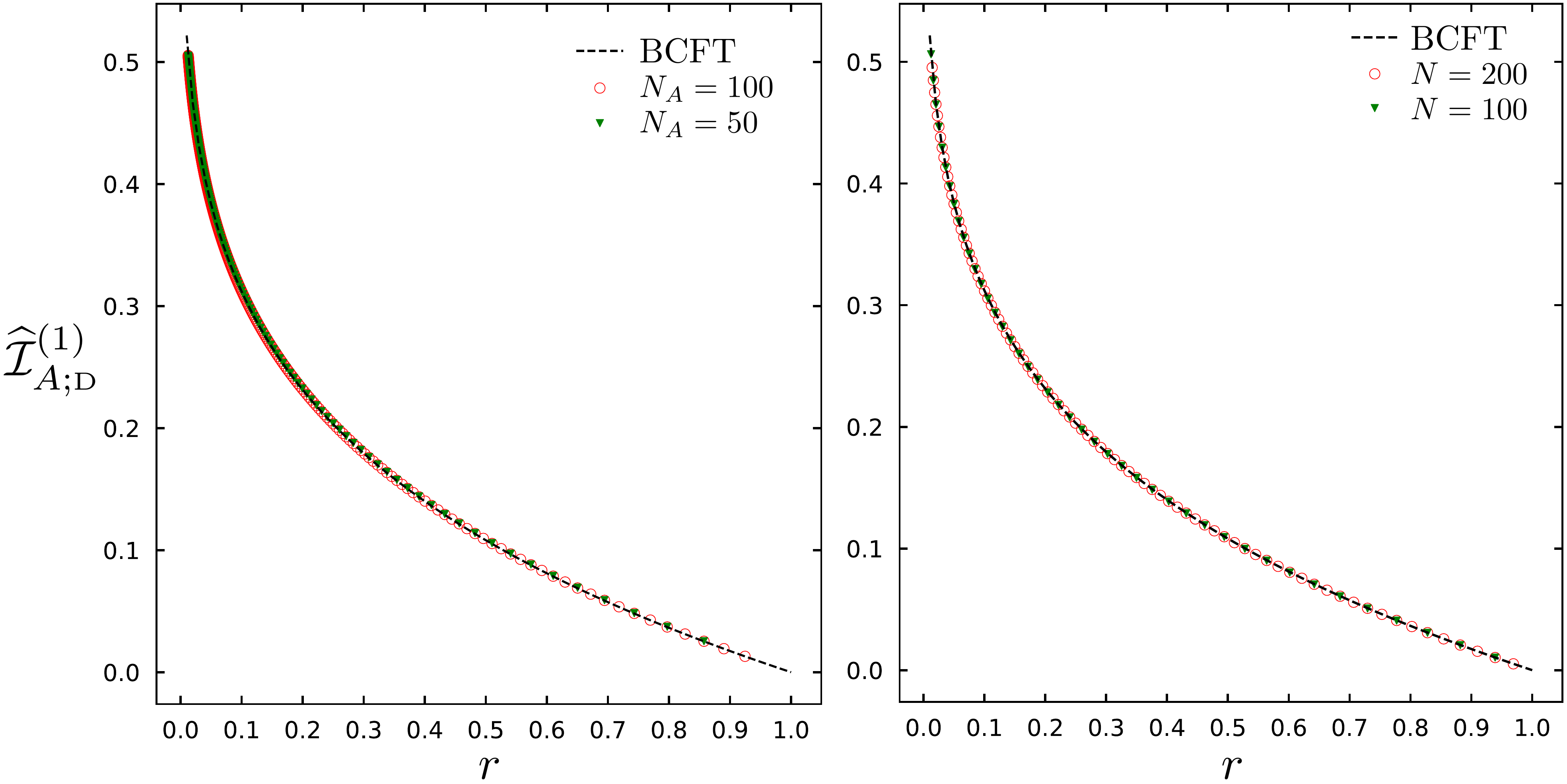}
  \caption{
  The UV finite quantity (\ref{eq:Ifunction-lattice}) for Dirichlet boundary conditions. 
  The dashed curves correspond to the BCFT prediction (\ref{MI-hat-def-n=1}). 
  The results for the semi-infinite chains and for the finite chains made by $N$ sites 
  are reported in the left and right panel respectively.
  }
  \label{fig:Mutual-Info-dec-Dir}
\end{figure}

\begin{figure}[b!]
\vspace{-.2cm}
\hspace{-1.6cm}
  \includegraphics[width=1.16\textwidth]{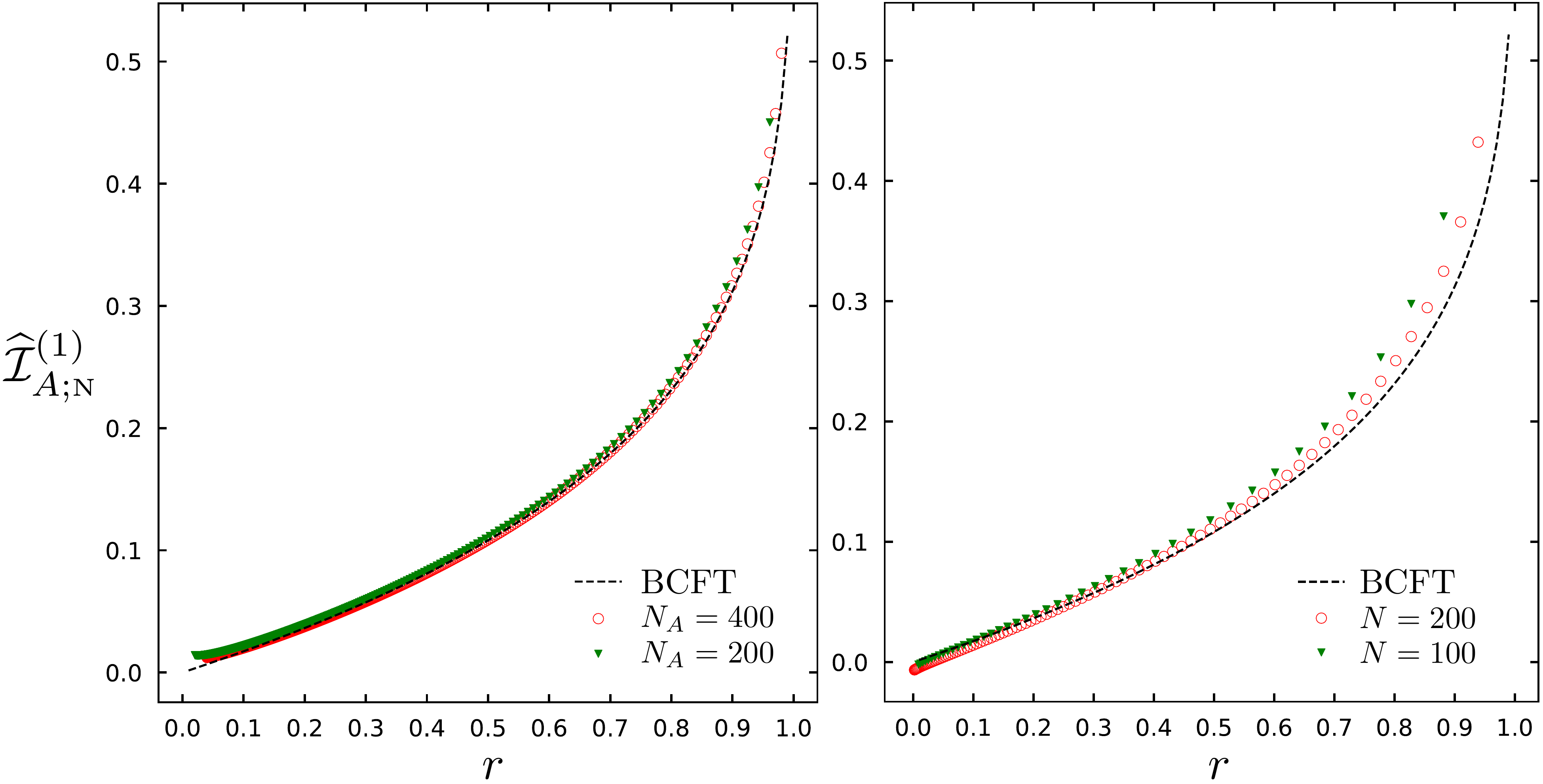}
  \caption{
  The UV finite quantity (\ref{eq:Ifunction-lattice}) for Neumann boundary conditions,
  with the same notation described in the caption of Fig.\,\ref{fig:Mutual-Info-dec-Dir}.
  }
  \label{fig:Mutual-Info-dec-Neu}
\end{figure}

In Sec.\,\ref{subsec:decompactification_derivations} we have introduced the UV finite quantity (\ref{MI-hat-def}), 
obtained from (\ref{R-ratios-cft-def}) and (\ref{MI-n-bdy-def}), 
which depends on the cross ratio $r$.
Its limit $n \to 1$ is given by  the following combination
\be 
\label{eq:Ifunction-lattice}
\widehat{\mathcal{I}}_{A;\alpha}^{(1)}
=
 \lim_{n\rightarrow 1} 
  \frac{ \log \! \big( r^{\Delta_n}  R^{(n)}_A  \big) }{1-n}
  \,=\,
  S_{A_u}+S_{A_v}-S_A +\frac{1}{6} \log (r) \,.
\ee
Our results for this UV finite quantity are reported 
in Fig.\,\ref{fig:Mutual-Info-dec-Dir} and Fig.\,\ref{fig:Mutual-Info-dec-Neu},
for Dirichlet b.c. and Neumann b.c. respectively. 
The corresponding BCFT expressions are given by (\ref{MI-hat-def-n=1}), 
which provide the dashed lines in these figures.

When imposing Dirichlet b.c., we observe an excellent agreement 
between the lattice data points and the BCFT prediction.
Instead, for Neumann b.c. a slight discrepancy is found.
It might be due to the occurrence of the zero mode, whose effect we are not able to characterize. 

\begin{figure}[t!]
\vspace{-.5cm}
\hspace{-1.6cm}
  \includegraphics[width=1.16\textwidth]{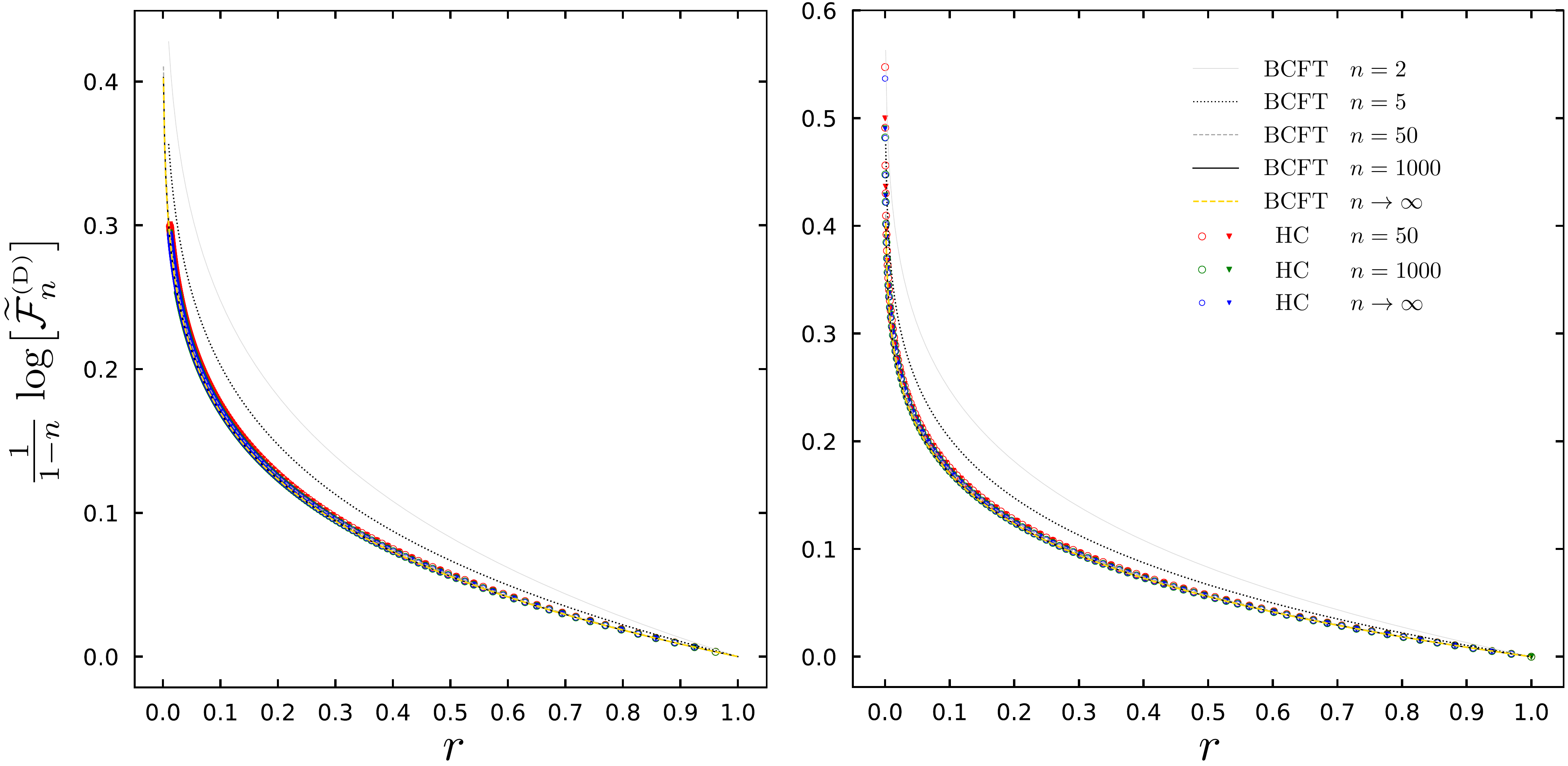}
 \caption{
 Single copy entanglement entropy for the bipartitions in Fig.\,\ref{figure-biparts-intro} (see (\ref{sc-hc-combination-figs}))
 when Dirichlet b.c. are imposed.
  The same harmonic chains of Fig.\,\ref{fig:FnDirichlet} have been employed.   }
  \hspace{-1cm}
  \label{fig:FnDirichletLargeRenyi}
\end{figure}

   The numerical analysis discussed above allow to evaluate also 
the corresponding single copy entanglement (\ref{single copy-def-intro}) 
by applying (\ref{SCE-HC-mu}).
The resulting numerical data in the thermodynamic limit can be compared with the BCFT predictions
given by (\ref{sc-cft-explicit-1}) and (\ref{sc-cft-explicit-0}).
These comparisons are shown in Fig.\,\ref{fig:FnDirichletLargeRenyi} and Fig.\,\ref{fig:FnNeumannLargeRenyi},
for Dirichlet and Neumann b.c. respectively
(semi-infinite chains and finite chains made by $N$ sites have been considered in the left and right panels respectively). 
In these figures, the harmonic chain data for $n \to \infty$ 
have been  obtained through (\ref{SCE-HC-mu}).

     \newpage
   
\clearpage

\clearpage

In particular, by using (\ref{R-ratios-cft-def}), one obtains 
\be
\label{sc-hc-combination-figs}
  \lim_{n\rightarrow \infty} 
  \frac{ \log \! \big( r^{\Delta_n}  R^{(n)}_A  \big) }{1-n}
  \,=\,  
  S_A^{(\infty)} - S_{A_u}^{(\infty)}- S_{A_v}^{(\infty)} -\frac{1}{12} \log (r) \,.
\ee

Also in this analysis we observe an excellent agreement between the lattice data points in the scaling limit and the 
corresponding BCFT predictions for Dirichlet b.c.,
while for Neumann b.c. some discrepancy occurs 
(see the right panel of Fig.\,\ref{fig:FnNeumannLargeRenyi}, for small values of $r$).

\section{Quantum quenches}
\label{sec-quenches}

A quantum quench is an efficient procedure to construct an out-of-equilibrium state. 
Given an initial state $| \psi_0\rangle$ (e.g. the ground state of some Hamiltonian $H_0$)
at $t=0$ and the Hamiltonian $H$ such that $| \psi_0\rangle$  is not among its eigenstates,   
the evolution $| \psi(t)\rangle \equiv \e^{\ri H t} | \psi_0\rangle$ for $t\geqslant 0$
provides an interesting out-of-equilibrium state to investigate. 
A global quench occurs e.g. when the Hamiltonian $H$ is defined through 
a sudden change of a parameter in $H_0$,
which therefore occurs everywhere in the system. 
Instead, in a local quench protocol $H$ is obtained 
through a sudden modification which is local in space \cite{eisler-peschel-07-local};
e.g. when two initially separated systems are connected at $t=0$.

We are interested in a class of quenches where $H$ is the Hamiltonian of a CFT,
which has been described by Calabrese and Cardy \cite{Calabrese:2005in, Calabrese:2009qy, Calabrese:2007mtj, Calabrese:2016xau}.
These authors also proposed an approach based on BCFT to investigate 
the temporal evolutions of the entanglement entropies of an interval of length $\ell$
after either a global or local quench. 
This approach employs the two-point function of twist fields in the presence of a boundary;
hence, it provides a natural setup to apply  the results presented in Sec.\,\ref{sec_cft_results}.
An important parameter in this approach is the extrapolation length $\tau_0$, 
which is related to the distance of the initial state $| \psi_0\rangle$ from a conformally invariant boundary state
(we refer the reader to \cite{Calabrese:2016xau} for a detailed discussion).
For the sake of completeness, 
in Appendix\;\ref{app-quenches} we report the derivation of the main formulas of this BCFT analysis
of quantum quenches  that are employed in the following.

For both the global and local quenches, 
the relevant regimes to consider in terms of the harmonic ratio $r = |x|^2$ 
are $r \to 0^+$ and  $r \to 1^-$.
In the case of the compactified massless boson, 
for both Dirichlet b.c. and Neumann b.c.,
we have that $\mathcal{F}_n^{(\alpha)}(r)\rightarrow g_{\alpha}^{2(1-n)}$ as $r\rightarrow 1$ (see \eqref{Fn-r=1-limit})
and $\mathcal{F}_n^{(\alpha)}(r)\rightarrow 1$ as $r\rightarrow 0$ 
(see Sec.\,\ref{subsec-finiteR-two-point-normalization}, in the text before \eqref{2pt-sphere-norm-cond}).
Instead, for the decompactified boson, 
the leading behaviours in the relevant regimes are
\bea 
& & \label{eq:divergence_noncompact_r_to0}
     \widetilde{\mathcal{F}}_n^{\textrm{\tiny (D)}}(r)\sim \frac{1}{\sqrt{n}} \left(\frac{2\pi}{|\log r|}\right)^{(n-1)/2}
     \hspace{.6cm}
      \widetilde{\mathcal{F}}_n^{\textrm{\tiny (N)}}(r)\sim 1
       \hspace{5cm}
           r \to 0
\\
& & \label{eq:divergence_noncompact_r_to1}
    \widetilde{\mathcal{F}}_n^{\textrm{\tiny (D)}}(r)\sim 1 
     \hspace{4.cm}
          \widetilde{\mathcal{F}}_n^{\textrm{\tiny (N)}}(r)\sim \frac{1}{\sqrt{n}} \left(\frac{2\pi}{|\log (1-r)|}\right)^{(n-1)/2}
       \hspace{.7cm}
           r \to 1
                             \hspace{1.5cm}
\eea
where the ones corresponding to Dirichlet b.c. can be read from \eqref{eq:twist-2-point-leading-decomp} and \eqref{eq:Dirichlet-leading-appendix}, 
while the ones for Neumann b.c. are implied by \eqref{eq:neumann-xgoes0-app} and \eqref{eq:Neumann-leading-appendix} .  
From these asymptotic behaviours for the decompactified boson,
we have that $S_A^{(n)}$ diverge like $\frac{1}{2}\log |\log r|$ as $r \to 0$ when Dirichlet b.c. hold,
while $S_A^{(n)}$ diverge like $\frac{1}{2}\log |\log (1-r)|$ as $r \to 1$ when Neumann  b.c. are imposed.

\subsection{Global quench}
\label{sec-global-quench}

In the case of a global quench, 
for the temporal evolution of $\textrm{Tr} \rho_A^n$ for an interval $A$ of length $\ell$,
the BCFT approach predicts the following result \cite{Calabrese:2005in, Calabrese:2009qy}
\begin{equation}
\label{ren-global-quench-final}
\textrm{Tr} \rho_A^n
\,\simeq\,
\left(\frac{\pi}{2\tau_0}\right)^{2\Delta_n} 
\left[ 
\frac{\e^{\pi\ell/(2\tau_0)} \big(\e^{\pi\ell/(2\tau_0)}  + \e^{-\pi\ell/(2\tau_0)}  +2  \cosh (\pi t/\tau_0) \big)
}{
2\big(\e^{\pi\ell/(2\tau_0)} -1\big)^2 \, \cosh(\pi t/\tau_0)}
\right]^{\Delta_n}
\mathcal{F}_n^{(\alpha)}(r)
\end{equation}
(whose derivation is discussed in the Appendix\;\ref{app:global_quench}),
where an $n$-dependent normalization constant has been neglected and for the harmonic ratio
we have\footnote{The harmonic ratio (\ref{r-global-t-dep}) is different from the one  chosen in \cite{Calabrese:2009qy}.}
\begin{equation}
\label{r-global-t-dep}
       r = 
       \frac{(\e^{\pi\ell/(2\tau_0)} - 1 )^2}{ \e^{\pi\ell/(2\tau_0)} \,\big( \e^{\pi\ell/(2\tau_0)} + 2 \cosh (\pi t/\tau_0)+ \e^{-\pi\ell/(2\tau_0)} \big)}
       \sim
       \frac{ \e^{\pi\ell/(2\tau_0)} }{ \e^{\pi\ell/(2\tau_0)} + \e^{\pi t / \tau_0}}
\end{equation}
where the last expression corresponds to the regime where both $\ell /\tau_0 \gg 1$ and $ t/\tau_0 \gg 1$.
The explicit expression of $\mathcal{F}_n^{(\alpha)}(r)$ are given in (\ref{Fn-main-res})
and the temporal dependence of (\ref{r-global-t-dep}) for a given value of $\ell/\tau_0$
is shown in the left panel of Fig.\,\ref{fig:FnAppRatio}.

The expression \eqref{ren-global-quench-final} provides the temporal evolution of the R\'enyi entropies $S_{A;\alpha}^{(n)}$ after the global quench we are considering. 
When the contribution of $\mathcal{F}_n^{(\alpha)}(r)$ is neglected,
which corresponds to set $\mathcal{F}_n^{(\alpha)}(r)=1$ in (\ref{ren-global-quench-final}),
the resulting expression for the R\'enyi entropies
is independent of the boundary conditions
and in the regime where both $\ell /\tau_0 \gg 1$ and $ t/\tau_0 \gg 1$ it reads \cite{Calabrese:2005in}
\begin{equation}
\label{ee-globalquench-known}
S_A^{(n)} \simeq \frac{ \pi\, c\, (n+1)}{12 \tau_0 n} \left[t+\frac{\ell}{2}- \textrm{max}(t,\ell/2)\right]
\end{equation}
i.e. a temporal dependence given by a linear growth followed by a plateau,
where the change between these two regimes occurs at $t/\ell = 1/2$.

\begin{figure}[t!]
\vspace{-.5cm}
\hspace{-1.4cm}
  \includegraphics[width=1.16\textwidth]{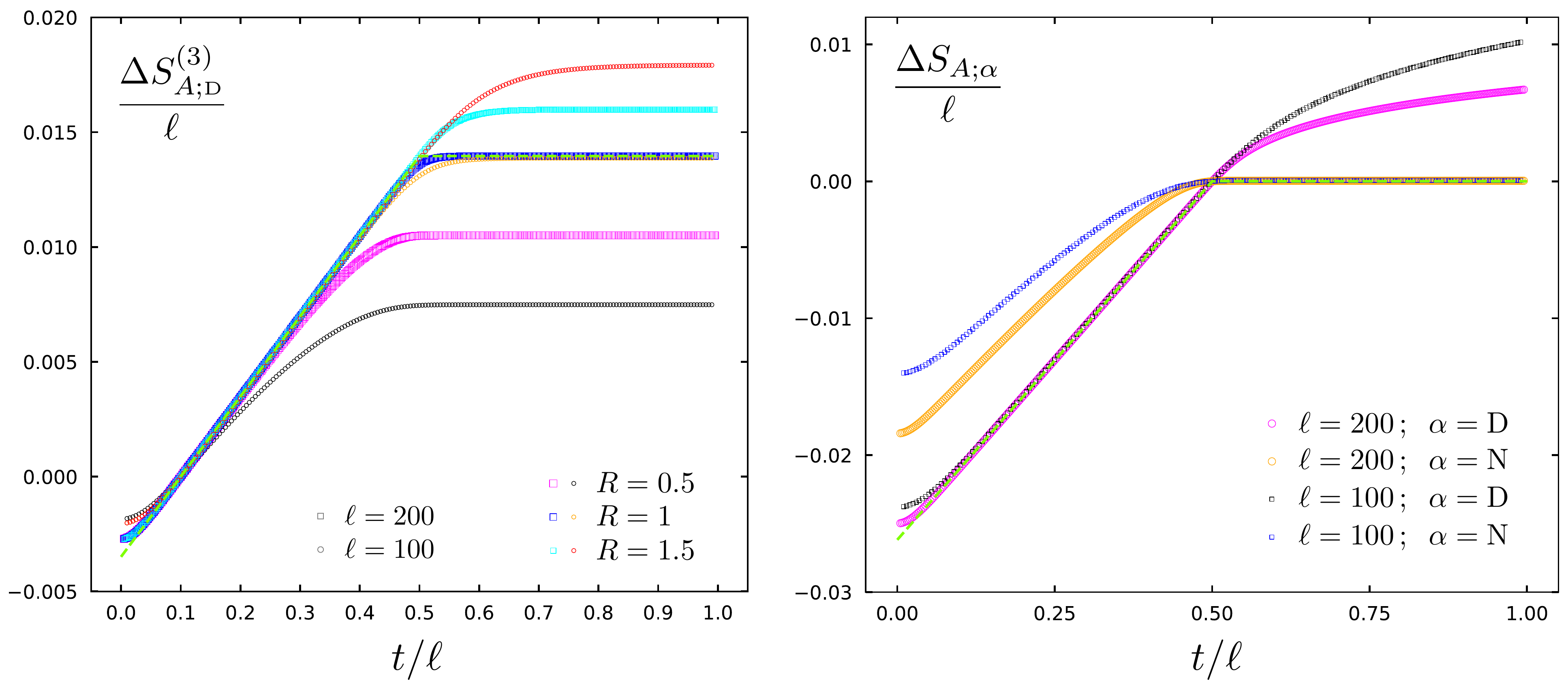}
 \caption{ 
 Temporal evolution of $\Delta S_{A; \alpha}^{n}$ after a global quench for an interval of length $\ell$ and $\tau_0=10$.
 Left: Compact boson with Dirichlet b.c. for $n=3$.
 Right: Decompactified boson with either Dirichlet or Neumann b.c.\,. 
 The dashed green line is given by  \eqref{ee-globalquench-known}  in both panels.
  }
  \label{fig:QuenchGlobal}
\end{figure}

In the left panel of Fig.\,\ref{fig:QuenchGlobal}, the temporal evolution  of 
$\Delta S_{A;\alpha}^{(n)} \equiv S_{A;\alpha}^{(n)} - S_{A;\alpha}^{(n)} \big|_{t/\ell=1/10}$ when $n=3$
is shown in the case of Dirichlet b.c., for $\ell\in\{100,200\}$, $\tau_0=10$ and some values of $R$.
The case of Neumann b.c. is obtained by replacing $R$ with $2/R$; hence the same qualitative behaviour is observed for the temporal dependence. 
The choice of the subtraction constant corresponding to $t/\ell=1/10$ is due to numerical issues that do not allow us to evaluate $S_{A;\alpha}^{(n)}$ at $t/\ell \simeq 0$.
We remark that, for the compact boson, in the regime given by $t\gg\tau_0$ and $\ell\gg\tau_0$,
the temporal evolution provided by \eqref{ren-global-quench-final} is compatible with \eqref{ee-globalquench-known} 
(see the dashed green curve in the left panel of Fig.\,\ref{fig:QuenchGlobal}), up to a constant.
Such compatibility is due to the fact that $\mathcal{F}_n^{(\alpha)}(r)$ becomes a constant
both when $r\rightarrow 0$ and  when $r\rightarrow 1$, as already highlighted above.
Thus, the factor  $\mathcal{F}_n^{(\alpha)}(r)$ makes smooth the transition between the linear growth and the plateau regime.
Notice that the height of the plateau in $\Delta S_{A;\alpha}^{(n)} $ depends on the value of the compactification radius $R$.

Instead, for the decompactified boson, the occurrence of a non-trivial $\mathcal{F}_n^{(\alpha)}(r)$
provides  a new qualitative feature for the temporal evolution of the entanglement entropies.
In the regime given by $t\gg\tau_0$ and $\ell\gg\tau_0$,
by using \eqref{r-global-t-dep}, \eqref{eq:divergence_noncompact_r_to0} and  \eqref{eq:divergence_noncompact_r_to1},
for Neumann b.c. we find the following early time divergent behaviour 
\begin{equation}
\label{log-t-global-N}
   \frac{1}{1-n}\log  \widetilde{\mathcal{F}}_n^{\textrm{\tiny (N)}}(r)
   \approx 
   \frac{1}{2}\log \! \left[\frac{\ell}{2\tau_0}\left(1-\frac{2 t}{\ell} \right)\right]
   + \textrm{const}
\;\;\qquad\;\;
    t/\ell \ll 1/2
\end{equation}
while for Dirichlet b.c. we observe the following late time divergent behaviour 
\begin{equation}
\label{log-t-global-D}
     \frac{1}{1-n}\log  \widetilde{\mathcal{F}}_n^{\textrm{\tiny (D)}}(r)
     \approx 
     \frac{1}{2}\log \! \left[ \frac{\ell}{2\tau_0}\left(\frac{2 t}{\ell}-1 \right) \right]
        + \textrm{const}
     \;\;\qquad\;\;
    t/\ell \gg 1/2\,.
\end{equation}
These divergencies are due to the fact that the limit of $\widetilde{\mathcal{F}}_n^{(\alpha)}$ is not a constant
both when $r \to 0$ and when $r \to 1$, as already highlighted above in 
\eqref{eq:divergence_noncompact_r_to0} and  \eqref{eq:divergence_noncompact_r_to1}.
It would be interesting to observe these corrections to the temporal evolution after a global quench
through numerical analyses in harmonic chains or other lattice models. 
In the right panel of Fig.\,\ref{fig:QuenchGlobal},
the difference  $ \Delta S_{A;\alpha}(t/\ell) \equiv  S_{A;\alpha} -  S_{A;\alpha}\big|_{t/\ell=1/2}$
is shown  for both the types of b.c.,
and the same values for $\ell $ and $\tau_0$ adopted in the left panel of the same figure. 
The deviations from the dashed green curve, which corresponds to (\ref{ee-globalquench-known}),
are due to the terms in (\ref{log-t-global-N}) and (\ref{log-t-global-D}).

\subsection{Local quench}
\label{sec-local-quench}

As for the local quench, in the following we consider 
the temporal evolution of $\textrm{Tr} \rho_A^n$ for an interval $A$ of length $\ell$,
in the case where the first endpoint of the interval $A$ coincides with the defect,
which corresponds to the case III in the classification of local quenches introduced in \cite{Calabrese:2007mtj}
(the case IV can be also studied but we do not report it here for simplicity).
For this bipartition the BCFT approach predicts the following result
\be
\label{eq:local_quench_full_final}
\textrm{Tr} \rho_A^n = 
C_n^2 \,\epsilon^{2\Delta_n}
     \left( \frac{ \sqrt{ \ell^2+\rho^2-t^2+2\,\ell\rho\cosh \varphi -2 \,t \rho \sinh \varphi  }  }{ 4\,\rho\, \big(1+(t/\tau_0)^2 \big)( \ell + \rho \cosh\varphi )}  \, \right)^{\Delta_n}
     \frac{ \mathcal{F}_n^{(\alpha)}(r) }{ r^{\Delta_n} }
\ee
(whose derivation is discussed in the Appendix\;\ref{app:local_quench}), where 
\begin{equation}
\label{r-local-t-dep}
    r = \frac{
    \big( \ell/\tau_0 - \sqrt{1+(t/\tau_0)^2}\, \big)^2 +(\rho/\tau_0)^2 +2\, \rho/\tau_0 \, \big( \ell/\tau_0 - \sqrt{1+(t/\ell)^2}\, \big)  \cosh\varphi
    }{
        \big( \ell/\tau_0 + \sqrt{1+(t/\tau_0)^2}\, \big)^2 +(\rho/\tau_0)^2 +2\, \rho/\tau_0 \, \big( \ell/\tau_0 + \sqrt{1+(t/\ell)^2}\, \big)  \cosh\varphi
    }
\end{equation}
with 
\begin{equation}
\label{rho-over-ell-sq}
   \rho^2
    =
    \ell^2\,\sqrt{\big[ (\ell /\tau_0)^{-2}+1+ (t/\ell)^2\big]^2- 4 (t/\ell)^2 }
    \;\;\qquad\;\;
           \varphi \equiv \frac{1}{4}   \left| \, \log \! \left[  \frac{(\ell / \tau_0 + t / \tau_0)^2+1 }{ (\ell / \tau_0 - t / \tau_0)^2 + 1 } \right] \, \right|
\end{equation}
hence the r.h.s. of (\ref{eq:local_quench_full_final}) depends on the ratios $t/\ell$, $\ell/\tau_0$ and $t/\tau_0$.
The explicit expression of $\mathcal{F}_n^{(\alpha)}(r)$ are given in (\ref{Fn-main-res})
and 
the temporal evolution of $r$ in terms of $t/\ell$, for some assigned values of $\ell/\tau_0$, 
is shown in the right panel of Fig.\,\ref{fig:FnAppRatio}.

In the regime given by $t\gg\tau_0$ and $\ell\gg\tau_0$, 
the expression \eqref{r-local-t-dep} simplifies to 
\be
    r \; \xrightarrow{t,\ell \,\gg\, \tau_0} \;
    \left\{\begin{array}{ll}
    \displaystyle 
    \frac{\ell-t}{\ell+t} \hspace{.7cm} & t<\ell 
    \\
    \rule{0pt}{.8cm}
    \displaystyle   
   \frac{ \tau_0 ^2\, \ell ^2}{4 t^2 \left(t^2-\ell ^2\right)} \hspace{.7cm}  & t>\ell 
    \end{array} \right.
\ee
and for the moments  (\ref{eq:local_quench_full_final}) in this regime one obtains 
\begin{equation}
\label{local-large-t-ell}
    \Tr\rho_A^n= 
    \left\{
    \begin{array}{ll}
    \displaystyle
    C_n^2\left(\frac{\epsilon^2}{t^2} \, \frac{(\ell+t)^2}{t(\ell-t)} \, \frac{\tau_0}{4 \ell}\right)^{\Delta_n} \mathcal{F}^{(\alpha)}_n \! \left(\frac{\ell-t}{\ell+t}\right) 
    \hspace{1cm}
    & t<\ell
    \\ 
    \rule{0pt}{.9cm}
        \displaystyle
    C_n^2\,\bigg(\frac{\epsilon}{\ell}\bigg)^{2\Delta_n} & t>\ell  \,.
    \end{array}
    \right.
\end{equation}
Hence, for the entanglement entropy in this regime we find
\begin{equation}
\label{ee-localquench-approx}
    S_A= 
        \left\{
    \begin{array}{ll}
    \displaystyle
    \frac{c}{3} 
    \log \frac{t}{\epsilon}
    +\frac{c}{6} \log \frac{\ell}{\tau_0}
    +\frac{c}{6} \log \frac{4(\ell-t)}{\ell+t}
    +\mathcal{G}^{(\alpha)}_1 \! \left(\frac{\ell-t}{\ell+t}\right)
    +   C'_1
    \hspace{.9cm}& t<\ell
    \\ 
    \rule{0pt}{.8cm}
       \displaystyle
    \frac{c}{3} \log \frac{\ell}{\epsilon}+2 C_1^{\prime} & t>\ell 
        \end{array}
    \right.
\end{equation}
where $C'_1 \equiv \lim_{n\rightarrow 1} \frac{\log C_n}{n-1}$ and $\mathcal{G}_1^{(\alpha)}(y)$ has been defined in \eqref{eq:mathcalG-1-intro-definition}.
While for the compact boson $\mathcal{G}_1^{(\alpha)}(y)$ is not known, 
he decompactified boson can be studied by replacing $\mathcal{G}^{(\alpha)}_1(y)$ 
with  $-D_1'(y_\alpha)/2$ in (\ref{ee-localquench-approx}),
where $D_1'(y)$ is given by \eqref{D1-def}.

\begin{figure}[t!]
\vspace{-.5cm}
\hspace{-1.4cm}
  \includegraphics[width=1.16\textwidth]{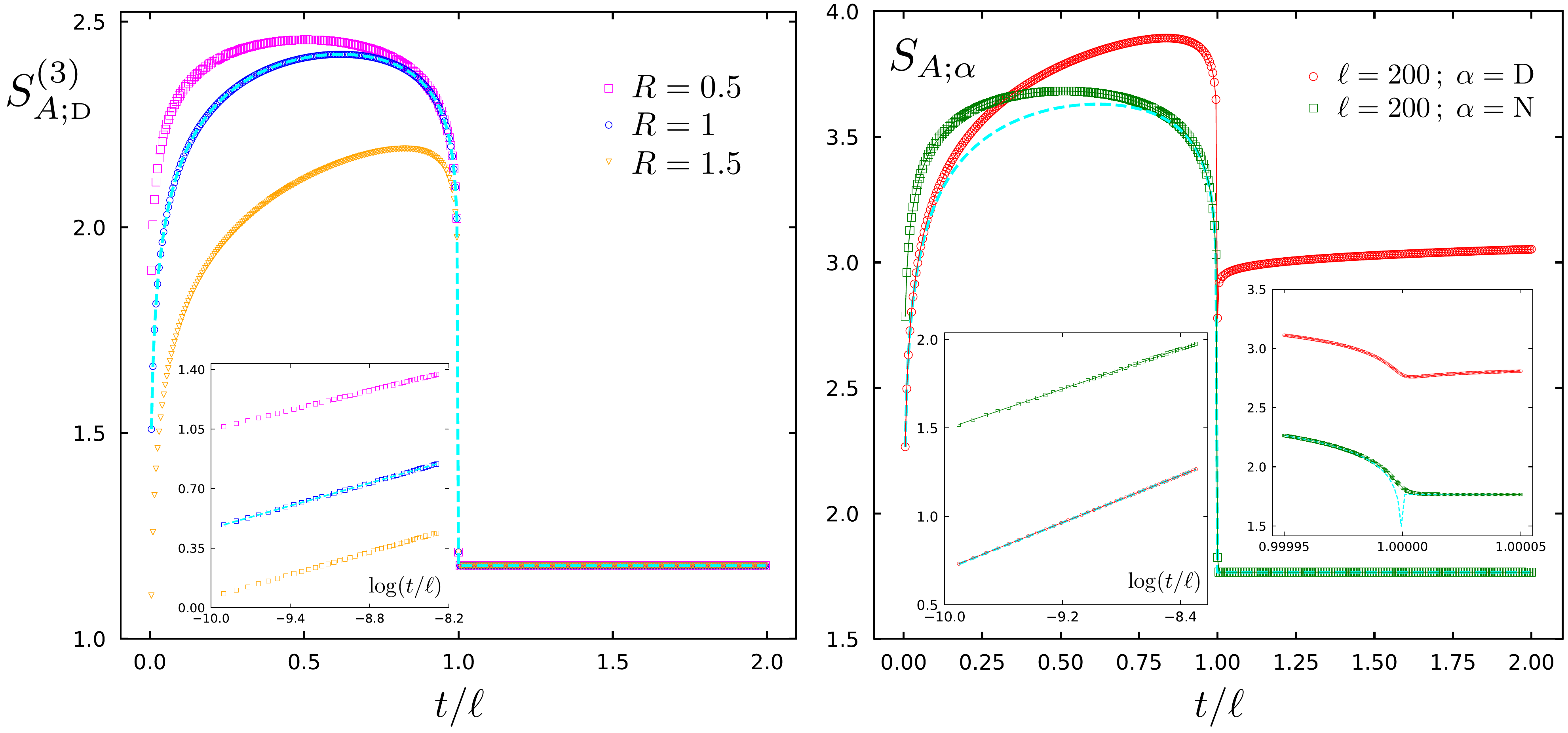}
 \caption{
 Temporal evolution of $S_{A; \alpha}^{n}$ after a local quench for an interval $A$ of length $\ell$ 
  whose first endpoint coincides with the defect, with $\tau_0=0.001$.
 Left: Compact boson with Dirichlet b.c. for $n=3$ (the inset zooms in on the early time regime).
 Right: Decompactified boson with either Dirichlet or Neumann b.c. (the two insets zoom in on the early time regime and on $t/\ell \approx 1$). 
 In both panels, the approximate results based on (\ref{ren-local-quench-known}) correspond to the dashed cyan line.
 }
  \label{fig:QuenchLocal}
\end{figure}

The expression \eqref{eq:local_quench_full_final} provides the temporal evolution of the R\'enyi entropies $S_{A;\alpha}^{(n)}$ 
after the local quench we are considering. 
In the left panel of Fig.\,\ref{fig:QuenchLocal} we have shown the case $n=3$ for the compact boson with Dirichlet b.c.,
$\ell=200$, $\tau_0=0.001$  and some assigned values of the compactification radius $R\in\{0.5,1,1.5\}$.
For Neumann b.c., the temporal evolution is qualitatively the same; hence we do not show it here.

In order to highlight the role of the factor $\mathcal{F}_n^{(\alpha)}$ 
in (\ref{eq:local_quench_full_final}), (\ref{local-large-t-ell}) and (\ref{ee-localquench-approx}),
we find it worth considering also 
 \begin{equation}
\label{ren-local-quench-known}
 \textrm{Tr} \rho_A^n
 \sim
    \left\{\begin{array}{ll}
    \displaystyle 
  \left(\frac{1}{t^2}\, \frac{\ell+t}{\ell-t} \,\frac{\tau_0}{4 \ell}\right)^{\Delta_n}  \hspace{1cm}  & t<\ell
          \\
    \rule{0pt}{.8cm}
    \displaystyle 
  \frac{1}{\ell^{2\Delta_n}}   \hspace{.7cm}  & t>\ell 
        \end{array} \right.
\end{equation}
which is obtained by setting $\mathcal{F}_n^{(\alpha)}$ equal to $1$ identically in (\ref{local-large-t-ell}) and dropping the factors given by the powers of $C_n$ and $\epsilon$.
The expression (\ref{ren-local-quench-known}) provides the dashed cyan curve in both the panels of Fig.\,\ref{fig:QuenchLocal}.
In the left panel, when $R\neq 1$ we observe a significant deviation with respect to \eqref{ren-local-quench-known}, 
i.e. from the dashed cyan curve.
Its  inset zooms in on the initial regime $\tau_0\ll t\ll\ell$, where $r\rightarrow 1$,
showing the expected perfect match with (\ref{ren-local-quench-known}) when $R=1$
and the need of a vertical shift when $R \neq 1$, due to the fact that $\mathcal{F}_n^{\textrm{\tiny (D)}}(r) \to R^{n-1}$ 
as $r\to 1$.

As for the decompactified boson, 
we have that  $\widetilde{\mathcal{F}}_n^{(\alpha)}$ diverges logarithmically when $r\rightarrow 0$ for Dirichlet b.c  
 and when $r\rightarrow 1$ for Neumann b.c. (see \eqref{eq:divergence_noncompact_r_to0} and  \eqref{eq:divergence_noncompact_r_to1}).
Hence, in the regime given by $t \gg\tau_0$ and $\ell\gg\tau_0$,
for Neumann b.c. we get the following early time divergence
\begin{equation}
\label{loc-quench-dec-N}
   \frac{1}{1-n}\log  \widetilde{\mathcal{F}}_n^{\textrm{\tiny (N)}}(r)\approx \frac{1}{2}\log\left|\log \frac{2t}{\ell}\right|
        +\textrm{const}
   \;\;\;\qquad\;\;\; 
   t/\ell\ll 1  
\end{equation}
while for Dirichlet b.c. we find a similar divergence, but at late time
\begin{equation}
\label{loc-quench-dec-D}
     \frac{1}{1-n}\log  \widetilde{\mathcal{F}}_n^{\textrm{\tiny (D)}}(r)
     \approx 
     \frac{1}{2}\log\left|\log \frac{2t^2}{\ell \tau_0}\right|
     +\textrm{const}
   \;\;\;\qquad\;\;\; \,
     t/\ell\gg 1   \,.
\end{equation}
It would be interesting to observe these corrections to the temporal evolution after a local quench
through numerical analyses in harmonic chains or other lattice models. 
In the right panel of Fig.\,\ref{fig:QuenchLocal}, 
we display the temporal evolution of the entanglement entropy after the local quench
given by (\ref{ee-localquench-approx}) (see the red and green curves)
and also the same quantity (\ref{ee-localquench-approx}) where the term $\mathcal{G}_1^{(\alpha)}$ has been removed
(see the dashed cyan curve).
Its first inset  zooms in on the early time regime,
where the curve corresponding to Dirichlet b.c. matches with the dashed cyan line,  
as expected from \eqref{eq:divergence_noncompact_r_to1};
while its second inset zooms in on $t/\ell \approx 1$,
displaying a smooth crossover between $t<\ell$ and $t>\ell$.

 
     \section{Conclusions}
     \label{sec:conclusions}

In this paper we have employed BCFT techniques to calculate the leading behaviour of the entanglement entropies 
of an interval in a critical one-dimensional system belonging to the Luttinger liquid universality class
in the presence of a boundary. 
We considered bipartitions characterised by an interval either on the half line 
or within a segment where the same b.c. are imposed at both its endpoints,
which is not adjacent to the boundary, as shown in Fig.\;\ref{figure-biparts-intro}.
Both Dirichlet b.c. and Neumann b.c. have been investigated. 

Our main results are the analytic expressions for the entanglement entropies reported in (\ref{SA-n-BCFT}),
with the functions $\mathcal{F}_n^{(\alpha)}(r)$ given in (\ref{Fn-main-res}),
which are written in terms of the Siegel theta function and of the period matrix (\ref{tau-matrix-element})
occurring also for the entanglement entropies of two disjoint intervals on the line for the compact massless scalar field \cite{Calabrese:2009ez}.
Our analysis extends to a generic value of the R\'enyi index $n$ the one performed in \cite{Estienne:2021xzo}, 
whose results are recovered when $n=2$. 
Furthermore, in the case of Dirichlet b.c.,
we have checked numerically that our analytical expression is compatible with the implicit result reported in \cite{Bastianello:2019yyc}.

In the decompactification regime, the analytic expressions found through BCFT and reported in Sec.\,\ref{subsec:decompactification_derivations}
have been compared with  the corresponding numerical results obtained in harmonic chains (see Sec.\,\ref{sec_HC_numerics}).
In the case of Dirichlet b.c. excellent agreement has been observed (see Fig.\,\ref{fig:FnDirichlet}, Fig.\,\ref{fig:Mutual-Info-dec-Dir} and  Fig.\,\ref{fig:FnDirichletLargeRenyi}), 
while for Neumann b.c. some discrepancies occur
(see Fig.\,\ref{fig:FnNeumann}, Fig.\,\ref{fig:logRnNeumann}, Fig.\,\ref{fig:Mutual-Info-dec-Neu} and Fig.\,\ref{fig:FnNeumannLargeRenyi})
that would be insightful to clarify in a quantitative way. 
However, we have also studied the UV finite quantity given by the difference between the entanglement entropy corresponding to different b.c. 
finding excellent agreement with the numerical data points from the harmonic chains (see Fig.\,\ref{fig:FnEntropyDiff}).

Finally, we have explored the consequences of our analytic results 
within the context of the BCFT approach to the quantum quenches \cite{Calabrese:2005in, Calabrese:2009qy, Calabrese:2007mtj}
(see Sec.\,\ref{sec-quenches}).
For global quenches (see Sec.\,\ref{sec-global-quench}) we find that, while in the case of the compact boson the main effect of the function $\mathcal{F}_n^{(\alpha)}$ is to smoothen 
the transition between the linear growth and the plateau regime (see the left panel of Fig.\,\ref{fig:QuenchGlobal})
for the decompactified boson it introduces a subleading logarithmic correction, either in the linear growth or in the plateau regime
(see (\ref{log-t-global-N}), (\ref{log-t-global-D}) and the right panel of Fig.\,\ref{fig:QuenchGlobal}).
For local quenches (see Sec.\,\ref{sec-local-quench}), the effect of $\mathcal{F}_n^{(\alpha)}$ is more relevant for the compact boson,
as highlighted in the left panel of  Fig.\,\ref{fig:QuenchLocal},
and, in the case of the decompactified boson, a subleading log-log correction occurs (see (\ref{loc-quench-dec-N}) and (\ref{loc-quench-dec-D})).
Both for global and local quenches, 
it would be insightful to observe the above mentioned subleading corrections 
also from the numerical data corresponding to quantum quenches in some lattice models.

Possible extensions of our analysis for the compact scalar
could involve mixed boundary conditions \cite{Estienne:2023tdw},
or non-vanishing temperature for the entire system,
or a non-vanishing mass \cite{Castro-Alvaredo:2008fni},
or a subsystem made by the union of a generic number of disjoint intervals
\cite{Casini:2009vk, Coser:2013qda, Rottoli:2022plr},
or spatially inhomogeneous backgrounds 
\cite{Dubail:2016tsc, Bastianello:2019yyc, Bastianello:2019ovv},
or defects \cite{Eisler:2010ep,Peschel:2012pe, Gutperle:2017enx, Mintchev:2020jhc, Roy:2021jus}.

Regarding the lattice calculations in harmonic chains discussed in Sec.\,\ref{sec_HC_numerics}, 
a critical task is to gain an analytical understanding of the effects of the zero mode. 
Such understanding could shed light on the discrepancies observed between BCFT expressions 
and the harmonic chain results when Neumann b.c. are imposed.

The spatial bipartitions presented in Fig.\,\ref{figure-biparts-intro} 
offer intriguing prospects for explorations in other compelling two-dimensional models, 
such as the Ising BCFT \cite{Cardy:1986gw} and more complex interacting BCFT models 
like the Liouville field theory \cite{Fateev:2000ik, Zamolodchikov:2001ah}.

Another intriguing direction of investigation involves going beyond the leading universal behavior of the entanglement entropy, which can be pursued for free fermions, 
like e.g. in the Schr\"odinger field theory at finite density \cite{eisler-peschel-13-spheroidal,Mintchev:2022xqh,Mintchev:2022yuo}. 
Another approach to compute subleading, finite-size corrections involves the use of excited twist-fields \cite{Estienne:2023tdw}.

For systems without boundaries, the application of Zamolodchikov's recursion relation \cite{Zamolodchikov:1987z} to twist fields has been used to study the entanglement entropy of disjoint intervals on a line \cite{Rajabpour:2011pt, Ruggiero:2018hyl}. Extending this analysis to the BCFT cases considered in our study would be a compelling endeavour.
Other interesting cases where it is worth studying the two-point functions of twist fields in the presence of boundaries
are suggested by further applications of the BCFT approach to quantum quenches
(see e.g. 
the geometry considered in \cite{Dubail-Stephan-11-local-quench}).

Furthermore, exploring the effect of physical boundaries on other entanglement quantifiers like the entanglement Hamiltonians and their spectra \cite{Casini:2011kv, Lauchli:2013jga, Cardy:2016fqc, Arias:2016nip, Tonni:2017jom, Alba:2017bgn, Eisler:2017cqi, Eisler:2018ez, Arias:2018tmw, Surace:2019mft, DiGiulio:2019cxv, DiGiulio:2019lpb, Eisler:2019rnr, Eisler:2020lyn, Javerzat:2021hxt, Eisler:2022rnp}, or the logarithmic negativity \cite{Vidal:2002zz, Calabrese:2012ew, Calabrese:2012nk, Calabrese:2014yza, Coser:2015eba, Coser:2015mta, Eisler:2016ez, DeNobili:2016nmj, Shapourian:2016cqu, Eisler:2015ez}, can provide further insights.

Lastly, exploring entanglement entropies in higher-dimensional BCFT models, where the shape of the subsystem plays a crucial role \cite{Klebanov:2012yf, Fonda:2014cca, Fonda:2015nma, Seminara:2017hhh, Seminara:2018pmr, Bueno:2021fxb}, is also relevant and opens new avenues for investigation.

     \subsection*{Acknowledgements}

     We are grateful to Viktor Eisler, Paul Fendley, Mihail Mintchev, Giuseppe Mussardo, Ivan Kostov, Gregory Schehr and Barton Zwiebach for useful discussions.
     We thank in particular Alvise Bastianello for helpful correspondence. 
     AR is grateful to SISSA for hospitality during part of this work.
     ET acknowledges the Galileo Galilei Institute (through the program {\it Reconstructing the Gravitational Hologram with Quantum Information}),
     the Institute Henri Poincar\'e, the Institut de Physique Th\'eorique at Universit\'e Paris-Saclay,
     the Laboratoire de Physique Th\'eorique et Hautes Energies at Sorbonne Universit\'e
     and the Center for Theoretical Physics at MIT
     for hospitality and financial support during part of this work.

     \appendix

\newpage
\section{An insight from the six-vertex model}
\label{subsec-XXZ-insight}

The Hamiltonian of the spin-$\tfrac{1}{2}$ XXZ spin chain with open b.c. is \cite{Affleck:1998a}
\begin{equation}
\label{XXZ-ham}
  H_{\textrm{\tiny XXZ}} 
  = 
  \sum_{j=1}^{N-1} \big(
  \sigma_j^x\sigma_{j+1}^x + \sigma_j^y\sigma_{j+1}^y + \Delta \,\sigma_j^z\sigma_{j+1}^z \big)
  -h_1 \sigma_1^x-h_N \sigma_N^x
\end{equation}
where $\sigma_j^{x,y,z}$ denote the Pauli matrices acting on the $j$-th site, $\{h_1,h_N\}$ are boundary fields
and the anisotropy parameter $\Delta$ lies in the critical regime $|\Delta| < 1$.
The model (\ref{XXZ-ham}) is a gapless one-dimensional quantum system belonging to the Luttinger liquid universality class. 
The related discrete 2D classical model is the six-vertex model on the square lattice with Boltzmann weights
\begin{center}
  \begin{tabular}{cccccc}
    \includegraphics{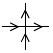} & \includegraphics[angle=180,origin=c]{Fig/6Va.pdf}
    & \includegraphics[angle=270,origin=c]{Fig/6Va.pdf} & \includegraphics[angle=90,origin=c]{Fig/6Va.pdf}
    & \includegraphics{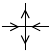} & \includegraphics[angle=90,origin=c]{Fig/6Vc.pdf} \\
    $a$ & $a$ & $b$ & $b$ & $c$ & $c$
  \end{tabular}
\end{center}
such that
\begin{equation}
  \frac{a^2+b^2-c^2}{2ab} = \Delta \,.
\end{equation}
The six-vertex model is mapped to a height model on the dual lattice, with height values $\varphi(p) \in \Zbb$, through the following simple rule:
for any pair on neighbouring faces $p,p'$, we set $\varphi(p')=\varphi(p)+1$ 
(resp. $\varphi(p')=\varphi(p)-1$) if $p'$ is above or to the right of 
(resp. below or to the left of) $p$, 
following the reasoning of \cite{Nienhuis:1984wm}. 
The arrow conservation around each vertex ensures that the height $\phi$ is well defined, up to an overall additive constant.

In the scaling limit, the height variable $\varphi(p)$ provides a real compact boson $\varphi(p) \to \phi(p)/(\pi R)$, 
with renormalized compactification radius $R=\sqrt{(2/\pi)\,\textrm{arccos}(-\Delta)}$  \cite{Affleck:1998a}. 
Note also that setting free b.c. in the XXZ chain ($h_1=h_N=0$) corresponds to Dirichlet b.c. in the compact boson \cite{Eggert:1992ur}, 
while turning on the boundary fields leads to Neumann b.c. \cite{Affleck:1998a}.

From the above mapping, we see that in any partition function the variation $\delta\phi$ along any trivial cycle is zero, 
whereas for non-trivial cycles or open paths joining two boundary points
this variation $\delta\phi$ is a multiple of $2\pi R$ 
(by convention, we only consider lattices such that these non-trivial cycles and open paths have even length, 
so that the configuration with $\delta\phi=0$ along each of these cycles and paths is allowed). In the scaling limit, this corresponds to (\ref{eq_winding}).

     \section{Period matrix} 
     \label{app_period_matrix}

In this Appendix we discuss the derivation of the period matrix \eqref{tau-matrix-element}.
This is a standard computation (see for instance \cite{Calabrese:2009ez}) that we report here for the sake of completeness.  

We are interested in the  period matrix of the Riemann surface $\Sigma_n$ 
     given by  the  $n$-sheeted covering surface over the Riemann sphere $\mathbb{CP}^1$ with four branch points at $0,x,1/\bar{x},\infty$
     and two branch cuts: one connecting $0$ to $x$ and another one connecting $1/\bar{x}$ to $\infty$. 
     Without loss of generality, we could assume $x$ real and positive.
     The Riemann surface $\Sigma_n$ can be defined as the algebraic curve  $w^n = z (z-x) (z- 1/\bar{x})^{n-1} $ with  $(z,w) \in \mathbb{C}^2$ 
     (up to compactification and resolution of the singularity at the origin by a blow-up).
     It is a compact Riemann surface with genus $n-1$. 
     This result can be obtained from the Riemann-Hurwitz theorem, which provides the Euler characteristics 
     of the $n$-sheeted covering $\mathscr{P}_n$ of a generic Riemann surface  $\mathscr{P}_1$ as 
     $\chi(\mathscr{P}_n) = n\, \chi(\mathscr{P}_1) - 4 (n-1)$,
     where $\chi (\mathscr{P}) = 2 - 2g$ is the Euler characteristic of a Riemann surface $\mathscr{P}$ without boundaries. 
     In our case $\Sigma_1 = \mathbb{CP}^1$; hence $\chi(\Sigma_n) = 4 -2n$.

     A compact Riemann surface of genus $g$ supports $g$ linearly independent holomorphic one-forms. 
     The ones for $\Sigma_n$ have been constructed explicitly and read \cite{Dixon:1986qv}
\be
       \omega_k = \frac{1}{z^{k/n} (z-x)^{(n-k)/n} (z-1/\bar{x})^{k/n}} \,dz 
     \;\;  \;\;\qquad\;\; \;\;
       k \in \{1 ,\cdots, n-1\} \,.
\ee
     This is the basis of holomorphic one-forms diagonalizing 
     the holomorphic deck transformation $f$ that sends the $j$-th sheet to the  $(j+1)$-th sheet.
     Indeed, we have $f^* \omega_k = \e^{-2 \pi \ri k/n} \,\omega_k$,
     being $f^*$ defined as the pullback of the deck transformation.

We work with the cycles $\mathcal{A}_j, \mathcal{B}_j$ and $\mathcal{C}_j$ described in Sec.\,\ref{sec_mirror_trick} 
and depicted in Fig.\,\ref{figure-cycles-disk}. 
Then, the period matrix $\boldsymbol{\tau}$ of $\Sigma_n$ is 
    \be
     \boldsymbol{\tau} =   \boldsymbol{A}^{-1}\!   \cdot \boldsymbol{B}         
     \;\;\;\;\;\;\;\qquad\;\;\;\;\;\;\;\;   
     \boldsymbol{A}_{k,j} = \int_{\mathcal{A}_j} \omega_k 
    \;\;  \qquad   \;\; 
     \boldsymbol{B}_{k,j} = \int_{\mathcal{B}_j} \omega_k \,.
    \ee
The integrals
    \be
       \oint_{\mathcal{C}_j}\omega_k 
       \;\;\; \qquad  \;\;\; 
       \oint_{\mathcal{B}_j}\omega_k
    \ee
     can be computed exactly \cite{Dixon:1986qv}. 
     Indeed, by using that $f^* \omega_k = \e^{-2 \pi \ri k/n} \,\omega_k$, we have 
    \be
       \oint_{\mathcal{C}_j}\omega_k \,=\, \e^{-2\pi \ri k (j-1)/n} \oint_{\mathcal{C}_1}\omega_k
       \;\;\;\qquad\;\;\;
        \oint_{\mathcal{B}_j}\omega_k \,=\, \e^{-2\pi \ri k (j-1)/n} \oint_{\mathcal{B}_1}\omega_k
    \ee
    which tell us that just two contour integrals in the r.h.s.'s must be evaluated. 
     These integrals can be computed by deforming the contours down to the branch cut
     and using the integral representation of the Gauss hypergeometric function. 
     This gives
\be
\label{eq_hyp_1}
       \oint_{\mathcal{C}_1}\omega_k 
       =\,
       2\, \ri   \sin (\pi k /n )  \int_0^{x} \!\!\omega_k 
       =\,
       2\pi \, \ri\, \bar{x}^{k/n}  F_{k/n}(|x|^2)
\ee
     and
\be
       \label{eq_hyp_2}
       \oint_{\mathcal{B}_1}\omega_k  
       =
        - \, 2\pi \ri \, \bar{x}^{k/n}  \,  \e^{-\pi \ri k/n }\, F_{k/n}(1-|x|^2)
\ee
     where we remind that $F_{a}(y) \equiv \, _2F_1(a, 1-a;1;y)$.
     The final expression for the generic element of the $(n-1)\times (n-1)$ period matrix reads
     \be
\label{tau-matrix-element-appendix}
\boldsymbol{\tau}_{i,j}\big(|x|^2\big)
=\,
\ri \, \frac{2}{n} \sum_{k=1}^{n-1} \sin(\pi k/n)\, \frac{F_{k/n}(1 -|x|^2)}{F_{k/n}(|x|^2)} \, \cos\!\big[2\pi k (i-j)/n\big]
\ee
     which is the result obtained in \cite{Calabrese:2009ez}. The period matrix (\ref{tau-matrix-element-appendix}) satisfies the following relation 
    \be
       \label{eq_modular_period_matrix}
       \boldsymbol{\tau} (1 - |x|^2) \,=\,  - \,\boldsymbol{T} \cdot  \boldsymbol{\tau}(|x|^2)^{-1} \cdot \boldsymbol{T}^{\textrm{t}}
    \ee
     where the generic element of the matrix $\boldsymbol{T}$ is $\boldsymbol{T}_{j,k} \equiv \delta_{j,k} - \delta_{j,k+1}$. 
     Similar relations have been found also in the Appendix C.3.3 of \cite{Coser:2013qda} in the case of two disjoint intervals on the line.


     \section{Green function in the presence of twist fields}
     \label{app_quantum_part}


     Consider a non-compact complex scalar field $\Phi$ on the unit disk $\mathbb{D}$
     satisfying the following condition after a rotation of the complex coordinate $z$ around a branch point at $z=0$ 
     \cite{Dixon:1986qv} 
     (see also (\ref{twist-field-condition}))
          \be
\label{twist-field-condition-app}
  \Phi(\e^{2i\pi} z, \e^{-2i \pi} \bar{z}) 
  = \e^{2\pi \ri\, k/n} \, \Phi(z, \bar{z}) 
\ee
     and a similar condition with opposite phase after a rotation around $z=x \in (0,1)$. 
     In order to have a simpler notation and to prevent any potential confusion with the mirror image, 
     we will use the notation $\Phi(z)$ instead of $\Phi(z, \bar{z})$.  
     However, it is important to note that this does not imply that the field $\Phi(z)$ depends holomorphically on the position $z$.

     These conditions define the occurrence of a twist field $\mathcal{T}_{k/n}$ at $z=0$
     and of its conjugate field $\mathcal{T}^{\dag}_{k/n}$ at $z=x$.
         The holomorphic part of the stress energy tensor $T(z)$ for the complex boson we are considering is
    \be
    \label{eq_T_point_splitting}
       T(z) =  -\frac{1}{2} : \! \partial_z \Phi(z) \, \partial_z  \overline{\Phi}(z)\! :
       \; \, =\;
        -  \lim_{w \to z} \left[  \frac{ \partial_z \Phi(z) \, \partial_w  \overline{\Phi}(w)}{2}   + \frac{1}{(z-w)^2} \right] .
    \ee

By adapting the analysis of \cite{Dixon:1986qv} to the case where a conformal boundary occurs, 
in the following we show that
\be
\label{ratio-2point-T-Twist-app}
        \frac{ 
   \langle T(z)\,\mathcal{T}_{k/n}(0)\, \mathcal{T}^{\dag}_{k/n}(x) \rangle_{_\mathbb{D}}
   }{  
   \langle \mathcal{T}_{k/n}(0)\, \mathcal{T}^{\dag}_{k/n}(x) \rangle_{_\mathbb{D}}
   }
   =
       \frac{h_{k/n}\, \big[x (1/\bar{x}-2 z)+z^2\big]^2}{\,z^2(z-x)^2(z-1/\bar{x})^2} 
       -  
       \frac{x(x-1/\bar{x})}{z(z-x)(z-1/\bar{x})} \,  \partial_x \log E^{(\alpha)}_{k/n}(x)
\ee
     where $h_{k/n}$ and $E^{(\alpha)}_{k/n}(x)$ have been defined in (\ref{h-E-k-defs}).
     Then, taking the residue of (\ref{ratio-2point-T-Twist-app}) as $z \to x$ leads to \eqref{ode-2point-twist}.

     Let us consider the following Green functions on the unit disk $\mathbb{D}$ 
    \bea
    \label{G-k-def-app}
       G_{k/n}(z, w) 
       &=&
       \frac{
       \langle \partial_z \Phi (z) \, \partial_w \overline{\Phi} (w)\,\mathcal{T}_{k/n}(0)\, \mathcal{T}^{\dag}_{k/n}(x) \rangle_{_\mathbb{D}}
       }{ 
       \langle \mathcal{T}_{k/n}(0)\, \mathcal{T}^{\dag}_{k/n}(x) \rangle_{_\mathbb{D}}
       } 
     \\
     \rule{0pt}{.8cm}
         \label{H-k-def-app}
        H_{k/n}(z, w) 
        &=&   
        \frac{
        \langle \partial_{\bar{z}} \Phi (z) \, \partial_w \overline{\Phi} (w)\,\mathcal{T}_{k/n}(0)\, \mathcal{T}^{\dag}_{k/n}(x) \rangle_{_\mathbb{D}}
        }{ 
        \langle \mathcal{T}_{k/n}(0)\, \mathcal{T}^{\dag}_{k/n}(x) \rangle_{_\mathbb{D}}
        }
    \eea
     where $|z| \leqslant 1$, $|w| \leqslant 1$ and the boundary condition on $\partial \mathbb{D}$ reads
    \be
       z \, \partial_z  \Phi = \pm  \,\bar{z} \, \partial_{\bar{z}}\Phi 
    \ee
     with $+$ and $-$ corresponding to Dirichlet b.c. and Neumann b.c. respectively. 
     As remarked above, the notation $H_{k/n}(z, w)$  does not mean that $H_{k/n}$ is holomorphic in $z$ (as a matter of fact, it is antiholomorphic in $z$). 
    On the other hand, the function $G_{k/n}$  is holomorphic in $z$ and $w$; 
     hence the Schwarz reflection principle can be employed to obtain its analytic continuation to the whole Riemann sphere, via
         \be
       G_{k/n}(z, w) 
        = 
        \left\{ 
        \begin{array}{ll}  
        \displaystyle
        \pm \, \frac{1}{z^2}  \; 
        \frac{
        \langle \partial_{\bar{z}} \Phi (1/\bar{z}) \, \partial_w \overline{\Phi} (w) \,  \mathcal{T}_{k/n}(0)\, \mathcal{T}^{\dag}_{k/n}(x) \rangle_{_\mathbb{D}}
        }{ 
       \langle \mathcal{T}_{k/n}(0)\, \mathcal{T}^{\dag}_{k/n}(x) \rangle_{_\mathbb{D}}
        }
        &  \qquad |z| \geqslant 1 , \, |w| \leqslant 1 
        \\
        \rule{0pt}{1cm}
         \displaystyle
         \pm \frac{1}{w^2} \;  
         \frac{
         \langle \partial_{z} \Phi (z) \, \partial_{\bar{w}} \overline{\Phi} (1/\bar{w}) \,  \mathcal{T}_{k/n}(0)\, \mathcal{T}^{\dag}_{k/n}(x) \rangle_{_\mathbb{D}}
         }{ 
         \langle \mathcal{T}_{k/n}(0)\, \mathcal{T}^{\dag}_{k/n}(x) \rangle_{_\mathbb{D}}
         }
         &  \qquad |z| \leqslant 1 , \, |w| \geqslant 1 
         \\
                 \rule{0pt}{1cm}
          \displaystyle
         \pm \frac{1}{w^2\, z^2}  \; 
         \frac{
         \langle \partial_{\bar{z}} \Phi (1/\bar{z}) \,\partial_{\bar{w}} \overline{\Phi} (1/\bar{w}) \,  \mathcal{T}_{k/n}(0)\, \mathcal{T}^{\dag}_{k/n}(x) \rangle_{_\mathbb{D}}
         }{ 
        \langle \mathcal{T}_{k/n}(0)\, \mathcal{T}^{\dag}_{k/n}(x) \rangle_{_\mathbb{D}}
         } 
         &  \qquad |z| \geqslant 1 , \, |w| \geqslant 1 \,.
         \end{array} 
         \right.
    \ee
     The Green function $H_{k/n}$, 
     which is antiholomorphic in $z$ and holomorphic in $w$, 
     can be  analytically continued in a similar way. 
Moreover, these two Green functions are related through a mirror relation as follows 
    \be
       \label{eq_mirror_relation}
       G_{k/n}(z, w)  =  \pm \frac{1}{z^2} \, H_{k/n}(1/\bar{z},w)
    \ee
     whenever they are well defined functions, namely for $z \neq w$ and  $z,w \notin \{ 0, x, 1/\bar{x},\infty\}$. 
     The r.h.s. of (\ref{eq_mirror_relation}) is indeed holomorphic in $z$ as $H_{k/n}(1/\bar{z},w)$ is the composition of two antiholomorphic functions, 
     namely $H_{k/n}(z,w)$ and $\sigma(z) = 1/\bar{z}$.

Now  one observes that $z^{k/n} (z-x)^{1-k/n} (z-1/\bar{x})^{k/n}\,G_{k/n}(z,w)$
is holomorphic on the whole Riemann sphere except for $z =w$, where a second order pole occurs.
    Hence, for $G_{k/n}(z,w)$ we must have 
    \be
        \label{G_k-f-expansion-app}
       G_{k/n}(z,w) = f_{k/n}(z)  \left( \frac{\alpha_{k/n}}{(z-w)^2} + \frac{\beta_{k/n}}{(z-w)} + \gamma_{k/n} \right) 
    \ee
     where $\alpha_{k/n},\beta_{k/n}$ and $\gamma_{k/n}$ are independent of $z$, while
    \be
    \label{fk-def-app}
       f_{k/n}(z) = \frac{1}{z^{k/n} (z-x)^{(1-k/n)} (z-1/\bar{x})^{k/n}} \,.
    \ee

     From the OPE  of $\partial_z \Phi (z) \,\partial_w \overline{\Phi} (w)$ as $z \to w$, we have that
      $ G_{k}(z,w)  =   - 2 /(z-w)^2 + O(1) $.
    This condition gives  $\alpha_{k/n}$ and $\beta_{k/n}$, which can be plugged into (\ref{G_k-f-expansion-app}), finding 
    \be
    \label{G-k-app-z}
       G_{k/n}(z,w) =  - \frac{2f_{k/n}(z)}{f_{k/n}(w)} 
       \left( \frac{1}{(z-w)^2} -  \frac{f'_{k/n}(w)/f_{k/n}(w)}{z-w} +  f_{k/n}(w) \,f_{1-k/n}(w) \,\mu_{k/n}(w) \right) 
    \ee
     for some unknown function $\mu_{k/n}$. 
     The same argument for the complex variable $w$ leads to
         \be
             \label{G-k-app-w}
       G_{k/n}(z,w) =  - \frac{2f_{1-k/n}(w)}{f_{1-k/n}(z)} 
       \left( \frac{1}{(z-w)^2} -  \frac{f'_{1-k/n}(z)/f_{1-k/n}(z)}{w-z} +  f_{1-k/n}(z) \,f_{k/n}(z) \,\mu_{1-k/n}(z) \right) .
    \ee
    Comparing (\ref{G-k-app-z}) and (\ref{G-k-app-w}), one obtains
    \be
    \label{mu-def_A}
       \mu_{k/n}(z) = \frac{k}{n} (z- x) + A_{k/n}(x) \;\;\; \qquad \;\;\;  A_{k/n} = A_{1-k/n} \,.
    \ee
Finally, combining \eqref{fk-def-app}-\eqref{mu-def_A}, we arrive to
    \bea
    \label{eq_G_kn}
       G_{k/n}(z,w)  
       &=& 
       \\
       & & \hspace{-2.2cm}
       =\, - \,2 \,f_{k/n}(z) f_{1-k/n}(w) 
        \left[\, \frac{k}{n} \;\frac{z (z-1/\bar{x}) (w - x)}{(z-w)^2}+ \left(1- \frac{k}{n} \right)  \frac{w (w-1/\bar{x})(z- x)}{(z-w)^2} + A_{k/n}(x) \right] .
       \nonumber
    \eea
    Then, from \eqref{eq_T_point_splitting} it follows that
    \be
        \frac{ 
   \langle T(z)\,\mathcal{T}_{k/n}(0)\, \mathcal{T}^{\dag}_{k/n}(x) \rangle_{_\mathbb{D}}
   }{  
   \langle \mathcal{T}_{k/n}(0)\, \mathcal{T}^{\dag}_{k/n}(x) \rangle_{_\mathbb{D}}
   }
   =
       \frac{h_{k/n}\, \big[x (1/\bar{x}-2 z)+z^2\big]^2}{\,z^2(z-x)^2(z-1/\bar{x})^2} 
       +
       \frac{A_{k/n}(x)}{z(z-x)(z-1/\bar{x})}
\ee
where the dependence on the boundary condition is encoded only in $A_{k/n}(x)$.

In the case of Neumann b.c., we can determine $A_{k/n}(x)$ by exploiting the fact that the field $\Phi(z)$ has no windings. 
In particular, the correlator $\langle \Phi (z) \,\partial_w \overline{\Phi}(w) \,\mathcal{T}_{k/n}(0)\, \mathcal{T}^{\dag}_{k/n}(x) \rangle_{_\mathbb{D}}$ 
must be a single-valued function of $z$. 
By using that $d \Phi = \partial_z \Phi \, dz +  \partial_{\bar{z}} \Phi (z) \, d\bar{z}$ 
and comparing with \eqref{G-k-def-app} and \eqref{H-k-def-app},
we find that  the above condition implies 
    \be
    \label{func-z-aux-appN}
 \oint_{\mathcal{C}_1}    \! \Big[\,  G_{k/n}(z,w) \, dz + H_{k/n}(z,w) \, d\bar{z} \,\Big] = 0 \,.
    \ee
To evaluate the l.h.s., one can first change variable to $\xi \equiv \sigma(z) = 1/\bar{z}$ and use \eqref{eq_mirror_relation} to get 
    \be
       \oint_{\mathcal{C}_1}  H_{k/n}(z,w) \,d\bar{z} 
       \, = \oint_{\mathcal{C}_1}  \! H_{k/n}(1/\bar{\xi},w) \,d(1/\xi) 
       \,=\,  \oint_{\mathcal{C}_1}  \! G_{k/n}(\xi,w) \,d \xi 
       \;\; \qquad \;\;  \textrm{Neumann b.c.} 
    \ee
  where $\sigma(\mathcal{C}_1) = \mathcal{C}_1$ has been employed. 
  Thus, the constraint \eqref{func-z-aux-appN} becomes
    \be
    \label{int-C1-neumann-app}
        \oint_{\mathcal{C}_1}  \! G_{k/n}(z,w) \,dz =0  
        \;\; \qquad \;\;  \textrm{Neumann b.c.} \,.
    \ee

In the case of Dirichlet b.c., the constraint \eqref{func-z-aux-appN} is automatically satisfied; indeed the above change of variable leads to 
    \be
       \oint_{\mathcal{C}_1}  H_{k/n}(z,w) \,d\bar{z} = -  \oint_{\mathcal{C}_1}  \! G_{k/n}(\xi,w) \,d \xi \,.
    \ee
Now, since  $\Phi(z)$ vanishes on all the components of the boundary, we have that
    \be
    \label{func-z-aux-appD}
 \int_{\mathcal{B}^+_1}    \! \Big[\,  G_{k/n}(z,w) \, dz + H_{k/n}(z,w) \, d\bar{z} \,\Big] = 0 \;\;\;\; \qquad \;\;\;\;  \textrm{Dirichlet b.c.} 
    \ee
     where $\mathcal{B}_1^+$ is the part of $\mathcal{B}_1$ located inside the white region in Fig.\,\ref{figure-cycles-disk}, which connects the two red points
     located on two different components of the boundary. 
     By using the above change of variable, the condition     \eqref{func-z-aux-appD} becomes
    \be
        \label{int-B1-dirichlet-app}
      \oint_{\mathcal{B}_1}  \! G_{k/n}(z,w) \,dz =0\;\;\;\; \qquad \;\;\;\;  \textrm{Dirichlet b.c.} 
    \ee 
   where we exploit the fact that mirror image of $\mathcal{B}_1^+$ is the remaining part of $\mathcal{B}_1$ with the opposite orientation.

From \eqref{eq_G_kn}, one finds that the constraints \eqref{int-C1-neumann-app} and  \eqref{int-B1-dirichlet-app} become 
    \be
    \label{eq_A_constraint}
       \oint_{\mathcal{C}}  f_{k/n}(z)  \left[\, \frac{k}{n} \; \frac{z (z- 1/\bar{x}) (w - x)}{(z-w)^2}+ \left(1- \frac{k}{n} \right)  \frac{w (w-1/\bar{x})(z- x)}{(z-w)^2} + A_{k/n}(x) \,\right] dz \, =\, 0
    \ee
where the contour is either $\mathcal{C} = \mathcal{C}_1$ for Neumann b.c.
or $\mathcal{C} = \mathcal{B}_1$ for Dirichlet b.c.;
hence
    \bea
    \label{A-int-derivation-app}
       A_{k/n}(x)  \oint_{\mathcal{C}}  f_{k/n}(z)\,dz 
      & = &
      \\
      & &
      \hspace{-1.5cm}
      =\;
        - \oint_{\mathcal{C}}  f_{k/n}(z) \left[\, \frac{k}{n} \; \frac{z (z- 1/\bar{x}) (w - x)}{(z-w)^2}+ \left(1- \frac{k}{n} \right)  \frac{w (w-1/\bar{x})(z- x)}{(z-w)^2} \,\right] dz \,.
        \nonumber
    \eea
The analysis of this equation has  been already carried out in \cite{Dixon:1986qv}.
However, in the following we report a detailed derivation for the sake of completeness.

The first important feature to highlight is that the r.h.s.  of \eqref{A-int-derivation-app} does not depend on $w$. This follows from the relation
    \bea
       \frac{\partial}{\partial w} \left[ f_{k/n}(z)  \left( \frac{k}{n} \frac{z (z- 1/\bar{x}) (w - x)}{(z-w)^2}+ \left(1- \frac{k}{n} \right)  \frac{w (w-1/\bar{x})(z- x)}{(z-w)^2}\right) \right]
       &=&
       \\
       \rule{0pt}{.8cm}
       && 
       \hspace{-3.4cm}
       =\,
       -        \frac{\partial}{\partial z} \left[  \frac{ z(z-x)(z-1/\bar{x}) f_{k/n}(z)}{(z-w)^2} \right] .
       \nonumber
    \eea
    Since (\ref{A-int-derivation-app}) is independent of $w$,
    we can choose convenient points for $w$ on the Riemann sphere.
    In particular, for $w= x$ we obtain
        \be
            \label{eq:w=x}
       A_{k/n}(x)  \oint_{\mathcal{C}}  f_{k/n}(z)\,dz 
        = 
       - \left(1- \frac{k}{n} \right) x \left(x - 1/\bar{x} \right) \oint_{\mathcal{C}}  \frac{ f_{k/n}(z) }{z- x} \,dz  \,.
    \ee
     Now one observes that the definition  of $f_{k/n}(z)$ in (\ref{fk-def-app}) straightforwardly leads to 
        \bea
\partial_x f_{k/n}(z) =  \left(1- \frac{k}{n} \right)\frac{ f_{k/n}(z) }{z- x}
    \eea
which can be employed in (\ref{eq:w=x}), finding that 
    \be
    \label{integ-A-final-gen}
      A_{k/n}(x)   = - \left(x - 1/\bar{x} \right) x \; \partial_x  \log \oint_{\mathcal{C}} f_{k/n}(z)  dz  \,.
    \ee
Since the integrals $ \oint_{\mathcal{C}} f_{k/n}(z)  dz $ have already been evaluated in \eqref{eq_hyp_1} and \eqref{eq_hyp_2}, we arrive to 
    \be
    \label{eq_A_k}
       A_k(x)  =  - \left(x - 1/\bar{x} \right) x \,\partial_x  \log E^{(\alpha)} _{k/n}(x)
    \ee
     where $E^{(\alpha)} _{k/n}(x)$ has been defined in (\ref{h-E-k-defs}). 
     This concludes the derivation of \eqref{ratio-2point-T-Twist-app}, whose residue at $z \to x$ leads to \eqref{ode-2point-twist}.

\section{Limiting regimes}
\label{app-limits}

\subsection{Compactified boson}
\label{app-limits-compact}

Let us consider the limiting regime where the interval is adjacent to the boundary. 

The entanglement entropies of the interval $A=[0,v]$ for a BCFT with central charge $c$
defined on the segment $[0,L]$ are
\cite{Calabrese:2004eu}  (see also (\ref{SA-adjacent}))
\begin{equation} 
\label{eq:EE_oneinterval}
  S_A^{(n)}
  = \frac{c}{12} \left( 1+\frac{1}{n} \right)
  \log \! \left[  \frac{2L}{\pi \epsilon} \, \sin \! \left( \frac{\pi v }{L} \right) \right]  +\log (g)
\end{equation}
up to subleading terms, where $g$ is the Affleck-Ludwig boundary entropy \cite{Affleck:1991tk}.
Considering the massless compact boson, which has $c=1$,
in the following we recover (\ref{eq:EE_oneinterval}) for this model
by taking the limit $u \to 0$
of $  \langle \mathcal{T}_{n}(u)\, \mathcal{T}^{\dag}_{n}(v) \rangle_{_{\mathbb{S}}}^{(\alpha)}$.
This provides an important consistency check of our BCFT results in (\ref{Fn-main-res}).

By employing (\ref{theta-prod-id}), the expressions in (\ref{Fn-main-res})  can be written respectively as follows 
\be
\label{Fn-app-version}
    \mathcal{F}_n^{\textrm{\tiny (D)}}(r)  = R^{n-1} \,\frac{\Theta\big(\! -\! R^2 \boldsymbol{\tau}(r)^{-1}\big)}{\Theta\big(\! -\! \boldsymbol{\tau}(r)^{-1}\big)} 
    \;\;\;\; \qquad \;\;\;\;
    \mathcal{F}_n^{\textrm{\tiny (N)}}(r)  = \left(\frac{2}{R}\right)^{n-1} \, \frac{\Theta\big(\! -\! 4 \boldsymbol{\tau}(r)^{-1}/R^2\big)}{\Theta\big(\! -\! \boldsymbol{\tau}(r)^{-1}\big)} 
\ee
where the prefactors can be expressed in terms of the ground state degeneracies \cite{Affleck:1991tk} for this model, which are given by  \cite{Oshikawa:1996dj,Affleck:1998a}
\begin{equation}
  g_{\textrm{\tiny D}} \equiv \sqrt{\frac{1}{R}} 
  \;\;\;\; \qquad \;\;\;\;
   g_{\textrm{\tiny N}} \equiv \sqrt{\frac{R}{2}}
\end{equation}
for  Dirichlet b.c. and Neumann b.c. respectively.

From (\ref{tilde-r-ratio-def}), 
we have that $r \to 1^-$ when $u\rightarrow 0^+$. 
Taking this limit in (\ref{Fn-app-version}), 
one finds
\begin{equation}
\label{Fn-r=1-limit}
\mathcal{F}_n^{\textrm{\tiny (D)}}(r) 
\to
 g_{\textrm{\tiny D}}^{2(1-n)} 
 \;\;\;\; \qquad \;\;\;\;
\mathcal{F}_n^{\textrm{\tiny (N)}}(r) 
\to
g_{\textrm{\tiny N}}^{2(1-n)} \,.
\end{equation}

The bulk-boundary Operator Product Expansion (OPE) of the twist field $\mathcal{T}_n(u)$ reads \cite{Sully:2020pza}
\begin{equation}\label{eq:bulkboundaryOPE}
\mathcal{T}_n(u)
\sim
\frac{ \mathcal{A}^{(\alpha)}_{n ; \mathbb{I}} }{ (2u)^{\Delta_n} }\; \mathbb{I}+\dots
\;\;\;\;\qquad\;\;\;
u \to 0^+
\end{equation}
where $\mathbb{I}$ denotes the identity operator on the boundary,
the dots indicate subleading contributions that have been neglected
and $\mathcal{A}^{(\alpha)}_{n ; \mathbb{I}}$ is the one-point structure constant,
which can be expressed in terms of the ground state degeneracy $g_{\alpha}$ 
as $\mathcal{A}^{(\alpha)}_{n ; \mathbb{I}} =g_{\alpha}^{1-n}$ \cite{Calabrese:2004eu, Sully:2020pza}.
Combining this observation with (\ref{Fn-r=1-limit}), 
we have that
$\mathcal{F}^{(\alpha)}_{n}(r)\to  \big(\mathcal{A}^{(\alpha)}_{n ; \mathbb{I}}\big)^{2} $ as $r\to 1$, for $\alpha \in \{\textrm{D}, \textrm{N}\}$.
Hence, by using also (\ref{Tr-n-segment}), one finds that
\begin{equation}\label{eq:ugoes0}
\langle \mathcal{T}_{n}(u)\, \mathcal{T}^{\dag}_{n}(v) \rangle_{_{\mathbb{S}}}^{(\alpha)}
    \sim
    \frac{ \big(\mathcal{A}^{(\alpha)}_{n ; \mathbb{I}}\big)^{2} }{
    (2u)^{\Delta_n} \,s(2v)^{\Delta_n} }
    +\dots \;.
\end{equation}
From (\ref{eq:bulkboundaryOPE}), it is straightforward to get that
\begin{equation}\label{eq:ugoes0_OPE}
  \langle \mathcal{T}_{n}(u)\, \mathcal{T}^{\dag}_{n}(v) \rangle_{_{\mathbb{S}}}^{(\alpha)}
    \sim
    \frac{ \mathcal{A}^{(\alpha)}_{n ; \mathbb{I}} }{(2u)^{\Delta_n}  }\;
      \langle  \mathcal{T}^{\dag}_{n}(v) \rangle_{_{\mathbb{S}}}^{(\alpha)} +\dots \;.
\end{equation}
Finally, consistency between (\ref{eq:ugoes0}) and (\ref{eq:ugoes0_OPE}) leads to 
$\langle \mathcal{T}^{\dag}_{n}(v) \rangle_{_{\mathbb{S}}}^{(\alpha)} = g_\alpha^{(1-n)} / s(2v)^{\Delta_n}$,
in agreement with (\ref{eq:EE_oneinterval}) and  (\ref{SA-adjacent}).

The limiting regime of small $x$ can be studied by employing the results of  \cite{Calabrese:2009ez} in a straightforward way. 
For Dirichlet b.c., this leads to 
\begin{equation}
\label{eq:subleadingDirichlet}
 \mathcal{F}^{\textrm{\tiny (D)}}_n(x) 
=
 1+n\left(\frac{x}{2n}\right)^{\frac{2}{R^2}}\,
 \sum_{j=1}^{n-1}\frac{1}{\left[\sin \left(\pi j/n\right)\right]^{2/R^2}}-\frac{x^2 }{12}\left(n-\frac{1}{n}\right)
 + \dots
 \qquad
 x\to 0
\end{equation}
while for Neumann b.c. one finds
\begin{equation}\label{eq:subleadingNeumann}
 \mathcal{F}^\textrm{\tiny (N)}_n(x)
 =
 1+n\left(\frac{x}{2n}\right)^{\frac{R^2}{2}}\,
 \sum_{j=1}^{n-1}\frac{1}{\left[\sin \left(\pi j/n\right)\right]^{R^2/2}}-\frac{x^2 }{12}\left(n-\frac{1}{n}\right)
  + \dots
  \qquad
 x\to 0
\end{equation}
where the dots correspond to subleading terms with respect to the ones reported in the r.h.s.'s, which have been neglected. 
These short distance expansions provide the two-point correlators of twist fields on the disk for small $x$;
indeed, $ \langle \mathcal{T}_{n}(0)\, \mathcal{T}^{\dag}_{n}(x) \rangle_{_\mathbb{D}}^{(\alpha)} 
\sim \mathcal{F}^{(\alpha)}_n(x) / x^{2\Delta_n}  $ as $x\to 0^+$.

\subsection{Decompactified boson}

\subsubsection{Normalization of two-point function of twist fields}
\label{app:normalization-decomp}

In the following we determine the overall normalization on the two-point function of twist fields for the decompactified boson. 
In the case of the compact boson, this constant has been fixed through the $x\rightarrow 0$ behaviour of this correlator
and such criterion will be employed also for the decompactified boson. 
The most important difference to take into account with respect to the case of the compact boson
is due to the continuous spectrum of primary operators, 
which is made by vertex operators $V_{\gamma}=\, : \! \e^{\ri  \gamma \phi(z,\bar{z})}\! :$
with scaling dimensions are $\Delta_{\gamma}=\gamma^2$, for $\gamma\in\mathbb{R}$. 
In the $\mathbb{Z}_n$ orbifold of this model \cite{Chen:2014ehg,dupic2018entanglement,Estienne:2022qpg}, 
the untwisted sector is built from the operators 
$\mathcal{V}_{\gamma}=V_{\gamma_1}\otimes \dots\otimes V_{\gamma_n}$
with $\gamma=\{\gamma_1,\dots,\gamma_n\}$ 
(only $\mathbb{Z}_n$ invariant linear combinations of such fields are local in the orbifold but this subtlety can be ignored at this level)
and the identity field $\mathbb{I}=\mathcal{V}_{0}$. 
Then, the leading $x\rightarrow 0$ behaviour in \eqref{two-point-twist-disk-ND-dec} will be obtained by considering the OPE of the twist fields, 
which can contain only fields of the above mentioned type because of the twist charge conservation 
and by their descendants, although the contribution of the latter ones is subleading the $x\rightarrow 0$ limit.

Since the spectrum of the untwisted primary fields is continuous, the OPE should be given by a weighted integral over
the $\mathcal{V}_{\gamma}$ operators. 
From previous works \cite{Chen:2014ehg}, we conjecture that the contribution of the untwisted primary operators to the OPE of conjugate twist fields reads
\begin{equation}
\label{eq:noncompactOPE}
    \mathcal{T}_n(0) \,\mathcal{T}^{\dagger}_n(x)
    \sim  \,
    \frac{1}{x^{2\Delta_n}}
    \int_{\mathbb{R}^n} \prod_{i=1}^{n} d \gamma_i \;
    \delta\bigg(\sum_i {\gamma}_i\bigg)  \, x^{\gamma^2} \,\mathcal{C}^{\mathcal{V}_{\gamma} }_{\mathcal{T}_n,\mathcal{T}^{\dagger}_n} \,  \mathcal{V}_{\gamma}(0)
\end{equation}
where $\mathcal{C}^{\mathcal{V}_{\gamma} }_{\mathcal{T}_n,\mathcal{T}^{\dagger}_n} $ are structure constants 
and the Dirac $\delta$ function appears as a consequence of  $U(1)$ charge conservation in the non-compact boson CFT.
We work with the usual  conventions where $\mathcal{C}^{\mathbb{I} }_{\mathcal{T}_n,\mathcal{T}^{\dagger}_n}=1$.

Plugging (\ref{eq:noncompactOPE}) into the two-point functions (\ref{two-point-twist-disk-ND-dec}) to find the primary fields contributions to the limit $x\rightarrow 0$, 
for Neumann b.c. and Dirichlet b.c. we find 
\begin{equation}\label{eq:correlatoregeneric}
    \langle  \mathcal{T}_n(0)\, \mathcal{T}^{\dagger}_n(x) \rangle^{(\alpha)}_{_\mathbb{D}} 
    \underset{x\rightarrow 0}{\sim}  
    \frac{1}{x^{2\Delta_n}}
    \int_{\mathbb{R}^n} \prod_{i=1}^{n} d \gamma_i \,  \delta\bigg(\sum_i {\gamma}_i\bigg)\, x^{\gamma^2} \,
    \mathcal{C}^{\mathcal{V}_{\gamma} }_{\mathcal{T}_n,\mathcal{T}^{\dagger}_n} \langle\mathcal{V}_{\gamma}(0) \rangle^{(\alpha)}_{_\mathbb{D}} \,.
\end{equation}
Conveniently, the correlators in the r.h.s. factorize into one-point functions of the non-compact boson BCFT, namely
\begin{equation}
    \langle\mathcal{V}_{\gamma}(0) \rangle^{(\alpha)}_{_\mathbb{D}} =\prod_{i=1}^n \langle V_{\gamma_i}(0)\rangle^{(\alpha)}_{_\mathbb{D}}\,.
\end{equation}
For Neumann b.c. and Dirichlet b.c. where $\phi=\phi_0$, we have respectively \cite{Recknagel:2013uja} 
\begin{equation}\label{eq:seed_1pointfunction}
    \langle \mathcal{V}_{\gamma}(0)\rangle^\textrm{\tiny (N)}_{_\mathbb{D}}
    =
    \prod^{n}_{i=1} \delta(\gamma_i) =
    \delta(\gamma) 
    \;\;\qquad\;\;
 \langle \mathcal{V}_{\gamma}(0)\rangle^\textrm{\tiny (D)}_{_\mathbb{D}}
        =  \prod^{n}_{i=1}  \e^{\textrm{i}  \phi_0\gamma_i}
     =
    \exp{\!\bigg(\textrm{i}  \phi_0\sum_{i=1}^n \gamma_i\bigg)}=1
\end{equation}
where, in the case of  Dirichlet b.c., the dependence on $\phi_0$ cancels for the fields that contribute to the OPE \eqref{eq:noncompactOPE},
because of the $U(1)$ charge neutrality condition $\sum_{i=1}^n {\gamma_i}=0$.

For Neumann b.c. we find 
\begin{equation}\label{eq:neumann-xgoes0-app}
       \langle  \mathcal{T}_n(0) \mathcal{T}^{\dagger}_n(x) \rangle^\textrm{\tiny (N)}_{_\mathbb{D}} \underset{x\rightarrow 0}{\sim} \frac{1}{x^{2\Delta_n}} +\dots
\end{equation}
which gives $C_n^\textrm{\tiny (N)}=1$
after a comparison with the  limit $x\rightarrow 0$  of (\ref{two-point-twist-disk-N}).

For Dirichlet b.c., we have 
\begin{equation}\label{eq:Dirichletxgoes0}
      \langle  \mathcal{T}_n(0) \,\mathcal{T}^{\dagger}_n(x) \rangle^{\textrm{\tiny (D)}}_{_\mathbb{D}}  
      \underset{x\rightarrow 0}{\sim}  
      \frac{1}{x^{2\Delta_n} }
      \int_{\mathbb{R}^n} \prod_{i=1}^{n} d \gamma_i \,  \delta\bigg(\sum_i {\gamma}_i\bigg)\, x^{\gamma^2} \mathcal{C}^{\mathcal{V}_{\gamma} }_{\mathcal{T}_n,\mathcal{T}^{\dagger}_n}
\end{equation}
where the structure constant $\mathcal{C}^{\mathcal{V}_{\gamma} }_{\mathcal{T}_n,\mathcal{T}^{\dagger}_n}$ depends smoothly on $\gamma$. 
Indeed, it can be unfolded to a $n$-point correlator of vertex operators on the Riemann sphere in the non-compact boson CFT as follows
\begin{equation}
    \mathcal{C}^{\mathcal{V}_{\gamma} }_{\mathcal{T}_n,\mathcal{T}^{\dagger}_n}
    =
    n^{-\sum_i ^2 \gamma_i^2/2}
    \big\langle V_{\gamma_1}\!\big(\e^{2 \pi \textrm{i} / n}\big)  \, V_{\gamma_2}\!\big(\e^{4 \pi \textrm{i} / n}\big) \cdots V_{\gamma_n}\!\big(1\big)\big\rangle_{\mathbb{C}} \,.
\end{equation}
By employing the following well known CFT result \cite{DiFrancesco:1997nk}
\begin{equation}
    \left\langle V_{\gamma_1}(z_1) \cdots V_{\gamma_n}(z_n)\right\rangle_{\mathbb{C}}=\prod_ {i<j} |z_i-z_j|^{2 \gamma_i \gamma_j}
\end{equation}
we find that the behaviour of the structure constants close to $\gamma_i=0$ is 
\begin{equation}
     \mathcal{C}^{\mathcal{V}_{\gamma} }_{\mathcal{T}_n,\mathcal{T}^{\dagger}_n}\underset{\gamma_i\rightarrow 0}{\sim} 1
\end{equation}
which provides the leading $x\rightarrow 0$ behaviour in (\ref{eq:Dirichletxgoes0}) as follows. 
Integrating the Dirac delta function, we arrive to 
\begin{equation}\label{eq:asymptotic-xgoes0-2twist}
    \langle  \mathcal{T}_n(0) \,\mathcal{T}^{\dagger}_n(x) \rangle^{\textrm{\tiny (D)}}_{_\mathbb{D}}  
    \underset{x\rightarrow 0}{\sim}  
    \frac{1}{x^{2\Delta_n} }
    \int_{\mathbb{R}^{n-1}} \prod_{i=1}^{n-1} d \gamma_i \,\exp\bigg\{ \!
    -\frac{1}{2} \big|\log x^2 \big| \sum^{n-1}_{i,j=1} \gamma_i A_{i,j} \gamma_j 
    \bigg\}
\end{equation}
where we have introduced the matrix $A_{i,j} \equiv 1+ \delta_{i,j}$, whose determinant is $\det A=n$. 
The Laplace's method  allows us to find
the leading $x \to 0$ asymptotic behaviour of this integral
\begin{equation}\label{eq:twist-2-point-leading-decomp}
   \langle  \mathcal{T}_n(0) \,\mathcal{T}^{\dagger}_n(x) \rangle^{\textrm{\tiny (D)}}_{_\mathbb{D}} 
   \underset{x\rightarrow 0}{\sim} 
   \frac{1}{x^{2\Delta_n} }\; \frac{1}{\sqrt{n}} \left(\frac{2\pi}{|\log x^2|}\right)^{(n-1)/2} .
\end{equation}
The normalisation constant is obtained by comparing this result with the limit $x\rightarrow 0$ of (\ref{two-point-twist-disk-D}),
which can be found by observing that 
\begin{equation}\label{eq:twist-2-point-Dirichlet-det-limit}
     \langle  \mathcal{T}_n(0) \,\mathcal{T}^{\dagger}_n(x) \rangle^{\textrm{\tiny (D)}}_{_\mathbb{D}} \underset{x\rightarrow 0}{\sim}  \frac{1}{x^{2\Delta_n} }\;  
     \frac{1}{\sqrt{\det( - \ri \boldsymbol\tau)}}\, \big(C_n^{\textrm{\tiny (D)}}+\dots \big) \,.
\end{equation}
By using the identities reported in Sec.\,4.5 of \cite{Calabrese:2009ez} for $\det( -\textrm{i}\boldsymbol\tau(x))$ 
and (\ref{theta-prod-identities}), we get
\begin{equation}\label{eq:smalldetxgoes0}
  \frac{1}{ \sqrt{\det( -\textrm{i}\boldsymbol\tau(x))}}=\prod_{k=1}^{n-1} \sqrt{\frac{F_{k/n}(x^2)}{F_{k/n}(1-x^2)}}\underset{x\rightarrow 0}{\sim}  \frac{1}{\sqrt{n}} \left(\frac{2\pi}{|\log x^2|}\right)^{(n-1)/2} .
\end{equation}
Finally, the comparison of (\ref{eq:twist-2-point-Dirichlet-det-limit}) and (\ref{eq:twist-2-point-leading-decomp}) gives $C_n^{\textrm{\tiny (D)}}=1$.

\subsubsection{The $x\rightarrow 1$ regime}
\label{app-x=1-regime}

It is worth investigating also the  behaviour of (\ref{two-point-twist-disk-ND-dec}) as $x\rightarrow 1$
because it provides information about the one-point structure constants of the twist fields. 
From the normalized two-point functions in (\ref{two-point-twist-disk-ND-dec-normalized}), 
since $\Theta(-\boldsymbol{\tau}^{-1}(x))\rightarrow 1$ as $x\rightarrow 1$,
  for Dirichlet b.c.  we have
\begin{equation}\label{eq:Dirichlet-leading-appendix}
      \langle  \mathcal{T}_n(0) \,\mathcal{T}^{\dagger}_n(x) \rangle^{\textrm{\tiny (D)}}_{_\mathbb{D}}
       \underset{x\rightarrow 1}{\sim} 
       \frac{1}{(1-x^2)^{\Delta_n} } \;.
\end{equation}
For Neumann b.c., by using (\ref{theta-prod-id}) in the second equation of (\ref{two-point-twist-disk-ND-dec-normalized}), we get
\begin{equation}\label{eq:Neumann-leading-appendix}
       \langle  \mathcal{T}_n(0) \mathcal{T}^{\dagger}_n(x) \rangle^\textrm{\tiny (N)}_{_\mathbb{D}} 
       \underset{x\rightarrow 1}{\sim} 
              \frac{\sqrt{\det( -\ri \boldsymbol\tau(x))}}{(1-x^2)^{\Delta_n} }
\end{equation}
where, from (\ref{eq:smalldetxgoes0}), we have that
\begin{equation}
\det( -\textrm{i}\boldsymbol\tau(x))=\prod_{k=1}^{n-1} \frac{F_{k/n}(1-x^2)}{F_{k/n}(x^2)}\underset{x\rightarrow 1}{\sim} \frac{1}{n} \left(\frac{2\pi}{|\log (1-x^2)|}\right)^{(n-1)} .
\end{equation}

Now we compare \eqref{eq:Neumann-leading-appendix} with the corresponding $x\rightarrow 1$ behaviour coming from the bulk-boundary OPE of $\mathcal{T}^{\dagger}_n(x)$ fields. 
In this case, the OPE has contributions only from  the untwisted boundary fields.
This comparison is done by using some facts about the boundary field spectrum  of the non-compact boson BCFT  \cite{Recknagel:2013uja},
which we denote $\psi^{(\alpha)}(\e^{\textrm{i}\theta})$, where $\theta\in(0,2\pi)$ is a coordinate parametrizing the boundary of the unit disk $\mathbb{D}$. 

For Dirichlet b.c., the spectrum of the boundary operators only contains the boundary identity operator $\psi_0^{\textrm{\tiny (D)}}$ and its descendants;
indeed, the constraint $\phi_0= \textrm{const}$ holds along the boundary. 
For Neumann b.c., any boundary vertex operator $\psi^\textrm{\tiny (N)}_{\gamma} =\; : \! \exp\! \big[\textrm{i}\gamma\phi(\e^{\textrm{i}\theta}) \big] \!\! :$ 
with charge $\gamma \in \mathbb{R}$ and scaling dimension $\Delta_{\gamma}=\gamma^2/2$ is allowed. 
In the $\mathbb{Z}_n$ orbifold of this BCFT, for Dirichlet b.c. the untwisted sector of primary boundary fields consists of the identity operator $\Psi^{\textrm{\tiny (D)}}_{0}$,
while for Neumann b.c. we have $ \Psi^\textrm{\tiny (N)}_{\gamma}=\psi^\textrm{\tiny (N)}_{\gamma_1}\otimes\dots  \otimes \psi^\textrm{\tiny (N)}_{\gamma_n}$.

For Dirichlet b.c., the leading contribution  to the bulk-boundary OPE of $\mathcal{T}^{\dagger}_n$ is 
\begin{equation}\label{eq:DirichletOPE}
    \mathcal{T}^{\dagger}_n(x)\sim 
    \frac{ \mathcal{A}^{\textrm{\tiny (D)}}_{n,\mathbb{I} } }{ (1-x^2)^{\Delta_n} }\; \,\mathbb{I}(1) \,.
\end{equation}
In the case of Neumann b.c., 
there are  issues with the continuous spectrum of boundary operators similar to the ones identified in the Appendix\;\ref{app:normalization-decomp} for Dirichlet boundary conditions. 
To bypass them, we conjecture the following  bulk-boundary OPE
\begin{equation}\label{eq:NeumannOPE}
    \mathcal{T}^{\dag}_n(x)\sim  
    \frac{1}{(1-x^2)^{\Delta_n}}
    \int^{+\infty}_{-\infty} \prod_{i=1}^{n} d \gamma_i \; 
    \delta\bigg(\sum_i {\gamma}_i\bigg)\, 
    (1-x^2)^{\gamma^2/2} \,\mathcal{A}_{n,\Psi_{\gamma} }^\textrm{\tiny (N)} \Psi^\textrm{\tiny (N)}_{\gamma}(1) 
\end{equation}
where $\mathcal{A}^\textrm{\tiny (N)}_{n,\Psi_{\gamma} } $ are the bulk-boundary structure constants
and the $U(1)$ charge neutrality on the boundary is manifest.

Plugging (\ref{eq:DirichletOPE}) and (\ref{eq:NeumannOPE}) 
into the corresponding correlators in \eqref{two-point-twist-disk-ND-dec-normalized},
for Dirichlet b.c. we find 
\begin{equation}
\label{app-TT-corr-x=1}
      \langle  \mathcal{T}_n(0) \, \mathcal{T}^{\dagger}_n(x) \rangle^{\textrm{\tiny (D)}}_{_\mathbb{D}} \underset{x\rightarrow 1}{\sim} 
      \frac{      \big(\mathcal{A}^{\textrm{\tiny (D)}}_{n,\mathbb{I} }\big)^2}{(1-x^2)^{\Delta_n} }
\end{equation}
where we used that \cite{Estienne:2023tdw}
\begin{equation}
\langle   \mathcal{T}_n(0)  \rangle^{(\alpha)}_{_\mathbb{D}}=\mathcal{A}^{(\alpha)}_{n,\mathbb{I} } \,.
\end{equation}
The comparison of (\ref{app-TT-corr-x=1}) and  (\ref{eq:Dirichlet-leading-appendix}) fixes $ \big(\mathcal{A}^{\textrm{\tiny (D)}}_{n,\mathbb{I} }\big)^2=1$.
For  Neumann b.c., the result is more involved and reads
\begin{equation}
   \langle  \mathcal{T}_n(0) \,\mathcal{T}^{\dagger}_n(x) \rangle^\textrm{\tiny (N)}_{_\mathbb{D}} 
   \underset{x\rightarrow 1}{\sim} 
   \frac{1}{(1-x^2)^{\Delta_n}}
   \int^{+\infty}_{-\infty} 
   \prod_{i=1}^{n} d \gamma_i \;
   \delta\Big(\sum_i {\gamma}_i\Big)\, 
   (1-x^2)^{\gamma^2/2} 
   \big(\mathcal{A}_{n,\Psi_{\gamma} }^{\textrm{\tiny (N)}} \big)^2
\end{equation}
where we have used that \cite{Estienne:2023tdw}
\begin{equation}\label{eq:structure-constant-bulk-boundary}
\langle   \mathcal{T}_n(0) \, \Psi^{(\alpha)}_{\gamma}(1) \rangle^{(\alpha)}_{_\mathbb{D}}=\mathcal{A}^{(\alpha)}_{n,\Psi_{\gamma} } \,.
\end{equation}
The derivation is analogous to the one described in the Appendix\;\ref{app:normalization-decomp}. 
The correlator in (\ref{eq:structure-constant-bulk-boundary}) unfolds to a $n$-point function of boundary vertex operators on the unit disk \cite{Estienne:2023tdw} which should be smooth in $\gamma_i$. 
Since we are only interested in the leading behaviour as $x\rightarrow 1$,
we can approximate $\mathcal{A}^{\textrm{\tiny (N)}}_{n,\Psi_{\gamma}}\approx \mathcal{A}^{\textrm{\tiny (N)}}_{n,\mathbb{I}}$ for $\gamma_i\rightarrow 0$ small.
Then, by evaluating the Gaussian integral as done in Appendix\;\ref{app:normalization-decomp}, one arrives to
\begin{equation}
      \langle  \mathcal{T}_n(0)\, \mathcal{T}^{\dagger}_n(x) \rangle^{\textrm{\tiny (N)}}_{_\mathbb{D}} 
      \underset{x\rightarrow 1}{\sim}
      \frac{   \big(\mathcal{A}_{n,\Psi_{\gamma} }^{\textrm{\tiny (N)}} \big)^2 }{(1-x^2)^{\Delta_n}}
       \left(\frac{2\pi}{|\log (1-x^2)|}\right)^{(n-1)/2}
\end{equation}
which can be compared with (\ref{eq:Neumann-leading-appendix}), finding that $    \big(\mathcal{A}_{n,\Psi_{\gamma} }^{\textrm{\tiny (N)}} \big)^2=1$. 

Finally, by employing the fact that the one-point structure constants of twist fields are related to the ground state degeneracies $g_{\alpha}$ as  $\mathcal{A}^{(\alpha)}_{n,\mathbb{I}}=g_{\alpha}^{1-n}$,
we get 
\begin{equation}
\label{eq:ludwig_affleck_noncompact}
    g_{\textrm{\tiny D}}=1 
\;\;\;\;\qquad\;\;\;\;
    g_{\textrm{\tiny N}}=1
\end{equation}
which is compatible with well-established results in the  literature \cite{Recknagel:2013uja}.


\section{Details about the BCFT approach to quantum quenches}
\label{app-quenches}

In this Appendix, we report the derivation of the main expressions on the quantum quenches reported 
and discussed  in Sec.\,\ref{sec-quenches}.

\subsection{Global quench}
\label{app:global_quench}

In \cite{Calabrese:2005in} the temporal evolution after a global quantum quench determined by a critical Hamiltonian has been studied through BCFT.
The setup is given by a Euclidean BCFT defined in the strip $\mathbb{S}_{\tau_0} \equiv \{ w  \in \mathbb{C}\,; \, 0< \textrm{Im}(w) < 2\tau_0 \}$,
with the same conformally invariant boundary condition $\alpha$ imposed both on $\textrm{Im}(w) =0$ and on $\textrm{Im}(w) = 2\tau_0 $.
We are interested in the two-point function of twist fields in $\mathbb{S}_{\tau_0}$
placed at $w_1=\ri \tau$ and $w_2=\ri \tau +\ell$, with $\ell>0$; i.e. (see (\ref{Tr-n-segment-00}) and (\ref{eq_scaling_tau}))
\be
\label{ren-app-global-strip}
\textrm{Tr} \rho_A^n \,= C^2_n \, \epsilon^{2\Delta_n}\,\,
\langle  \mathcal{T}_n(w_1)\,\mathcal{T}^{\dag}_n(w_2) \rangle_{_{\mathbb{S}_{\tau_0} }}  \,.
\ee
The strip $\mathbb{S}_{\tau_0}$ is sent into the upper half plane 
$\mathbb{H} \equiv \{ z  \in \mathbb{C}\,; \, \textrm{Im}(z) \geqslant 0 \}$
through the map $z=\e^{\pi w/(2\tau_0)}$.
As a consequence, the positions of the twist fields  in $\mathbb{H}$ are $z_1= \e^{\ri \theta}$ and $z_2=f \e^{\ri \theta} $,
where $f \equiv \e^{\pi\ell/(2\tau_0)}$.

In order to employ the two-point correlation function of twist fields in the unit disk $\mathbb{D}$
given in (\ref{two-point-twist-disk-final-finiteR})  and (\ref{two-point-twist-disk-ND-dec-normalized}),
first we map the upper half plane $\mathbb{H} $ into the unit disk (parameterised by the complex coordinate $\zeta$)
through $z \to \zeta=\tfrac{z-z_1}{z-\bar{z}_1}$ 
and then we perform a global rotation by $\textrm{arg} \big( \tfrac{z_2-z_1}{z_2-\bar{z}_1} \big)$.
This procedure leads to the following expression for the harmonic ratio  $r=|x|^2$
\be
\label{r-global-fin}
   r    =\left| \frac{z_2-z_1}{z_2-\bar{z}_1}\right|^2
   =\, \frac{(f-1)^2}{f^2-2 f \cos (2 \theta)+1}
      =\, 
      \frac{2 \big[ \sinh\!\big(\pi\ell / (4 \tau_0)\big) \big]^2}{
      \cosh\!\big(\pi\ell / (2 \tau_0)\big) - \cosh(\pi\tau / \tau_0)}
\ee
(the last expression agrees e.g. with the cross ratio $\eta_{1,2}$ in \cite{Coser:2014gsa})
and the following expression for (\ref{ren-app-global-strip})
\be
\label{ren-app-global-disk}
\textrm{Tr} \rho_A^n  = 
C^2_n \, \epsilon^{2\Delta_n}\;  
\frac{ J }{ [r(1-r)]^{\Delta_n}  }\; \mathcal{F}_n^{(\alpha)}(r) 
\ee
in terms of the functions given in \eqref{Fn-main-res} for the compact boson
and in \eqref{F-DN-dec-def} for the decompactified boson,
where $J$ includes all the jacobian terms associated to the various mappings and reads
\begin{equation}
\label{J-complex-app}
    J=     
   \left(\frac{\pi}{2\tau_0}\right)^{2\Delta_n}  
   \left| \frac{(z_2-\bar{z}_2)(z_1-\bar{z}_1)}{(z_2-\bar{z}_1)^2\,(z_1-\bar{z}_1)^2} \right|^{\Delta_n} 
   \big|z_1 z_2\big|^{\Delta_n}\,.
\end{equation}

\begin{figure}[t!]
\vspace{-.5cm}
\hspace{-1.1cm}
  \includegraphics[width=1.08\textwidth]{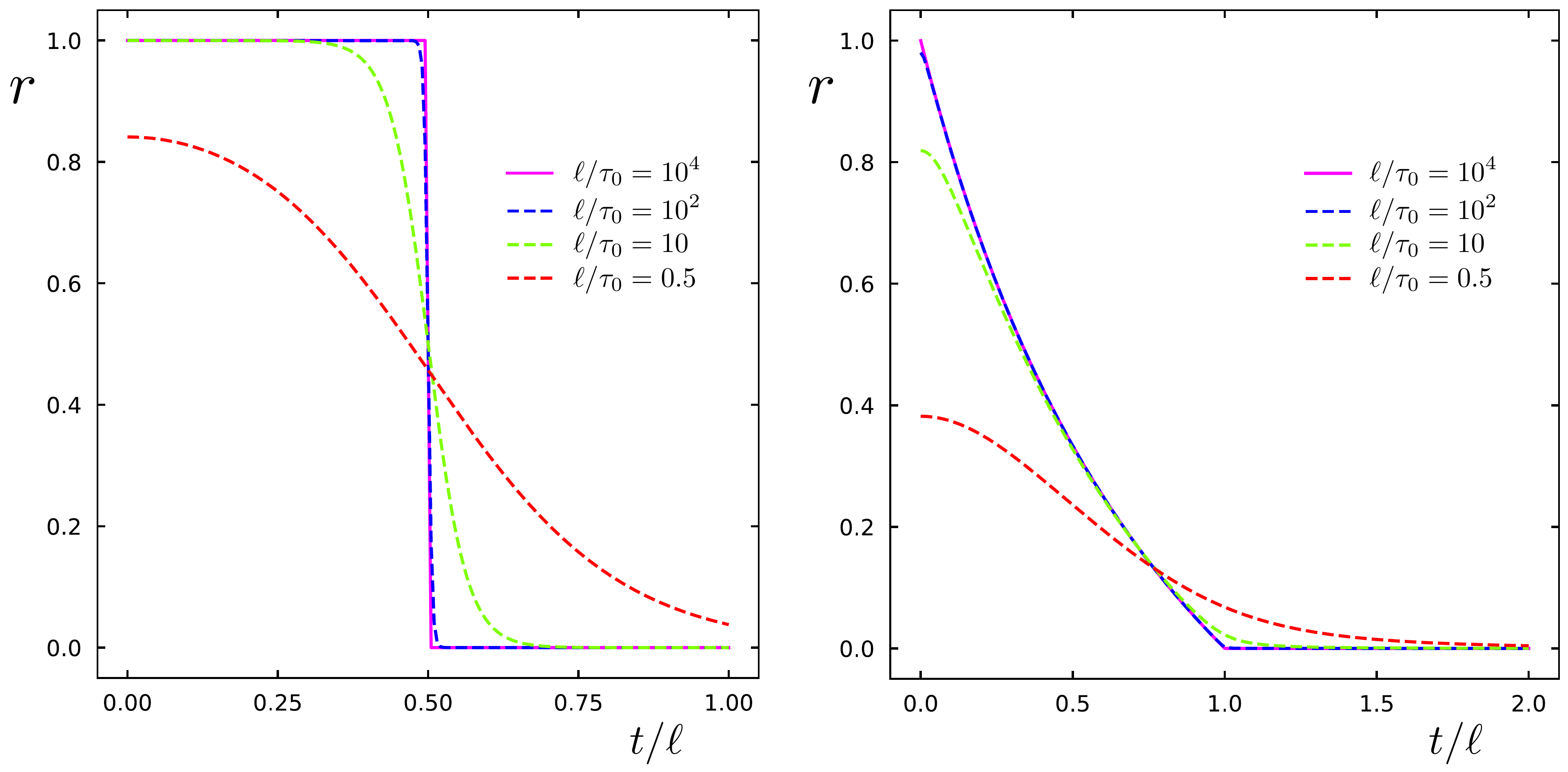}
 \caption{
 Temporal evolution of the harmonic ratio $r$
 for the global quench (left panel, see (\ref{r-global-t-dep}))
 and  for the local quench (right panel, see (\ref{r-local-t-dep})),
 for different values of $\ell/\tau_0$.
  }
  \label{fig:FnAppRatio}
\end{figure}

The crucial step introducing the physical time $t$ is the analytic continuation $\tau\rightarrow \tau_0+\ri t$.
This brings (\ref{r-global-fin}) into (\ref{r-global-t-dep}), which is shown in the left panel of Fig.\,\ref{fig:FnAppRatio} for different values of $\ell / \tau_0$.
Notice that a step function is obtained when $\ell / \tau_0 \gg 1$.
Performing the analytic continuation $\tau\rightarrow \tau_0+\ri t$ in (\ref{J-complex-app}), after some algebra one finds
\begin{equation}
     J = 
     \left(\frac{\pi}{2\tau_0}\right)^{2\Delta_n}  
     \left(\frac{f}{f^2+2 f \cosh (\pi t/\tau_0)+1}\right)^{\Delta_n} .
\end{equation}
By using this expression and \eqref{r-global-t-dep} in \eqref{ren-app-global-disk}, we arrive to (\ref{ren-global-quench-final}).

\subsection{Local quench}
\label{app:local_quench}

As for the local quench, in the following we review the BCFT analysis performed in \cite{Calabrese:2007mtj}
and adapt it in order to employ the analytic expressions for the two-point functions of the twist fields reported in Sec.\,\ref{sec_cft_results}.

In \cite{Calabrese:2007mtj} the temporal evolution after a local quench of
the moments of the reduced density matrix $\rho_A$ of an interval of length $\ell$ are investigated by considering
the two-point correlator of twist fields
in the Euclidean geometry $\mathbb{C}_{\tau_0}$ defined as the complex plane (parameterised by the complex coordinate $w$)
where boundaries are introduced by removing the slits corresponding to $(-\ri \infty,-\ri\tau_0]$ and $[\ri \tau_0, +\ri\infty)$.
In the case we are interested in, the defect coincides with the first endpoint of the interval;
hence the twist fields are located at $w_1=\ri \tau$ and $w_2=\ell+ \textrm{i} \tau$. 
Thus we have (see (\ref{Tr-n-segment-00}) and (\ref{eq_scaling_tau}))
\be
\label{ren-app-local-cutplane}
\textrm{Tr} \rho_A^n \,=\,C^2_n \, \epsilon^{2\Delta_n}\,\,
\langle  \mathcal{T}_n(w_1)\,\mathcal{T}^{\dag}_n(w_2) \rangle_{_{\mathbb{C}_{\tau_0} }}  \,.
\ee
The domain $\mathbb{C}_{\tau_0}$ is mapped into the right half plane 
$\widetilde{\mathbb{H}} \equiv \{ z  \in \mathbb{C}\,; \, \textrm{Re}(z) \geqslant 0 \}$
 through the following conformal map
\begin{equation}
z=\frac{w}{\tau_0}+\sqrt{\left(\frac{w}{\tau_0}\right)^2+1}
\;\;\;\qquad\;\;\; 
w=\tau_0\,\frac{z^2-1}{2z} 
\end{equation}
hence the positions of the twist fields in $\widetilde{\mathbb{H}} $ are
\begin{equation}
   z_1= \ri \,\tau/\tau_0 + \sqrt{1-(\tau/\tau_0)^2}
   \;\;\;\qquad\;\;\;
       z_2 = \frac{\ell }{\tau_0} + \frac{\ri \,\tau}{\tau_0} + \frac{\rho}{\tau_0}\, \e^{\ri \theta}
\end{equation}
where 
\begin{equation}
\label{rho-theta-euc}
    \rho^2 \equiv \sqrt{\left(\tau_0^2+\ell^2-\tau^2\right)^2+4 \ell^2 \tau^2}
     \;\;\; \qquad \;\;\; 
     \theta \equiv \frac{1}{2} \arctan \frac{2 \ell \tau}{\tau_0^2+\ell^2-\tau^2} \,.
\end{equation}
The factor coming from the Jacobians occurring in the transformation of the two-point correlator reads
\be
\label{calJ-def-app}
    \mathcalJ
    \equiv 
    \left|\frac{dz}{dw}\right|^{\Delta_n}_{w_1}\left|\frac{dz}{dw}\right|^{\Delta_n}_{w_2}
        =\,
        \left( \frac{\sqrt{(\ell+\rho \cos \theta)^2+(\tau+\rho \sin \theta)^2}}{\tau_0\rho\sqrt{\tau_0^2-\tau^2}}\, \right)^{\Delta_n} .
\ee

Then, one employs the conformal map $z\rightarrow \zeta=\frac{z-z_1}{z+\bar{z}_1}$,
which sends the right half plane $\widetilde{\mathbb{H}} $ into the unit disk $\mathbb{D}$
parameterised by the complex variable $\zeta$.
Hence the location of the twist fields in $\mathbb{D}$ are the points $\zeta_1=0$ and $\zeta_2=\tfrac{z_2-z_1}{z_2+\bar{z}_1}$ 
and the corresponding factor occurring in the transformation of the two-point function reads
\be
\label{tilde-calJ-def-app}
      \widetilde{\mathcalJ}
      \equiv 
     \left|\frac{d\zeta}{dz}\right|^{\Delta_n}_{z_1}\left|\frac{d\zeta}{dz}\right|^{\Delta_n}_{z_2}
            = \, \frac{1}{|z_2+\bar{z}_1|^{2\Delta_n}}
            \,=
            \left[\frac{\tau_0^2}{(\ell+\sqrt{\tau_0^2-\tau^2})^2+\rho^2+2\rho(\ell+\sqrt{\tau_0^2-\tau^2})\cos\theta}\right]^{\Delta_n} \! .
\ee
Performing also a rotation by $\textrm{arg}\big( \tfrac{z_2-z_1}{z_2+\bar{z}_1}\big)$, 
one arrives to the correlators in (\ref{two-point-twist-disk-final-finiteR}) in $\mathbb{D}$ with
the harmonic ratio given by 
\begin{equation}
\label{r-app-local}
    r=|x|^2=\left| \frac{z_2-z_1}{z_2+\bar{z}_1}\right|^2
    =
    \frac{\big( \ell-\sqrt{\tau_0^2-\tau^2} \,\big)^2 + \rho^2+2\rho\,\big(\ell-\sqrt{\tau_0^2-\tau^2}\, \big)\cos\theta}{ \big(\ell+\sqrt{\tau_0^2-\tau^2}\, \big)^2+\rho^2+2\rho\,\big(\ell+\sqrt{\tau_0^2-\tau^2}\, \big)\cos\theta} \,.
\end{equation}
By using also that $1-r=\tfrac{(z_1+\zb_1)(z_2+\zb_2)}{|z_2+\bar{z}_1|^2} $, we find that (\ref{ren-app-local-cutplane}) can be written as 
\begin{equation} 
\label{eq:localquench_fullresult} 
\aver{ \mathcal{T}_n(w_1)\,\mathcal{T}_n^{\dagger}(w_2 ) }_{\mathbb{C}_{\tau_0}}
=   
\frac{ \mathcalJ \, \widetilde{ \mathcalJ} }{ \big[r(1-r)\big]^{\Delta_n} }\;
\mathcal{F}_n^{(\alpha)}(r) \,.
\end{equation}

The crucial step in the analysis of \cite{Calabrese:2007mtj} which introduces the physical time $t$
is the analytic continuation $\tau\rightarrow \ri t$.
By employing the identity $\arctan(z)=\tfrac{1}{2\ri}\log\tfrac{\ri-z}{\ri+z}$, 
we have that the expressions in (\ref{rho-theta-euc}) after the analytic continuation become respectively 
\be
\label{rho-theta-app-t}
    \rho^2 = \sqrt{\left(\tau_0^2+\ell^2+t^2\right)^2-4 \ell^2 t^2},
\;\;\;\qquad\;\;\;
\theta =\frac{\ri }{4} \log  \! \left[  \frac{(\ell / \tau_0 + t / \tau_0)^2+1 }{ (\ell / \tau_0 - t / \tau_0)^2 + 1 } \right]  .
\ee
From these expressions and (\ref{r-app-local}), one obtains (\ref{r-local-t-dep}) and (\ref{rho-over-ell-sq}).
The temporal evolution of (\ref{r-local-t-dep}) is shown in the right panel of Fig.\,\ref{fig:FnAppRatio}
for some assigned values of $\ell/\tau_0$.

After the analytic continuation, (\ref{calJ-def-app}) and (\ref{tilde-calJ-def-app}) become respectively 
\bea
\label{app-local-R}
     \mathcalJ
     & = & 
     \left[\,\frac{\sqrt{ \ell^2+\rho^2-t^2+2\ell\rho\cosh|\theta|-2 \rho t\sinh |\theta|  }}{\tau_0\rho\,\sqrt{\tau_0^2+t^2}}\,\right]^{\Delta_n}
     \!\! \equiv 
     \mathcal{R}^{\Delta_n}
\\
\label{app-local-tilde-R}
  \widetilde{\mathcalJ}
  & = &
  \left[\, \frac{\tau_0^2}{(\ell+\sqrt{\tau_0^2+t^2})^2+\rho^2+2\rho(\ell+\sqrt{\tau_0^2+t^2})\cosh |\theta|}\,\right]^{\Delta_n}
  \!\! \equiv 
     \widetilde{\mathcal{R}}^{\Delta_n} \,.
\eea
Combining the above results, we arrive to (\ref{eq:local_quench_full_final}).

The result of \cite{Calabrese:2007mtj} for the quench we are considering, 
i.e. the case III in their classification, 
is obtained from \eqref{eq:local_quench_full_final} 
by first setting $\mathcal{F}_n^{(\alpha)}(r)$ equal to $1$
and then considering the regime where $ t \gg\tau_0$ and $\ell\gg\tau_0$.
Taking these limits in (\ref{r-local-t-dep}) and (\ref{rho-over-ell-sq}), one obtains
\be
\rho \xrightarrow{t,\ell \,\gg\, \tau_0} \sqrt{|\ell^2-t^2|}
\;\;\qquad\;\;
 \rho \cosh |\theta| \xrightarrow{t,\ell \, \gg\, \tau_0} \max{(t,\ell)} 
\;\;\qquad\;\;
 \rho \sinh |\theta| \xrightarrow{t,\ell \,\gg\, \tau_0} \min{(t,\ell)}
 \ee
 and 
\be
    r \; \xrightarrow{t,\ell \,\gg\, \tau_0} \;
    \left\{\begin{array}{ll}
    \displaystyle 
    \frac{\ell-t}{\ell+t} \hspace{.7cm} & t<\ell 
    \\
    \rule{0pt}{.8cm}
    \displaystyle   
   \frac{ \tau_0 ^2\, \ell ^2}{\left(4 t^2\right) \left(t^2-\ell ^2\right)} \hspace{.7cm}  & t>\ell 
    \end{array} \right.
\;\;\;\;\qquad\;\;\;\;\;
    1-r \; \xrightarrow{t,\ell \,\gg\, \tau_0} \;
        \left\{\begin{array}{ll}
    \displaystyle 
    \frac{2t}{\ell+t} \hspace{.7cm}   & t<\ell
    \\
    \rule{0pt}{.7cm}
    \displaystyle  
    1 & t>\ell \,.
        \end{array} \right.
\end{equation}
Similarly, in this limiting regime, from (\ref{app-local-R}) and (\ref{app-local-tilde-R}) we get respectively
\be
   \mathcal{R}
    \; \xrightarrow{t,\ell \,\gg\, \tau_0} \;
    \left\{\begin{array}{ll}
    \displaystyle 
    \frac{2}{\tau_0\, t}  \hspace{.7cm} & t<\ell
    \\
    \rule{0pt}{.8cm}
    \displaystyle  
    \frac{ 1}{t(t-\ell)} \hspace{.7cm} & t>\ell 
        \end{array} \right.
\;\;\;\;\qquad\;\;\;\;
\widetilde{\mathcal{R}}
      \; \xrightarrow{t,\ell \,\gg\, \tau_0} \;
    \left\{\begin{array}{ll}
    \displaystyle 
  \frac{\tau_0^2}{4\ell(\ell+t)} \hspace{.7cm} & t<\ell
      \\
    \rule{0pt}{.8cm}
    \displaystyle 
  \frac{ \tau_0^2}{4t(\ell+t)} \hspace{.7cm} & t>\ell \,.
        \end{array} \right.
\ee
Combining these results, we arrive to 
\begin{equation}
 \frac{ \mathcal{J}\,  \widetilde{\mathcalJ }  }{[r(1-r)]^{\Delta_n} }
  \; \xrightarrow{t,\ell \,\gg\, \tau_0} \;
    \left\{\begin{array}{ll}
    \displaystyle 
  \left(\frac{1}{t^2} \, \frac{\ell+t}{\ell-t} \,\frac{\tau_0}{4 \ell}\right)^{\Delta_n}  \hspace{1cm}  & t<\ell
        \\
    \rule{0pt}{.8cm}
    \displaystyle 
  \frac{1}{\ell^{2\Delta_n}}   \hspace{.7cm}  & t>\ell 
        \end{array} \right.
\end{equation}
which is the expression reported in \cite{Calabrese:2007mtj}.
%

\bibliographystyle{nb}
\bibliography{refsBdyBoson}

\end{document}